\documentclass[pra,reprint,longbibliography,aps,onecolumn,notitlepage]{revtex4-1}

\usepackage{amsmath}
\usepackage{amssymb}
\usepackage{amsfonts}

\usepackage{graphicx}

\usepackage[
	pdftitle={Spheroidal and conical shapes of ferrofluid-filled capsules in magnetic fields},
         pdfauthor={Christian Wischnewski, Jan Kierfeld}
         ]{hyperref}

\let \vec \mathbf

\usepackage{xcolor}

\begin{document}

\title{Spheroidal and conical shapes of ferrofluid-filled capsules in magnetic fields}

\author{Christian Wischnewski}
\email{christian.wischnewski@tu-dortmund.de}
\author{Jan Kierfeld}
\email{jan.kierfeld@tu-dortmund.de}
\affiliation{Department of Physics,
  TU  Dortmund University, 44221 Dortmund, Germany}

\date{\today}
\begin{abstract}

We investigate the deformation of soft spherical 
elastic capsules filled with a ferrofluid  in external
uniform magnetic fields at fixed volume by a combination 
of numerical and analytical approaches. 
We develop a numerical  iterative  solution strategy based on nonlinear
elastic shape equations to calculate the  stretched capsule shape
numerically and  a coupled finite element and boundary element method to
solve the corresponding magnetostatic problem,
and employ analytical linear response theory, approximative 
energy minimization, and slender-body theory. 
The observed deformation behavior is qualitatively 
 similar to the deformation of ferrofluid droplets in uniform
magnetic fields.  Homogeneous magnetic fields
elongate the capsule, and a  discontinuous shape transition 
from a spheroidal shape  to a  conical
shape takes place at a critical field strength.  
We investigate how capsule elasticity modifies 
this  hysteretic shape transition.
We show that conical capsule shapes are possible but 
involve diverging stretch factors at the tips, which 
gives rise to rupture for real capsule materials.
In a slender-body approximation we find that the  
critical susceptibility above which conical shapes occur
for ferrofluid capsules is the same as for droplets.
At small fields capsules remain spheroidal, and we  characterize
the deformation  of spheroidal 
capsules  both analytically and numerically.
 Finally, we determine whether
 wrinkling of a spheroidal 
 capsule occurs during elongation in a magnetic field
 and how it modifies the  stretching behavior. We find the nontrivial
dependence between the extent of the wrinkled region  and capsule
elongation.
Our results can be helpful in quantitatively 
determining capsule or  ferrofluid material 
properties from magnetic deformation experiments. 
All results also  apply to elastic capsules filled 
with a dielectric liquid in an external uniform electric field. 
%

\end{abstract}
\maketitle

\section{Introduction}

Elastic capsules consist of a  thin elastic shell enclosing
 a fluid inside.  Elastic microcapsules are found in nature,
 for example, as red
blood cells or virus capsids. They can also be produced
artificially by various methods, for example,
interfacial polymerization at liquid-liquid interfaces or multilayer 
polyelectrolyte deposition \cite{Neubauer2013}. 
Artificially produced microcapsules are attractive systems
 for encapsulation and transport, for example, in 
  delivery and release systems.
 Their overall shape is often nearly spherical,
and the shell can be treated
as a two-dimensional elastic  solid with a curved equilibrium shape.
 In experiments and for applications, 
 elastic properties of capsules can be
tuned by varying size, thickness,  and shell materials.
For applications involving delivery by rupture of capsules
it is necessary to understand and 
characterize  the mechanical properties and elastic instabilities  
of capsules. 

The mechanical properties of  elastic capsules are governed
by the elastic shell, which is curved (typically spherical)
in its equilibrium shape.
 This gives rise to 
different characteristic instabilities in response to external
forces \cite{Fery04,Vinogradova2006,Neubauer2013}.
Thin elastic membranes bend much more easily than stretch.
This protects a curved equilibrium shape against deformations 
changing its Gaussian curvature and that is the reason for the 
stability of capsules under uniform compression. 
 In contrast to fluid drops, elastic capsules under uniform compression
fail in a buckling instability 
below a critical volume or critical internal pressure 
\cite{Gao01, Sacanna11, Datta2012, Knoche11, Knoche14, Knoche2014a}.
Buckling-type instabilities 
 can also be triggered by external
forces, for example, in  electrostatically 
driven buckling transitions of charged shells \cite{Jadhao14} or in 
hydrodynamic flows \cite{Boltz15}.
Under point force loads, for example, exerted by atomic force microscopy tips,
elastic capsules indent  linearly at small forces and assume 
buckled shapes in the  nonlinear 
regime at higher forces \cite{Fery04,Zoldesi08,Vella2012}.
As opposed to fluid droplets, elastic capsules can also 
develop wrinkles upon deformation 
\cite{Rehage2002,Vella2011,Aumaitre2013,Knoche13} 
if compressive hoop stresses arise.

Microcapsules can be manipulated and deformed in
 hydrodynamic flow \cite{Pieper1998,Rehage2002,Barthes-Biesel2011}, 
by micromanipulation using an atomic force 
microscope  \cite{Fery04,Vinogradova2006}
or micropipettes or capillaries \cite{Aumaitre2013,Knoche13}.
Another  promising route to exert 
mechanical forces and to actuate elastic capsules in a noninvasive manner 
 is via magnetic  or electric fields 
\cite{Degen08,Karyappa2014}. For magnetic fields 
this requires the presence of magnetizable material 
either in the shell or in the capsule interior. 
The whole capsule then acquires a magnetic  dipole moment,
which can be manipulated in external magnetic  fields. 
For actuation by electric fields the capsule has to contain polarizable 
dielectric material such that the capsule acquires an electrostatic dipole
moment, which can be manipulated by an electric field. 
Homogeneous  fields orient dipole moments but 
 also induce 
capsule deformations, which increase the size of the dipole moment
after orientation. 
Therefore, homogeneous fields always  lead to 
stretching and elongation of the capsule. 
Inhomogeneous  fields can also exert a net force 
on the capsule and induce directed motion  at 
fixed magnetic dipole moment along the field gradient. 

 In the following we focus on 
 spherical elastic capsules that are filled with a
(quiescent) magnetic fluid and deformed
 in homogeneous external magnetic fields. 
As  magnetic fluid we consider
a ferrofluid, which is a  liquid that is magnetizable by external
magnetic fields because it  consist of ferromagentic or ferrimagnetic 
nanoparticles
suspended in a carrier fluid. Because of the small particle size, ferrofluids
are stable against phase separation and show superparamagnetic behavior
\cite{Rosensweig85}.  Ferrofluids are used in technical and
medical applications \cite{Voltairas01, Holligan03, Liu07}. 
All our results also apply to elastic capsules filled with a 
(quiescent) dielectric fluid which are placed in a 
homogeneous external electric field.

The  problem of ferrofluid  droplets in 
uniform  external  magnetic fields has already been theoretically 
studied  in the
literature. Also, a  spherical ferrofluid droplet
 is elongated in the direction  of the magnetic field
for increasing field strength; the resulting elongated shape 
 was observed to be nearly spheroidal
\cite{Arkhipenko79}. Bacri and  Salin \cite{Bacri82} 
used the assumption of a spheroidal shape
for  a quite precise approximation of  the
elongation by minimizing the total energy. 
Although the droplet is only elongated by the field,  an 
abrupt shape transition is possible \cite{Bacri82}:
Beyond a threshold magnetic field strength 
the spheroidal droplet becomes unstable and 
  elongates discontinuously into a shape with 
conical tips. 
The conical shape is stabilized by a positive feedback between 
shape and magnetic field distribution:
 A sharp tip gives rise to a diverging field
strength at the tip, which  in turn  generates strong 
stretching forces stabilizing the sharp tip. 
The mechanism of forming sharp tips is
  reminiscent of the normal field instabilities (Rosensweig
  instabilities) of free planar ferrofluid surfaces in a perpendicular
  homogeneous magnetic field, which were first described by Cowley and
  Rosensweig \cite{Rosensweig67} and later extended by 
   a nonlinear stability analysis to study 
   subsequent pattern formation \cite{Boudouvis1987}.

The discontinuous shape transition to a conical shape exhibits
  hysteresis and only occurs above a critical susceptibility $\chi_c$
  of the ferrofluid.  In Refs.\ \cite{Li1994,Ramos1994} a value 
 $\chi_c \simeq 16.59$ was found below which no conical shape can exist; 
 a slender-body approximation in Ref.\ \cite{Stone99} gives 
 $\chi_c \simeq 14.5$.  Using the approximative energy minimization for
  spheroidal shapes of Bacri and Salin \cite{Bacri82} gives 
 $\chi_c \simeq 19.8$. The jump in droplet elongation at the transition to a
  conical shape depends on the magnetic susceptibility: Large
  elongation jumps are possible for high susceptibilities.  This
  behavior was investigated in more detail in several numerical
  studies \cite{Guillaume92, Lavrova04, Afkhami10}.  Apart from free
  ferrofluid droplets, the deformation behavior of sessile droplets on
  a plate \cite{Zhu11} or sedimenting ferrofluid drops in external
  fields \cite{Korlie08} have also been investigated for homogeneous
  external magnetic fields.  

Dielectric droplets  in a 
homogeneous external electric field exhibit the same 
shape transition from a spheroidal to a conical shape. 
For the  electric field, however,  free charges  exist, and 
conducting droplets are  easily realized experimentally. In fact, the first 
experimental observations of conical droplet shapes were
made for water droplets \cite{Zeleny1917} and soap bubbles
\cite{Wilson_Taylor1925}. In Ref.\ \cite{Ramos1994} it was
shown that also a conducting liquid droplet surrounded by 
an outer  conducting liquid in a homogeneous electric field
exhibits conical shapes  above  a corresponding  critical conductivity 
ratio $\sigma/\sigma_{\rm out} = 1+ \chi_c \simeq 17.59$. 
In the limit of an ideally  conducting droplet  with 
infinite  susceptibility  or 
infinite conductivity (both resulting in zero electric field 
inside the droplet), 
the conical solutions in Refs.\ \cite{Li1994,Ramos1994,Stone99} approach
Taylor’s cone solution with a
half opening angle $\simeq 49.3^\circ$ \cite{Taylor1964}.
Both for liquid metal (i.e., ideally conducting)  and 
dielectric droplets, the conic cusp formation has been studied dynamically
and dynamic self-similar solutions have been obtained
\cite{Zubarev2001,Zubarev2002}.
Fluid droplets, which are neither perfect 
conductors nor perfect insulators, 
disintegrate at higher external electric fields 
by emitting jets of fluid at the tip, from which 
small droplets pinch off. Also for this process, 
scaling laws for droplet sizes 
could be theoretically obtained \cite{Collins2008,Collins2013}.

In a ferrofluid-filled elastic capsule the 
 ferrofluid drop is enclosed by a thin elastic membrane,
which will modify the 
 transition from a spheroidal to a conical shape observed for droplets.
Such  ferrofluid-filled capsules 
  have  already been realized experimentally. 
   Neveu-Prin {\it et al.}\ \cite{NeveuPrin93} 
encapsulated ferrofluids by polymerization and analyzed
the  magnetization  behavior of the magnetic capsules. 
 Degen {\it et al.}\ \cite{Degen08} 
investigated experimentally elastic capsules filled with a magnetic
liquid in an external magnetic field. They used a 
 magnetic liquid 
consisting  of micrometer-sized magnetic particles 
that do not show the special
properties of ferrofluids but form long chains in the presence of
 external magnetic fields. 
These magnetite-filled elastic capsules  could be actuated 
to deform in a  magnetic field. 
A quantitative theoretical description of  their 
deformation is still missing.  
In Ref.\ \cite{Karyappa2014}, capsules
 filled with a dielectric liquid in an external 
electric field were investigated experimentally and theoretically 
with a focus on  small deformations.

We will describe the elastic shell by a   nonlinear
elastic model  based on a Hookean elastic energy density
for thin shells,
 assume axisymmetric capsules,  and
calculate the shape at  force equilibrium  by solving shape
equations as they have been derived in Refs.\ 
 \cite{Knoche11, Knoche13}. 
As stated above, homogeneous magnetic fields acting on ferrofluid-filled
 capsules
give rise to stretching and elongation of the capsule in order 
to increase the total dipole moment. Therefore, stretching tensions
are dominant in the elastic shell. This is why  we will 
consider the limiting case of vanishing 
 bending modulus and bending  moments 
 for most of the present work, which is commonly called 
the elastic membrane limit (as opposed to the elastic shell case). 

In our numerical approach, the magnetic field inside the capsule
is calculated using a coupled finite element and  boundary element method.
The capsule shape provides the geometric boundary for the 
field calculation. 
Vice versa, the magnetic field distribution 
couples to the shape equations via the magnetic surface stresses.
We solve the full coupled problem numerically by an iterative method.

We combine this numerical approach with several analytical approaches to
investigate the capsule deformation in a homogeneous 
magnetic field as a function of the magnetic field strength 
and  Young modulus of the capsule material.
First we characterize the linear deformation regime of spheroidal 
capsules for small fields both numerically and analytically.
Then we   answer the  question  to what extent 
the elastic shell will suppress the discontinuous 
spheroidal to conical 
shape transition of a ferrofluid droplet and 
whether elastic properties such as the Young modulus 
of the shell material can be used to tune and control 
the instability.
 We  show that  conical shapes can also  occur for capsules with
nonlinear Hookean membranes
but require diverging strains at the conical tips.
As a real elastic material is not able to support arbitrarily high strains,
we expect that diverging  local stretch factors at the capsule poles
indicate that real capsules tend to rupture close to 
the poles as soon as the conical shape is assumed.
Then the existence of a sharp discontinuous shape transition
into a conical shape provides an interesting route to trigger 
capsule rupture at the poles at rather well-defined magnetic
 (for ferrofluid-filled capsules)
or electric (for dielectric-filled capsules) field values. 
The subsequent rupture  process has some analogies to the onset of the 
disintegration of droplets in electric fields 
\cite{Collins2008,Collins2013}, but our 
static  approaches based on nonlinear Hookean material 
laws are not suited to model  the rupture process itself.

We find that the discontinuous  shape  transition 
between spheroidal and conical shapes 
with hysteresis effects and shape bistability 
is also present for  elastic ferrofluid-filled capsules.
Numerically, we obtain a complete classification of the shape transition
 in the parameter
plane of dimensionless magnetic field strengths (magnetic Bond number)
and the dimensionless ratio of the Young modulus of the shell material
and the surface tension of the ferrofluid. 
These findings are partly corroborated by an analytical 
approximative energy minimization extending 
 the  spheroidal shape  approximation 
of  Bacri and Salin  \cite{Bacri82}  
to ferrofluid-filled capsules.
For conical shapes we generalize the slender-body approximations 
of Stone {\it et al.}\ \cite{Stone99}, which allows us to 
quantify the divergence of local stretch factors at the capsule poles 
and to show that the same analytic formula as for ferrofluid droplets
governs the dependence of the cone angle on the magnetic 
susceptibility $\chi$ (or the dielectric susceptibility 
${\varepsilon}/{\varepsilon_{\rm out}}-1$ for a dielectric droplet with 
dielectric constant
$\varepsilon$  in a surrounding liquid with $\varepsilon_{\rm out}$).
In particular, we predict  the critical susceptibility $\chi_c$, above 
which the  hysteretic shape transition between  spheroidal and 
conical capsule shapes can be observed, to be 
{\it identical} to the critical value for ferrofluid or dielectric droplets.
We also find that, for elastic capsules,  
magnetic stretching can give rise to 
 wrinkling along the capsule equator region. 
 We predict the parameter range for the 
appearance of wrinkles and the extent of the wrinkled 
region on a spheroidal capsule 
depending on its elastic properties and its elongation.

\section{Theoretical model and numerical methods}
\label{sec:theory}

We start with   a ferrofluid drop suspended  in an external
nonmagnetic liquid of  the same density as the ferrofluid, which eliminates 
gravitational forces.  
Thus the drop is force-free except for the surface
tension $\gamma$, which forces the drop to be spherical and is balanced 
by internal pressure.  
If the drop is enclosed by an elastic shell,  for example, after 
a polymerization reaction at the liquid-liquid interface, 
we  have a spherical elastic capsule. We assume 
that the relaxed rest shape  of this capsule is  spherical 
with a rest radius $R_0$, which is given by the fixed 
volume $V_0 = 4\pi R_0^3/3$ of the droplet or capsule. 

\begin{figure}[hbtp]
\begin{center}
\includegraphics[width=0.25\textwidth,clip]{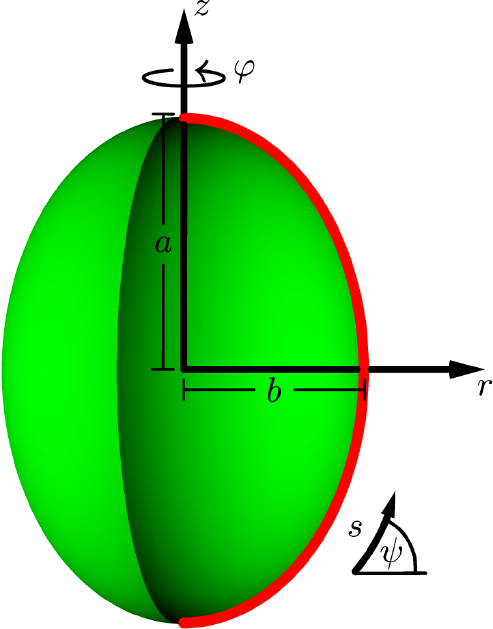}
\caption{ 
Illustration of the parametrization in cylindrical coordinates
  ($r,z,\varphi$) and the contour line with arc length $s$.  The
  complete capsule is obtained by revolution of the red contour line, while the
  angle $\psi$ describes its slope. This contour line is  calculated
  numerically.  The polar radius is called $a$, while $b$ denotes the
  equatorial radius.
}
\label{fig:geometry}
\end{center}
\end{figure}

After applying a uniform magnetic field $H_0\vec{e}_\text{z}$ in  
the $z$ direction,  
the resulting shape of the capsule becomes stretched in the 
$z$ direction, but the capsule shape and magnetic field 
distribution remain  axisymmetric  around
the $z$ axis. 
A uniform external magnetic field  causes  mirror-symmetric forces on the
capsule, resulting in a shape  with reflection  symmetry with respect 
to the plane $z=0$ (see  Fig.\ \ref{fig:geometry}).

\subsection{Geometry}
\label{sec:geometry}

We describe the axisymmetric  shell  using cylindrical
coordinates $r$, $z$, and $\varphi$. The capsule's shell is thin compared to its
diameter, so we consider the shell to be  a two-dimensional elastic 
surface. Because of  the axial  symmetry, 
we only need the contour line $r(z)$
to describe the whole capsule shape.

For our calculations, we parametrize the surface by the arc length $s_0$ of the
undeformed spherical contour 
 with $s_0 \in [0, L_0 = \pi R_0]$, starting at the lower apex
and ending at the upper apex.  Using the reflection symmetry, 
we only need half of
that interval, $s_0 \in [0, {L_0}/{2}]$, to describe the capsule's shape
completely.  In addition to the coordinates $r(s_0)$ and $z(s_0)$, we 
define a 
slope angle $\psi(s_0)$ by the unit vector
$\vec{e}_s$ following the contour line via $\vec{e}_s = (\cos\psi,
\sin\psi)$.

\subsection{Magnetostatics}

\subsubsection{Forces by the ferrofluid}

In order to calculate the shape of the capsule in an 
external magnetic field, we have
 to take the magnetic forces that are  caused by the ferrofluid 
on the capsule surface into
 account.  
Because we are interested in a static solution, we can 
assume that the
 fluid is at rest. Then the fluid  can only exert hydrostatic 
 forces  {\it normal}
 to the surface, while tangential components are zero. 
In order to calculate the  normal 
 magnetic  force density $f_m(r,z)$ on the surface, 
we use the magnetic stress tensor by Rosensweig,
\cite{Rosensweig85} 
\begin{equation}
  \label{eq:stress_tensor}
  f_m(r,z) = \mu_0 \int\limits_0^{H(r,z)}M(r,z)\text{d}H(r,z) 
   + \frac{\mu_0}{2}M_n^2(r,z).
\end{equation}
Here $M = |\vec{M}|$ is the absolute value 
of the magnetization and $M_n =\vec{M}\cdot \vec{n}$ 
its normal component ($\vec{n}$ is the outward unit 
normal to the capsule surface).  
Magnetization $M$ and magnetic field $H$ are taken 
on the inside of the capsule surface.

We assume a linear magnetization law
\begin{equation}
\vec{M} = \chi\vec{H}
\label{eq:linear}
\end{equation}
 with a susceptibility $\chi$ for the ferrofluid ($\chi = \mu-1$ in terms of
 its magnetic permeability $\mu$), which is justified for small fields 
$H\ll M_s/3\chi$, where $M_s$ is the saturation magnetization of the 
ferrofluid. 
References \cite{Boudouvis1988,Basaran1992} 
studied the behavior of drops with a nonlinear Langevin magnetization 
 (polarization) law. 
The saturation of the magnetization or polarization forbids sharp 
tips and leads to more rounded drops. 
It was shown, on the other hand, that the linear law is a very good 
approximation for small and even medium fields.
 This typically requires the maximum magnetic flux density
   $B_{\text{max}}=\mu_0H_{\text{max}}$ to be in a range of $50-100\,{\rm
   mT}$, depending on the specific fluid \cite{Chantrell1978,Zhu11}.
 For a linear magnetization we can rewrite Eq.\ \eqref{eq:stress_tensor} as
\begin{equation}
\label{fm}
 f_m(r,z) = \frac{\mu_0\chi}{2}\left[H^2(r,z) + \chi H_n^2(r,z)\right]
\end{equation}
(assuming $\chi_{\rm out}=0$ for the  external
non-magnetic liquid or using $\chi = {\mu}/{\mu_{\rm out}}-1$ 
in terms of the magnetic permeabilities $\mu$ of the ferrofluid and the 
$\mu_{\rm out}$ of the external liquid),
where $H = |\vec{H}|$  and $H_n$  is the 
normal component of the magnetic field.
We will use this position-dependent normal  magnetic 
force density to modify the pressure in our elastic
equations in Sec.\ \ref{sec:shape_eqns}.

\subsubsection{Calculation of the magnetic field}
\label{fieldcalculation}

To calculate the total magnetic field, i.e., the 
superposition of the external uniform
field and the field from  the ferrofluid magnetization,
 we use the fact that
ferrofluids are generally non-conducting  \cite{Rosensweig85}. 
 Then Maxwell's
equations give $\nabla \times {\bf H} = 0$, which allows us to introduce 
 a scalar
magnetic potential $u$ with  $\vec{\nabla} u = \vec{H}$.
From Maxwell's equation $\vec{\nabla} \cdot \vec{B} = \vec{\nabla}
    \cdot \mu_0(\vec{H} + \vec{M})=0$ 
we  get  Poisson's equation in magnetostatics
\begin{equation}
\label{eq:Poisson}
 \vec{\nabla}^2 u(r,z) = - \vec{\nabla}\cdot \vec{M}(r,z).
\end{equation}
For the linear magnetization law \eqref{eq:linear},
 Poisson's equation  simplifies to the Laplace equation
$\vec{\nabla}^2 u(r,z) = 0$.

For the numerical solution of this partial differential equation we use a
coupled axisymmetric 
finite element -- boundary element method \cite{Costabel87,
  Wendland88, Arnold83, Arnold85} with a cubic spline interpolation for the
boundary \cite{Ligget81}. This combination of methods was also 
used by Lavrova {\it et al.}\  for 
free ferrofluid drops \cite{Lavrova04, Lavrova05,Lavrova06} 
and earlier for electric drops, e.g., 
by Harris and Basaran \cite{Harris1993}.  
The finite element method (FEM) is used to solve
Eq.\ \eqref{eq:Poisson} in the magnetized domain inside the capsule
 and the boundary element method (BEM) for the nonmagnetic domain outside.
Both domains are coupled by the continuity
conditions of magnetostatics for 
$u$ and its normal derivative on the boundary of the capsule,
\begin{equation}
\label{eq:magnetic_boundary}
u_{\text{in}} = u_{\text{out}},~~
\mu\frac{\partial u_\text{in}}{\partial n} 
= \frac{\partial u_\text{out}}{\partial n},
\end{equation}
with $\mu_{\rm out}=1$ for the external nonmagnetic liquid.
Both the FEM and BEM exploit axial symmetry and effectively
operate in the two-dimensional
 $rz$ plane, where the axisymmetric capsule shape 
is described by a contour line $(r(s),z(s))$.  
For the FEM we
use a standard Galerkin method with linear elements on a 
triangular two-dimensional 
grid in the $rz$ plane that is created with a Delauney
triangulation using the Fade2D software package \cite{Fade2D}, where 
 we set a
fixed number of grid points on the capsule's boundary.

In the BEM we express solutions $u(\vec{r}_0)$ of the 
Laplace equation $\vec{\nabla}^2 u = 0$ for $\vec{r}_0$ on the 
outside or the boundary 
of the capsule in terms of integrals over the boundary of the 
capsule.
Using fundamental solutions with rotational
symmetry \cite{Wrobel02}, we have to solve a set of one-dimensional integrals
over the whole boundary of the capsule
\begin{equation}
\label{eq:BEM_integral_eq}
cu(\vec{r}_0) - \int\limits_0^L \left[ 
   u(\vec{r})
   \frac{\partial u_{\text{ax}}^*(\vec{r}_0, \vec{r})}{\partial {n}} 
   -  \frac{\partial u(\vec{r})}{\partial {n}}
   u_{\text{ax}}^*(\vec{r}_0, \vec{r}) \right]
  r \text{d}s = z_0.
\end{equation} 
Here $u_{\text{ax}}^*(\vec{r}_0, \vec{r})
     \equiv  \int_0^{2\pi}u^*(\vec{r}_0, \vec{r})\text{d}\varphi$ 
 is the axially symmetric fundamental solution of Laplace's equation, 
which is obtained from the 
fundamental solution 
$u^*(\vec{r}, \vec{r}_0) = {1}/{4\pi|\vec{r} - \vec{r}_0|}$ 
of Laplace's equation, 
 $\Delta u^*(\vec{r}, \vec{r}_0) = -\delta(\vec{r} - \vec{r}_0)$.
In the integral equation (\ref{eq:BEM_integral_eq}), 
$u$ and its normal derivative 
are evaluated on the  outside of the capsule surface. 
The point $\vec{r}_0$ 
is the point where  $u$ is to be calculated,
while the integrals are taken over points $\vec{r}(s)$ on the capsule 
contour. Both $\vec{r}$ and $\vec{r}_0$ lie in the same $rz$-plane.
For the geometric factor $c$, we have $c={1}/{2}$ for
points $\vec{r}_0$ on the boundary $\Gamma$ and $c = 1$ for points 
$\vec{r}_0$ in the exterior domain.
The vector $\vec{n}$ denotes the outward unit
normal vector and $z_0$ describes the $z$-component of $\vec{r}_0$. 
On the right-hand side of \eqref{eq:BEM_integral_eq}, 
$z_0$ can be interpreted as the potential of the external 
electric field.
For numerical evaluation, the integrals in Eq.\ \eqref{eq:BEM_integral_eq}
are discretized by a point
collocation method and solved by applying Gaussian quadrature for nonsingular
integrands and a midpoint rule for weakly singular integrands.

\begin{figure*}[htbp]
\begin{center}
\includegraphics[width=0.8\textwidth,clip]{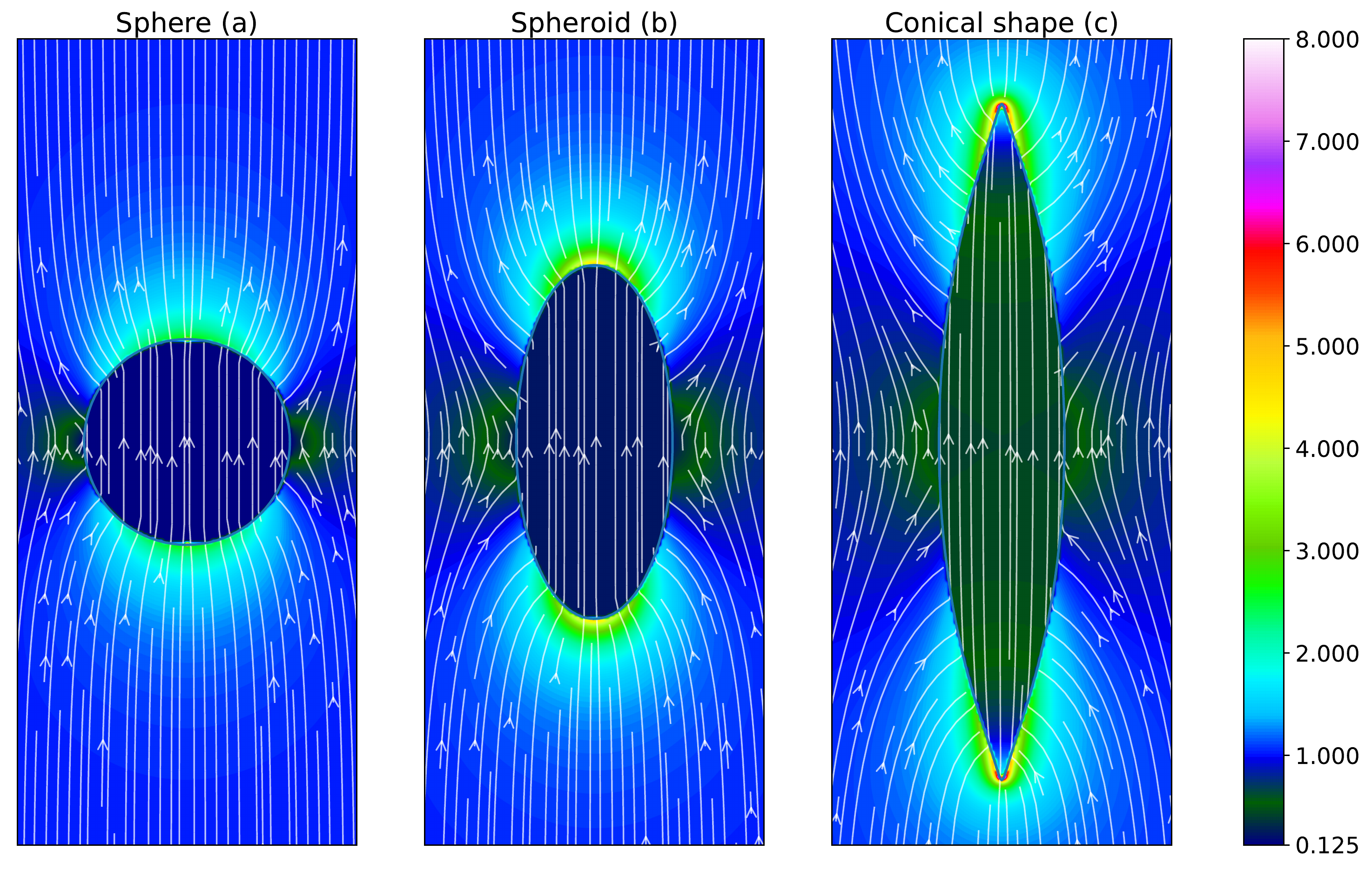}
\caption{
 Numerical results for the magnetic field distribution and capsule 
shape (two-dimensional projection) 
  for a capsule filled with a ferrofluid with a susceptibility of $\chi
  = 21$. The ratio of Young's modulus and  surface tension is 
 $Y_{2\text{D}}/\gamma = 100$.
The external magnetic field $H_0$ is uniform and points in the upward
  direction.
 Arrows indicate the local direction of $H$; the color codes for 
  the absolute value of $H$ in units of $H_0$. The (a) spherical capsule and
  the (b) spheroidal capsule have uniform fields inside, while the field in
  the (c) conical-shaped capsule  increases strongly in the tips. 
  The elongations $a/b$ 
  (ratio of the polar radius to the equatorial radius) are 
  (a) $a/b = 1$, (b) $a/b = 2.26$,
  and (c) $a/b = 5.38$.
   The magnetic Bond numbers $B_m$ 
  [see definition in Eq.\eqref{BmDefinition}]
  are (a) $B_m = 0$ , (b) $B_m = 262.4$, and (c) $B_m = 702.2$. 
}
\label{fig:fieldplot}
\end{center}
\end{figure*}

The FEM and BEM are coupled at the boundary by the continuity conditions 
(\ref{eq:magnetic_boundary}). 
The FEM provides values 
for $u$ on every finite element grid point inside the capsule 
including values $u_{\text{in}}$ on 
the inner side of the boundary; in addition,  the normal derivatives 
${\partial u_\text{in}}/{\partial n}$ 
 on the inside of the discretized capsule boundary are needed 
for the FEM but remain  {\it a priori} unknown. 
Values for these normal derivatives on the boundary points of the 
FEM grid are obtained by 
the BEM method.  
Our BEM uses linear interpolation for $u$ between the discretized 
boundary points.
We use the continuity conditions \eqref{eq:magnetic_boundary}
to write  the  boundary integral equation
\eqref{eq:BEM_integral_eq} in terms of quantities on the 
inner capsule boundary. 
Using one BEM equation \eqref{eq:BEM_integral_eq}
for each boundary point (with $c=1/2$),
we obtain a set of equations 
that allows us to calculate the 
unknown derivatives  ${\partial u_\text{in}}/{\partial n}$
for given $u_\text{in}$ and to get a 
closed system of equations for  $u$ everywhere inside 
the capsule.
After solving the resulting system of FEM equations 
 we know $u$ everywhere inside 
the capsule.
For the calculation of $u$ inside the capsule and thus
for the calculation of the magnetic force density $f_m(r,z)$ acting 
on the capsule using  (\ref{fm}), which is also calculated
with  the magnetic field on the inside,  it is 
 not necessary to calculate  $u$ in the 
entire external domain explicitly. This is done implicitly by the BEM.
If needed (for example, in order to calculate the field in the exterior 
regions in Fig.\ \ref{fig:fieldplot}), 
 $u$ can be calculated  by 
solving \eqref{eq:BEM_integral_eq} for 
points $\vec{r}_0$ in the exterior  with $c=1$.

  In a ferrofluid capsule or drop with sharp edges, very high
field strengths can arise [see Fig.\ \ref{fig:fieldplot}(c)]. 
Also  field gradients can be large, which makes pointed shapes prone
to discretization errors caused by the grid. 
This effect can be countered to
some degree by placing more FEM 
grid points at the tip in order to improve
the precision there, which is, however,  limited by the BEM part of the 
solution scheme:
The collocation points must not come too close to the symmetry axis because
the weakly singular integrals become strongly singular 
on the $z$ axis \cite{Lavrova2006}. This
leads to massively increasing numerical errors near the axis and a decrease of
the overall precision.
Overall, our numerical scheme to calculate singular 
BEM integrals is not the most advanced as a trade-off for simplicity.
There are more elaborated schemes for the integration of singular integrals
as, for example,  developed over many years by Gray
{\it et al.} \cite{Gray1990,Gray2004}, which could 
 provide a more elegant way to deal with the problem.
We use the  following compromise for the discretization:
We place $N=250$ elements on the boundary such that  the
length $L_i$ of the $i$th boundary 
element (beginning at the equator) is given by
\begin{align}
  L_i = c_0\exp\left(\ln(l_0)\frac{i-1}{N} \right).
\label{eq:Li}
\end{align} 
The constant $c_0$ is chosen in order to obtain the 
correct  total arc length $L$, which is given by 
the meridional stretch factors $\lambda_s = {\text{d}s}/{\text{d}s_0}$
of the deformed capsule [see Eq.\ (\ref{eq:lambda}) below],
$\sum_{i=1}^N L_i = L/2 = \int_0^{{L_0}/{2}} \lambda_s ds_0$.
We choose $l_0 = 0.1$ ($l_0 = 1$ gives a constant element length and $l_0 <
1$ leads to a higher element density at the capsule's tip).  
Increasing $N$ 
beyond 250  does not improve the precision
  significantly. A  higher density of points at the
  capsule's tip (lower $l_0$) leads to stronger oscillations in the iterative
  solution scheme (see Sec.\ \ref{sec:iterative_solution} below).

 \subsubsection{Electric  fields and dielectric liquid}
\label{sec:electric}

Our approach  to elastic capsules filled with a ferrofluid in a 
magnetic field also applies to capsules filled with a dielectric fluid 
in an electric field. 
The generic situation for a capsule filled with a 
fluid with dielectric constant $\varepsilon$ is to be suspended 
in a dielectric liquid with a different $\varepsilon_{\rm out}\neq
\varepsilon$, 
 which does not equal unity $\varepsilon_{\rm out} \neq 1$. 
Then the dielectric force density in a linear medium is 
\begin{equation}
 f_e(r,z) = \frac{\varepsilon_0\varepsilon_{\rm out}  \chi_\varepsilon}{2}
  \left( E^2(r,z) +  \chi_\varepsilon
       E_n^2(r,z)\right)~,~~
  \chi_\varepsilon \equiv \frac{\varepsilon}{\varepsilon_{\rm out}}-1,
\label{fmepsilon}
\end{equation}
which is completely analogous to (\ref{fm}) with 
$\chi_\varepsilon$ playing the role of the susceptibility $\chi$. 
For the general case, Poisson's equation becomes 
\begin{equation}
 \vec{\nabla}^2 \phi(r,z) = -\frac{1}{\epsilon_0}\vec{\nabla}\cdot \vec{P}(r,z),
\end{equation}
with the electric potential $\phi$ and the polarization $\vec{P}$. 
For a linear polarization law, it simplifies to the Laplace equation
$\vec{\nabla}^2 \phi(r,z) = 0$.

\subsection{Equilibrium shape of the capsule}
\label{sec:elastic_modell}

\subsubsection{Elasticity and shape equations}
\label{sec:shape_eqns}

The capsule is deformed by the normal magnetic 
stresses $f_m$ from  the ferrofluid. We
have to calculate the resulting deformed equilibrium 
shape, where all
elastic stresses, surface tension and magnetic stress are 
balanced everywhere on the capsule.
 Every point of the reference shape $[r_0(s_0), z_0(s_0)]$ 
is mapped onto a new point $[r(s_0), z(s_0)]$.  The deformed
shape $[r(s_0), z(s_0)]$ is calculated by solving shape equations, which are
derived from  nonlinear 
theory of thin shells \cite{Libai98,Pozrikidis03, Knoche11,
  Knoche13}.  We use a Hookean elastic energy density with a spherical
rest shape.  The Hookean 
elastic  energy density (defined as energy per undeformed unit area) 
is given by
\begin{equation} 
w_s = \frac{1}{2}\frac{Y_{2\text{D}}}{1-\nu^2}
   (e_s^2 +2 \nu e_s e_\varphi+e_\varphi^2)
   +\frac{1}{2}E_{\text{B}}(K_s^2+2\nu K_sK_\varphi + K_\varphi^2).
   \label{eq:ws}
\end{equation}
Here $e_s$ and $e_\varphi$ are meridional and circumferential strains that
contain the stretch factors $\lambda_s$ and $\lambda_\varphi$:
\begin{equation}
 e_s = \lambda_s - 1, \quad e_\varphi = \lambda_\varphi -1~,~~
\lambda_s = \frac{\text{d}s}{\text{d}s_0}, \quad
\lambda_\varphi = \frac{r}{r_0}.
\label{eq:lambda}
\end{equation}
Here and in the following, 
quantities with subscript 0  refer to the undeformed
spherical reference shape and  quantities without 0 describe 
 the  deformed shape. 
Analogously, the 
 bending strains  $K_s$ and $K_\varphi$ are generated by the
curvatures $\kappa_s$ and $\kappa_\varphi$:
\begin{equation*}
  K_s= \lambda_s\kappa_s - {\kappa_s}_0, \quad
  K_\varphi = \lambda_\varphi \kappa_\varphi - {\kappa_\varphi}_0~,~~
  \kappa_s = \frac{\text{d}\psi}{\text{d}s}, \quad 
 \kappa_\varphi = \frac{\sin\psi}{r}.
\end{equation*}
In the elastic energy (\ref{eq:ws}),  $Y_{2\text{D}}$ is the
two-dimensional Young modulus governing stretching deformations,
 $E_\text{B}$ is  the bending modulus, and  $\nu$ is the
two-dimensional Poisson ratio. 
Elastic properties are usually only weakly
$\nu$ dependent; we use $\nu = 1/2$, which is the 
typical value for an incompressible polymeric material.
The arc length of the deformed capsule's contour is given by
$ L = \int_0^{L_0} \lambda_s ds_0$,
while $L_0 = \pi R_0$ is the fixed arc length 
of the undeformed spherical capsule.

In experiments, the capsule's shell is constructed by polymerization 
on the surface of a drop. Therefore, the undeformed reference shape, 
which is spherical in the absence of gravity, 
is also a solution of the Laplace-Young equation
\begin{equation}
\label{LaplaceYoung}
  \gamma(\kappa_s+\kappa_\varphi) = p,
\end{equation}
where $\gamma$ is the surface tension of the droplet.
The solution of the Laplace-Young equation 
 will be discussed in detail in 
Sec.\ \ref{sec:DropletTheorie} below.

In the following, we will neglect the bending energy, which means we set
$E_\text{B} = 0$. The characteristic length scale  of the problem is the 
radius $R_0$ of the undeformed sphere, such that 
the neglect of the bending energy corresponds to  the limit of large 
F{\"o}ppl-von K\'{a}rm\'{a}n numbers 
$\gamma_{\rm FvK}\equiv Y_{2\text{D}}R_0^2/E_B$. 
This is the limiting case of an  elastic Hookean membrane and 
is a good approximation for two reasons. First,
we will only consider capsules with thin shells as they were prepared in
experiments \cite{Knoche13, Degen08}. The shell thickness $D$ is very
small compared to the capsule size, $D \ll R_0$. With $Y_{2\text{D}}
\propto D$ and $E_\text{B} \propto D^3$ it follows that 
$\gamma_{\rm FvK}\sim (R_0/D)^2 \gg 1$ and 
stretching energies are typically larger  than  bending energies.
The second argument
is that  the homogeneous magnetic field acting on the ferrofluid-fluid
 capsule predominantly stretches and elongates  the capsule in order 
to increase its total dipole moment. 
This increases stretching energies, whereas 
the capsules develop high curvatures only 
at the conical tips. However, 
we show below that stretch factors diverge at conical tips, so 
the stretching energy dominates over the 
bending energy associated with these high curvatures 
also in the tip regions.

Elastic tensions in the shell (defined as force per
deformed unit length) derive from the surface elastic energy density 
by
\begin{align}
\begin{split}
\tau_s &= \frac{1}{\lambda_\varphi}\frac{\partial w_s}{\partial e_s}
    = \frac{Y_{2\text{D}}}{(1-\nu^2)\lambda_\varphi}
   \left[(\lambda_s-1) + \nu(\lambda_\varphi-1)\right],\\
  \tau_\varphi &= \frac{1}{\lambda_s}\frac{\partial w_s}{\partial
  e_\varphi} = \frac{Y_{2\text{D}}}{(1-\nu^2)\lambda_s}
   \left[(\lambda_\varphi-1) + \nu(\lambda_s-1)\right].
\end{split}
\label{eq:taulambda}
\end{align}
Although we use a Hookean 
elastic energy density, the constitutive relation (\ref{eq:taulambda}) 
is {\it nonlinear} because 
of the additional $1/\lambda$ factors, which arise for purely geometrical 
reasons: The Hookean elastic energy density is defined per 
undeformed unit area such that $\partial w_s/\partial e_s$ is 
the force per undeformed unit length, whereas the Cauchy stresses $\tau_s$
and $\tau_\varphi$ are defined per deformed unit length.

In addition to the elastic tensions $\tau_s$ and $\tau_\varphi$,
there is also  a contribution
from an isotropic effective surface tension $\gamma$
between the outer liquid and the capsule.
Such a contribution arises 
either as the sum of surface tensions of the liquid outside 
with the outer capsule surface and the liquid inside with the 
inner capsule surface or, if
the capsule shell is porous such that there is still contact 
between the liquids outside and inside the capsule,  
with additional contributions 
from the surface tension between outside and inside liquids. 
In the absence of elastic tensions, the surface tension $\gamma$ 
also gives rise  to the spherical rest shape of the capsule. 
For macroscopic capsules the  surface tensions should be negligible,
but for microcapsules with weak walls they should not be neglected. 
We expect the effective surface tension $\gamma$ to be somewhat 
smaller than typical liquid-liquid surface tensions, which are 
around $\gamma = 50\,{\text{mN}}/{\text{m}}$; we will use 
$\gamma = 10\,{\text{mN}}/{\text{m}}$ below.

The equilibrium of forces in the deformed elastic membrane  is described by
\begin{align}
  0&=\tau_s\kappa_s + \tau_\varphi\kappa_\varphi +
  (\kappa_s+\kappa_\varphi)\gamma -p,
  \label{eq:equilibrium_eqns_norm}\\
  0&=\frac{\cos \psi}{r}\tau_\varphi -
  \frac{1}{r}\frac{\text{d}(r\tau_s)}{\text{d}s},
\label{eq:equilibrium_eqns_tang}
\end{align}
where Eq.\ (\ref{eq:equilibrium_eqns_norm})
 describes the normal force equilibrium  and Eq.\ 
(\ref{eq:equilibrium_eqns_tang})
 tangential force equilibrium (in the $s$ direction, 
equilibrium in the $\varphi$ direction is always fulfilled  by axial 
symmetry). 
In the presence of  magnetic forces, the pressure 
\begin{equation}
p(s) = p_0 + f_m(s)
	\label{eq:pfm}
\end{equation}
 is modified by the magnetic stress $f_m$, which is
a position-dependent normal stress
 pointing outwards and thus stretching the capsule and given 
by the magnetic field at the capsule surface [see Eq.\ (\ref{fm})]. 
It is important to note that magnetic forces are always 
normal to the surface such that they do no enter the tangential 
force equilibrium  (\ref{eq:equilibrium_eqns_tang}).
The (homogeneous) pressure $p_0$ is the Lagrange multiplier for the 
volume constraint $V=V_0=4\pi R_0^3/3$.

The equations of force equilibrium and geometric relations 
can be used to derive 
 a system of four first-order differential equations with the 
arc length $s_0$ of the undeformed spherical contour as 
an independent variable, 
which are called shape equations in the following:
\begin{align}
\label{eq:shape_eqns}
\begin{split}
r'(s_0)&=\lambda_s\cos\psi~~,~~
z'(s_0)=\lambda_s\sin\psi,\\
\psi'(s_0)&=\frac{\lambda_s}{\tau_s + \gamma}
  \left[-\kappa_\varphi(\tau_\varphi+\gamma) + p(s_0)\right],\\
\tau_s'(s_0)&=\lambda_s\frac{\cos\psi}{r}(\tau_\varphi-\tau_s).
\end{split}
\end{align}
In these shape equations, the surface tension $\gamma$ gives an
 isotropic and constant 
stress contribution, in addition  to the elastic stresses
$\tau_s$ and $\tau_\varphi$. This is because 
we assume that the undeformed rest state, 
where the elastic stresses $\tau_s$ and $\tau_\varphi$ vanish, is  
identical to the shape of a ferrofluid droplet of surface tension $\gamma$.
We neglect that  $\gamma$ could change during capsule preparation and 
during elastic deformation. 

The system of shape equations 
is closed by the constitutive relation 
(\ref{eq:taulambda}) for $\tau_\varphi$ and the 
 relations
\begin{align*}
\lambda_s &= (1-\nu^2)\lambda_\varphi\frac{\tau_s}{Y_{2\text{D}}}
  -\nu(\lambda_\varphi-1)+1~~
\mbox{with}~
\lambda_\varphi = \frac{r}{r_0},~~
\kappa_\varphi = \frac{\sin\psi}{r},
\end{align*}
where the first relation derives from 
the constitutive relation  (\ref{eq:taulambda}) for $\tau_s$ 
and the second relation is geometrical.
For further details on the derivation of the shape equations, we refer 
the reader to
Refs.\ \cite{Libai98, Knoche13, Knoche11}.

\subsubsection{Numerical solution of the shape equations}

The system of shape equations \eqref{eq:shape_eqns} has to be solved
numerically. The integration starts at the pole with $s_0 = 0$
and runs to the capsule's equator at $s_0 = {L_0}/{2}$.  To integrate the four
first-order differential equations we have three boundary conditions 
at $s_0 = 0$:
\begin{equation}
r(0) = 0, \quad z(0)~{\rm arbitrary}, \quad \psi(0)=0.
\label{eq:shape_eqns_bc}
\end{equation}
The condition for $r(0)$   follows from the absence of
 holes in the capsule. We can choose $z(0)$ arbitrarily because
the external magnetic field does not depend on the $z$ coordinate. 
The  boundary condition $\psi(0)=0$  at the pole 
seems to exclude possible conical capsule shapes with
$\psi(0)>0$. We discuss this issue  below in Sec.\ \ref{sec:conical}
and in Appendix \ref{app:cone_tip2}.
There we derive the  boundary condition  $\psi(0)=0$ 
for finite  stretches $\lambda_s$ and $\lambda_\varphi$ at the poles. 
The boundary condition  $\psi(0)=0$  also arises if the magnetic forces
 $f_m$ remain finite at the poles such that the normal 
force equilibrium  requires finite 
curvatures at the poles. 
Conical shapes, however,  have 
divergent stretches $\lambda_s$ and $\lambda_\varphi$ 
and  divergent magnetic normal forces  $f_m$  at their 
conical tips. In the numerical calculation of capsule shape and 
magnetic field we have to 
discretize the capsule surface such that divergences are cut off
(this numerical issue is discussed in more detail also 
in  Appendix \ref{app:errors})
and the boundary condition   $\psi(0)=0$ 
for finite  stretches $\lambda_s$ and $\lambda_\varphi$ 
or finite magnetic force $f_m$  is appropriate. 
Then the right-hand side
 of the shape equation for $\tau_s$ in \eqref{eq:shape_eqns}
vanishes, $\tau_s'(0)=0$  for $s_0=0$ [see also Eq.\ (\ref{eq:tau1_app})], 
which can be used to start the 
integration at the pole. 
{\it A priori}, a fourth boundary condition for 
the tension  $\tau_s(0)$ at the pole is unknown. 
 On the other hand, we have $\psi\left({L_0}/{2}\right)=\pi/2$ as a
matching condition at $s_0 = {L_0}/{2}$ to prevent kinks there.  With the
help of this matching condition, we can use a shooting method to determine
$\tau_s(0)$. 
To increase numerical stability, we expand the shooting
method to a multiple shooting method, where we use several integration
intervals with several matching points.

To keep the volume of the capsule constant, we have to use the internal
pressure $p_0$ as the Lagrange multiplier, which is  adjusted during the
calculation. In order to do so, $p_0$ becomes another shooting parameter
with $V-V_0$ as the corresponding residual.  In this work, we use a fourth
order Runge-Kutta scheme with a step size of $\Delta s_0=10^{-6}$ in the first
integration interval starting at the apex and $\Delta s_0=10^{-4}$ in all
other intervals, while there is a total of 250 integration intervals.

\subsubsection{Wrinkling}
\label{sec:WrinkleTheorie}

A  ferrofluid-filled capsule is stretched 
in a uniform external magnetic field  in the 
direction of the magnetic field.   
As opposed to a ferrofluid droplet, a capsule can develop 
wrinkles if circumferential compressive stresses arise as a result 
of this stretching. 

\begin{figure}[htbp]
\begin{center}
\includegraphics[width=0.2\textwidth,clip]{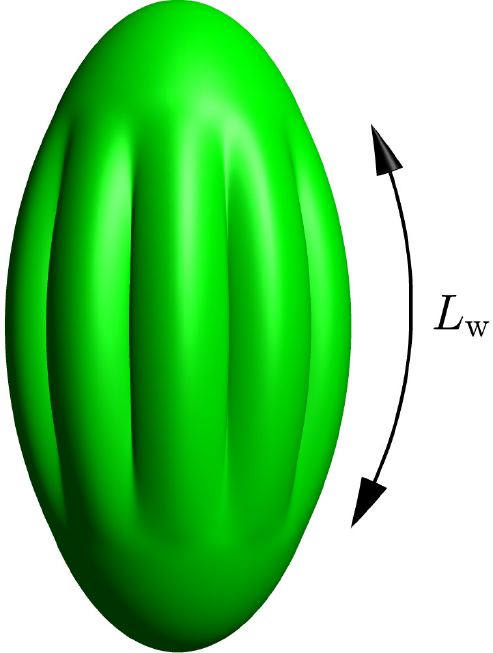}
\caption{
Three-dimensional illustration of a wrinkled capsule. The length
  $L_{\text{w}}$ of the wrinkles is measured as the length of the 
  region, where $\tau_\varphi + \gamma < 0$. 
The wrinkling wavelength is not determined explicitly here.
}
\label{fig:3Dwrinkling}
\end{center}
\end{figure}

Because of volume conservation, the circumferential 
radius of the capsule  has to decrease in the equator region
giving rise to compression with $\lambda_\varphi < 1$ in this region
and a region of  negative elastic stress $\tau_\varphi < 0$ develops.
 In
contrast to a droplet with a liquid surface and constant 
surface tension $\gamma >0$,  regions of negative 
total hoop stress
$\gamma+ \tau_\varphi < 0$ can develop for capsules if the
negative elastic hoop stress exceeds the surface tension.
Then the elastic shell can reduce its total energy by
developing wrinkles in the circumferential direction (see
Fig.\ \ref{fig:3Dwrinkling} for illustration). 
These wrinkles cost stretching energy in the  meridional $s$ direction
 and bending energy, 
 but this is compensated by a 
 release of compressional stresses and a reduction of 
elastic compression energy in the $\varphi$ direction. 
Strictly speaking, $\gamma + \tau_\varphi<0$
is only an approximation neglecting the bending energy, which will 
also increase upon  wrinkling,  and the negative stress has to exceed
a small Euler-like threshold value. 
We expect the wrinkles to occur in a region near the capsule
equator. Thus they will be roughly parallel to the external magnetic field and
therefore we assume that they do not effect the magnetic
properties of the capsule.

 In order to introduce wrinkling  in the shape
equations, we will use the same approach that has been used 
 for pendant capsules in Ref.\ \cite{Knoche13}.
 The wrinkles will break the axial symmetry. 
In the wrinkled regions, where 
 $\gamma + \tau_\varphi<0$, we approximate the shape by 
an  axisymmetric pseudomidsurface 
$(\overline{r}(s_0), \overline{z}(s_0))$ for which we 
use  modified axisymmetric shape equations, where we set 
$\gamma + \tau_\varphi=0$. This condition states that 
the total circumferential hoop stress is completely relaxed by 
fully developed
wrinkles \cite{Davidovitch11}.
This leads us to a new set of equations (see also Ref.\ \cite{Knoche13}),
 which read
\begin{align}
\label{eq:wrinkle_eqs}
\overline{r}'(s_0)&= \lambda_s\cos\overline{\psi},~~
z'(s_0) =\lambda_s\sin\overline{\psi},~~
\overline{\psi}'(s_0) =\frac{\lambda_s}{\overline{\tau}_s 
  + \overline{\gamma}}p,~~
\overline{\tau}_s'(s_0) = -\lambda_s\frac{\cos\overline{\psi}}{\overline{r}}
    (\overline{\tau}_s+\overline{\gamma}).
\end{align}
We also have to introduce a modified effective surface tension
$\overline{\gamma}= {\lambda_\varphi}/{\overline{\lambda}_\varphi}$,
because the real surface area exceeds  the pseudosurface area, and
we have to model this increase of $E_\gamma$ by increasing $\gamma$
instead. This new system of 
differential equations is closed by the relations
\begin{equation*}
\lambda_s =\frac{\overline{\tau}_s\overline{\lambda}_\varphi
+Y_{2\text{D}}}{Y_{2\text{D}}- \nu \gamma}, ~~
\overline{\lambda}_\varphi =\frac{\overline{r}}{r_0}.
\end{equation*}
In order to calculate $\overline{\gamma}$, the circumferential stretch factor 
$\lambda_\varphi$ of the \textit{real, wrinkled} surface has to be calculated 
via the constitutive relations \eqref{eq:taulambda}.
To calculate wrinkled  capsule shapes we start to solve the
shape equations \eqref{eq:shape_eqns} as described before.  As soon as the
condition $\tau_\varphi + \gamma < 0$ is valid, we continue the calculations
by solving the modified system  \eqref{eq:wrinkle_eqs}.  
By following the
solution of the modified system, we can calculate 
 the length  $\L_\text{w}$ of the wrinkled region
\begin{align}
 L_{\text{w}} = \int\limits_{\tau_\varphi + \gamma < 0} \text{d}s.
\label{eq:Lw}
\end{align}
At this point, it is also possible to calculate the wavelength of the
wrinkles using the same methods as in Ref.\ \cite{Knoche13}. 
Here we will mainly be interested in the extent $L_\text{w}$
of the wrinkled region. 

\subsubsection{Ferrofluid droplet}
\label{sec:DropletTheorie}

The special case $Y_{2\text{D}}=0$ describes a ferrofluid droplet
 without an
elastic shell and has been treated  in the literature before. 
The balance of forces on the
surface is given by the Laplace-Young equation
\eqref{LaplaceYoung}.
Using the definitions of $\kappa_s$ and $\kappa_\varphi$, 
this equation can be translated into
\begin{equation*}
  \frac{\text{d}\psi}{\text{d}s} = \frac{p}{\gamma} - \frac{\sin\psi}{r}.
\end{equation*}
In order to have a parametrization in the reference arc length $s_0$ and a
fixed integration interval, we introduce a constant stretch factor
$\lambda_s$,
 which is adjusted as a shooting parameter. The boundary and matching
conditions are the same as in the case of the elastic shape
equations. Together with the already known geometrical relations for $r$ and
$z$ we get a system of three shape equations for a droplet:
\begin{align}
\label{dropsystem}
\begin{split}
r'(s_0)&=\lambda_s\cos\psi,~~
z'(s_0)=\lambda_s\sin\psi,~~
\psi'(s_0) = \lambda_s\left(\frac{p_0 + f_m}{\gamma} 
- \frac{\sin\psi}{r}\right).
\end{split}
\end{align}
This system is solved in the same way as the shape equations 
for elastic capsules in the previous
sections. The basic shooting parameters are given by $\lambda_s$ and $p_0$.
Our solution scheme for the Laplace-Young equation is chosen 
such that it is 
completely analogous and comparable to the elastic shape equations. 
There are several other ways 
to solve this equation with a volume constraint, 
for example, by employing finite elements \cite{Brown1980}.


\subsection{Iterative numerical solution of the coupled problem}
\label{sec:iterative_solution}

The magnetostatic and the elastic problem are coupled: The 
capsule shape determines the boundary conditions for the magnetic 
field via the continuity conditions (\ref{eq:magnetic_boundary}), while 
the  normal  magnetic  force density $f_m(r,z)$
 acting on the capsule surface  [see Eq.\ (\ref{fm})]  enters 
the shape equations (\ref{eq:shape_eqns}) via the pressure [see
Eq.\ (\ref{eq:pfm})]. 
To find a joint solution we use an iterative numerical 
solution scheme.  We start
with the reference shape and calculate the corresponding magnetic field
$\vec{H}(r,z)$ for a given external field $\vec{H}_0$. Then, we
can calculate a deformed shape of the capsule using this magnetic field. Now we
recalculate the magnetic field and so on until the iteration converges. 
At this fixed point, the solution of the shape equations and the
magnetic field are self-consistent. This iterative coupling of elastic shape
equations to an external field calculated by a boundary element method is
similar to the iterative scheme used in Ref.\ \cite{Boltz15} to
calculate the shape of sedimenting capsules in an external flow
field. For the problem of ferrofluid droplets, an analogous 
iterative strategy has been introduced in 
Refs.\ \cite{Lavrova04,Lavrova05,Lavrova06,Lavrova2006}.

 The iteration can cause numerical problems in the  solution of  the
 the nonlinear elastic shape equations.  If the
capsule shape changes rapidly during the iteration, 
the shooting method used to solve 
the shape equations does not find a solution.
This problem can be reduced by  slowing down the
iteration.  To solve the elastic shape equations in the $n$th step, we 
use a convex linear combination of the 
 updated magnetic field $\vec{H}'_n$ and 
 the magnetic field $\vec{H}_{n-1}$ from
the previous iteration step instead of  $\vec{H}'_n$ itself
\cite{Lavrova2006,Boltz15}:
\begin{align}
\vec{H}_n = \vec{H}_{n-1} 
 + \alpha(\vec{H}'_n - \vec{H}_{n-1}).
\end{align}
The parameter $\alpha$ ranges between 0 and 1 and has to be lowered in
situations of quickly changing shapes of the capsule. Finally, it is 
switched back to 1 to ensure real convergence.
To track a solution as a function of the magnetic field 
strength, 
it is  helpful to increase the external magnetic field
$\vec{H}_0$ in small steps $\Delta\vec{H}_0$ and let the
capsule's shape converge after each step.  This  slows down the
calculation speed drastically but increases numerical 
stability  and helps to track 
 a specific branch of stable
solutions (see Sec.\  \ref{sec:hysteresis_effects}).

 A problem with the
iterative solution scheme can arise if the capsule shape becomes nearly
conical with a very sharp tip of high curvature. Then the numerical error in 
the calculation of the magnetic field (see 
Sec.\  \ref{fieldcalculation}), makes it difficult or even prohibitive 
to reach a fixed point of the
iterative scheme.  Instead the iteration gives
oscillations of the capsule shape around the required fixed point, which 
worsens the quality of the results.
The iterative strategy used here directly converges to stationary 
shapes without simulation of the real dynamics.

An alternative to our iterative scheme  is to directly 
simulate the dynamics for the
fluid from the electromagnetic, elastic, and hydrodynamic
forces. Then the fluid motion is simulated over 
time until it reaches a steady state. 
This method was used by Karyappa {\it et al.}\
 for elastic capsules in electric fields \cite{Karyappa2014}.
For liquid droplets, 
there are comparable problems with sharp tips and numerical singularities,
where the full dynamics could by solved to great accuracy,
such as  the emission of fluid jets at the tip of drops 
in electric fields \cite{Collins2008}, pinch-off dynamics \cite{Suryo2006},
and coalescence phenomena \cite{Anthony2017}. 
The errors of the field calculation with finite elements at such sharp tips
can also be reduced by using advanced mesh algorithms,
 such as the elliptic mesh generation \cite{Christodoulou1992}.

\subsection{Control parameters and non-dimensionalization} 
\label{sec:control_parameters}

In order to identify the relevant control parameters 
 and reduce the parameter space, we introduce dimensionless quantities. 
We measure lengths in units of the radius $R_0$ of the spherical 
rest shape,  energies in units of $\gamma R_0^2$, i.e., tensions 
in units of the surface tension $\gamma$ of the ferrofluid, and 
magnetic fields in units of the external field $H_0$. 
The problem is then governed by essentially three dimensionless control 
parameters. 
 
The magnetic Bond number $B_m$, 
\begin{equation}
\label{BmDefinition}
 B_m \equiv \frac{\mu_0 R_0\chi H_0^2}{2\gamma},
\end{equation}
is the dimensionless strength of the magnetic force density.
With this dimensionless number, the Laplace-Young equation 
(\ref{LaplaceYoung})
for a ferrofluid droplet can be written in dimensionless form 
\begin{align*}
 \tilde{\kappa}_s + \tilde{\kappa}_\varphi
 &=\widetilde{p}_0 + B_m \left( \tilde{H}^2 + \chi \tilde{H}_n^2 \right),
\end{align*}
with $\tilde{H} \equiv H/H_0$, $\tilde{\kappa} \equiv R_0 \kappa$, and 
$\tilde{p} \equiv  pR_0/\gamma$. 
The scaled droplet shape  described by this Laplace-Young equation then only 
depends  on the two dimensionless 
parameters $B_m$ and $\chi$.

 The  dimensionless Young modulus 
 ${Y_{2\text{D}}}/{\gamma}$ is the control parameter for elastic properties
of the capsule shell. Another dimensionless control parameter for 
elastic properties is  Poisson's ratio $\nu$, 
which is set to $\nu=1/2$ and thus
fixed throughout this paper.
  The limit
${Y_{2\text{D}}}/{\gamma}=0$ describes a droplet without an elastic shell 
 while
${Y_{2\text{D}}}/{\gamma} \gg 1$ describes a system dominated by the
shell elasticity. 

The three dimensionless parameters
$B_m$, ${Y_{2\text{D}}}/{\gamma}$, and the magnetic susceptibility 
 $\chi$ of the ferrofluid
uniquely determine the capsule shape 
 (apart from its overall size $R_0$).
In the following we consider Bond numbers $B_m$ between $0$ and $10^3$ (see 
Sec.\ \ref{sec:results}).
For a  typical ferrofluid-filled capsule with  $\chi=21$, $R_0 = 1\,$mm 
\cite{Karyappa2014,Zwar2018}, 
and 
$\gamma = 0.01\,{\text{N}}/{\text{m}}$, these Bond numbers
correspond to magnetic field strengths
$H$ between $0$ and about $500\,$kA/m 
(or fields $B=\mu_0H$ between $0$ and $0.5\,$T).
We consider dimensionless Young moduli  ${Y_{2\text{D}}}/{\gamma}$
 from $10^{-2}$ (nearly no elasticity) 
to $100$ (elastically dominated) 
and the purely elastic  limit ${Y_{2\text{D}}}/{\gamma} = \infty$ 
(where the definition of $B_m$ is not useful anymore).

For the analogous problem of a dielectric droplet in an external 
electric field  $E_0$ 
we can introduce a dielectric Bond number $B_e$ by
$B_e = {\varepsilon_0\varepsilon_{\rm out}  R_0\chi_\varepsilon E_0^2}/{2\gamma}$,
where $\chi_\varepsilon$ is the analog of the 
magnetic susceptibility $\chi$ and has been defined in (\ref{fmepsilon}).

\section{Analytical approaches}

In this section we introduce three approximative analytical approaches 
to the problem, which describe ferrofluid-filled elastic capsules 
in three different deformation regimes. 
The first approach is the analysis of the linear response of the 
capsule to small magnetic forces. 
The second approach applies to spheroidal shapes at moderate magnetic forces 
and is an  approximative minimization of the 
total magnetic and elastic energy under the assumption of a spheroidal 
shape and uniform stretch factors. This extends 
the  approximative energy minimization  
of  Bacri and Salin  \cite{Bacri82} for ferrofluid droplets to capsules. 
Finally, we investigate conical capsule shapes as they can arise 
under strong  magnetic forces. We investigate the existence of 
conical shapes 
and derive the governing equations 
in  a slender-body approximation by extending 
the approach of Ref.\ \cite{Stone99} from conical droplets 
to conical capsules.

\subsection{Linear shape response at small fields}
\label{sec:lin_def}

In this section we  derive the  linear response of the 
spherical  capsule shape to small magnetic forces. In particular,
we derive the 
elongation $a/b$ of the capsule, 
where $a$ denotes the capsule's polar radius and $b$ its
equatorial radius (see Fig.\ \ref{fig:geometry}).
 Details of the derivation
are given in Appendix \ref{app:lin_def}; here we present the main 
results. 

At small fields displacements change linearly in the magnetic force
density $f_m$. Therefore, radial and tangential displacements 
$u_R(\theta)$ and $u_\theta(\theta)$ (using spherical coordinates with a 
polar angle $\theta$ and assuming axisymmetry) 
are of $O(H^2)$. 
In order to calculate the displacements we consider 
the force equilibria in normal direction, i.e., the Laplace-Young equation
 (\ref{eq:equilibrium_eqns_norm}),
and in  tangential direction, i.e., Eq.\ (\ref{eq:equilibrium_eqns_tang}). 
For a liquid 
ferrofluid droplet with an isotropic surface tension $\gamma$ 
both force-equilibria give equivalent results. 
Expanding to linear order 
in the displacements around the spherical shape, we obtain two coupled
differential equations for the functions $u_R$ and $u_\theta$. 

These linearized force-equilibrium equations can be solved exactly.
The solution takes the form 
\begin{equation}
   u_R= A+B\cos^2\theta,~~u_\theta = C\sin\theta\cos\theta,
\label{eq:AnsatzuRcap}
\end{equation}
where 
$A$, $B$, and $C$ are determined in Appendix  \ref{app:lin_def}
explicitly.  
We find 
$B  = {\mu_0(5+\nu)\chi^2 H^2R_0^2}/{8[Y_{2\text{D}}+(5+\nu)\gamma]}$
from the normal force equilibrium, and  
$C = - 2(1+\nu)B/(5+\nu)$ from the tangential force equilibrium, 
and the pressure is adjusted such that 
$A=-B/3$ in order to fulfill  the volume constraint. 

The functional form $u_R = A+B\cos^2\theta$  of the 
 normal displacement  leads to a spheroidal shape in linear response. 
For a spheroid we can use the relation $H=3H_0(3+\chi)$ and 
obtain $B= R_0 B_m 9(5+\nu)\chi/4(3+\chi)^2[Y_{2\text{D}}/\gamma+(5+\nu)]$.
The linear response approach remains valid as long as $A,B,C\ll R_0$
or $B_m/[Y_{2\text{D}}/\gamma+(5+\nu)] \ll (3+\chi)^2/\chi \approx \chi$.

\begin{figure}[htbp]
\begin{center}
\includegraphics[width=0.5\textwidth,clip]{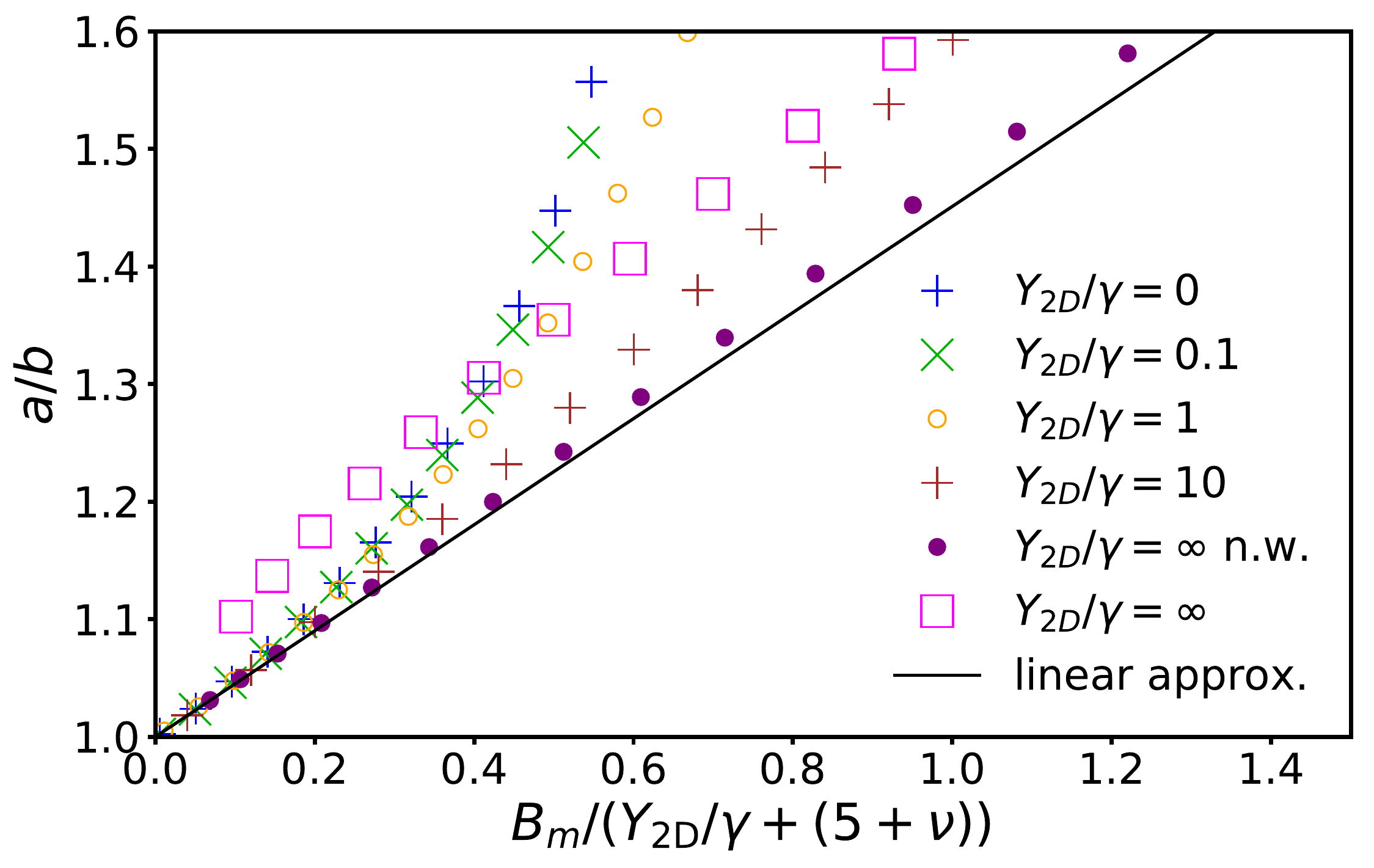}
\caption{
  Elongation ${a}/{b}$ of a capsule filled with a ferrofluid with 
 $\chi = 21$ as a function of
  $B_m /[Y_{2\text{D}}/\gamma + (5+\nu)]$ for different values of
   $Y_{2\text{D}}/\gamma$ in the region of small deformations.
  The solid line describes the linear 
  approximation from Eq.\ \eqref{eq:lin_approx}.
  The best agreement between the numerical data and the linear
   approximation is given for a 
  purely elastic system without wrinkling effects (closed purple circles).
  Wrinkling effects lead to considerable deviations (squares).
}
\label{fig:lin_region}
\end{center}
\end{figure}

From the displacement $u_R(\theta)$ we can calculate its 
elongation 
\begin{equation*}
  \frac{a}{b}  \approx 1 + \frac{u_R(0)-u_R(\pi/2)}{R_0} = 1+ \frac{B}{R_0}
\end{equation*} 
 in linear order in the displacement. 
For a ferrofluid droplet with surface tension
$\gamma$ and without any elastic tensions, i.e., $Y_{2\text{D}}/\gamma = 0$,
 we get, for the elongation $a/b$ 
in linear order [see Eq.\ (\ref{eq:droplinear_app})],
\begin{align*}
  \frac{a}{b} &= 1 + \frac{9\mu_0R_0\chi^2}{8\gamma(3+\chi)^2}H_0^2.
\end{align*}
For the general case $Y_{2\text{D}}/\gamma > 0$, we find 
[see Eq.\ (\ref{eq:capdroplinear_app})]
\begin{equation}
  \frac{a}{b} = 1 +\frac{9\mu_0R_0\chi^2(5+\nu)}
   {8 [Y_{2\text{D}}+\gamma(5+\nu)](3+\chi)^2}
  H_0^2
    = 1 +\frac{9}{4}\frac{\chi}{(3+\chi)^2}
    \frac{B_m}{{Y_{2\text{D}}}/{\gamma}(5+\nu) + 1},
\label{eq:lin_approx}
\end{equation}
which  gives a precise prediction of the capsule's
elongation for small fields, as a comparison with the numerical results 
in  Fig.\ \ref{fig:lin_region} shows. 
To leading order in $B_m$ 
Eq.\ (\ref{eq:lin_approx}) 
 agrees with the results from a similar small deformation approach 
in Ref.\ \cite{Karyappa2014} for capsules  filled with a 
dielectric liquid in electric fields.

\subsection{Approximative energy minimization  for spheroidal shapes}
\label{sec:AnalyticalApprox}

In this section we  derive an analytical approximation for the 
elongation $a/b$ of the capsule at moderate magnetic forces 
  by minimizing an 
approximative  total energy, which assumes a spheroidal shape 
for  magnetic and elastic contributions.
For ferrofluid droplets, the spheroidal  approximation is based on
the experimental observation that the droplet shape in uniform
magnetic fields is very similar to a prolate spheroid \cite{Arkhipenko79,
  Bacri82, Afkhami10} for sufficiently small 
magnetic Bond numbers before a transition into a conical shape 
can take place.  
Our numerical results show that this behavior remains 
qualitatively unchanged  with an additional elastic shell (see
Sec.\ \ref{sec:spheroid}).

 Therefore,  we consider a capsule with 
prolate spheroidal shape. Analogously 
 to Bacri and Salin \cite{Bacri82}, 
we use an energy argument by minimizing the total energy of
the capsule at fixed volume $V= (4\pi/3) ab^2 = V_0$. 
 The total energy consists of three different contributions.
First is the surface energy $E_\gamma$, which is caused by the surface tension
$\gamma$. It is proportional to the surface area $A$ and given by 
\begin{equation}
  E_\gamma = \gamma A = 2 \pi ab\left[\frac{b}{a} + \frac{1}{\epsilon}
   \arcsin{\epsilon}\right]\gamma,
\label{Egamma}
\end{equation}
where $\epsilon \equiv \sqrt{1-{b^2}/{a^2}}$ is the eccentricity. 

 The
second energy contribution is the magnetic field energy
$E_\text{mag}$. According to Ref.\ \cite{Stratton41}, $E_{\text{mag}}$ can be
written as
\begin{equation}
  E_{\text{mag}} = -\frac{V\mu_0}{2} \frac{\chi}{1 + n\chi} H_0^2
\label{Emag}
\end{equation}
 for $\mu_{\rm out}=1$  and with 
the demagnetization factor $n =  ({b^2}/{2a^2\epsilon^3})\left[-2\epsilon +
\ln{\left( ({1+\epsilon})/({1-\epsilon})\right)} \right]$.

The third energy contribution is
the elastic stretching energy $E_\text{el}$, which we construct by taking the
energy density $w_s$ from Sec.\ \ref{sec:elastic_modell},
\begin{equation*}
E_\text{el} = \int w_s \text{d}A_0 
	    = \int\frac{1}{2}\frac{Y_{2\text{D}}}{1-\nu^2}
  (e_s^2 +2 \nu e_se_\varphi+e_\varphi^2)\text{d}A_0,
\end{equation*}
with $e_\text{s} = \lambda_s - 1$ and $e_\varphi = \lambda_\varphi-1$, as
defined in Sec.\ \ref{sec:shape_eqns}. 
 At this point, 
the stretch factors $\lambda_s$ and $\lambda_\varphi$ are unknown and 
we need further approximations. An acceptable approximation
  for spheroidal shapes, which is checked below by comparison 
with the numerics (see Fig.\ \ref{fig:lambda_s})
is constant stretch factors throughout the shell, i.e.,  $\lambda_s,
\lambda_\varphi = \text{const}$, which leads to 
\begin{equation}
  E_\text{el} =\frac{1}{2}\frac{Y_{2\text{D}}}{1-\nu^2}
  (e_s^2 +2 \nu e_s e_\varphi+e_\varphi^2)A_0.
\label{Eel}
\end{equation}
We approximate the circumferential stretch factor 
$\lambda_\varphi$ by  the stretching 
of a fiber at the capsule equator and set
\begin{equation*}
  \lambda_\varphi = \frac{b}{R_0}.
\end{equation*}
In meridional direction we approximate $\lambda_s$ by taking the ratio of the
perimeter $P_\text{ellipse}$ of the corresponding ellipse, which generates the
prolate spheroid by rotation, and the perimeter
 $P_\text{circle}=2\pi R_0$ of a great 
circle on the initial sphere.  The perimeter of the
ellipse is given by an elliptic integral. Therefore, we use Ramanujan's
approximation \cite{Ramanujan14}, which leads us to
\begin{equation*}
  \lambda_s = \frac{P_\text{ellipse}}{P_\text{circle}} 
\approx \frac{a+b}{2R_0}
\left(1 + \frac{3\eta^2}{10 + \sqrt{4-3\eta^2}}\right),
\end{equation*}
with $\eta \equiv {(b-a)}/{(b+a)}$.

 As the last step,
we have to minimize the total energy 
$E_\text{tot} = E_\gamma + E_{\text{mag}} + E_\text{el}$ 
with respect to the elongation ratio ${a}/{b}$ at fixed volume 
$V= (4\pi/3) ab^2=V_0$
in order to get the equilibrium elongation as a function 
of the magnetic Bond number $B_m$ for spheroidal shapes. 
Details of the calculation are presented in Appendix \ref{app:Bm}.
We obtain a closed but quite complicated analytical expression 
for the inverse relation $B_m=g(b/a)$, i.e.,  
the magnetic Bond number $B_m$ as a function of the inverse 
elongation $b/a<1$ for spheroidal shapes in Eq.\ (\ref{eq:BMab}). 
The function $g(k)$ in Eq.\ (\ref{eq:BMab})
 still depends on  three dimensionless parameters:
 the susceptibility $\chi$, the dimensionless
Young modulus $Y_{2\text{D}}/\gamma$,   and Poisson's ratio $\nu$.
This relation reduces to the results of Bacri and Salin \cite{Bacri82}
for ferrofluid droplets in the limit $Y_{2\text{D}}= 0$.

\subsection{Conical membrane shapes with normal magnetic forces}
\label{sec:conical}

For ferrofluid-filled droplets a shape transition into a 
stable conical shape   with $\psi(0)> 0$ 
is possible above a critical susceptibility $\chi_c$ and 
at high magnetic fields \cite{Bacri82,Wohlhuter92,Li1994,Stone99}.
We want to show that  a conical shape 
with a strictly conical tip can also exist for an elastic 
capsule with spherical rest shape and normal 
magnetic stretching forces if the constitutive relation
 is of the nonlinear  form 
(\ref{eq:taulambda}). 
Details of the argument are presented in
 Appendix \ref{app:cone_angle}.

The existence of sharp cones in   deformed membranes is an 
important issue in deformations of membranes with planar rest shape
\cite{Witten2007}.
A membrane of thickness $D$ 
prefers bending deformations (energy proportional to $D^3$) 
over stretching deformations  
 (energy proportional to $D$). If external forcing or constraints 
are such that stretching can be avoided, the membrane responds by pure
bending. 
Any deformation of such an  unstretched membrane
has to preserve the metric and thus the vanishing  Gaussian curvature 
of a plane. This results in so-called developable cones, which 
have zero Gaussian curvature everywhere except at the tip of the cone. 
 Cones only develop in response to external 
forces or constraints, typically under compressional constraints or forcing
as in the crumpling of paper. 
Then unstretched membranes develop  folds or wrinkles 
around the  developable cones in order to accommodate the excess 
area that occurs under compression \cite{BenAmar1997,Cerda1998,Witten2007}.

Our ferrofluid elastic membranes differ in several respects.
The magnetic forces are always {\it stretching}
forces and they are always
{\it normal} to the surface  such that the tangential  force 
equilibrium (\ref{eq:equilibrium_eqns_tang})
only involves internal stresses of the membrane. 
Under stretching  forces the membrane cannot respond by 
pure bending and changes in the metric are unavoidable. 
However, 
the forcing depends on the magnetic field distribution [see Eq.\ (\ref{fm})]
and becomes concentrated in points of high fields, which are typically 
points of high curvature. This  establishes a 
positive feedback between 
shape and magnetic field distribution that 
can stabilize conical tips. 
Moreover,  we consider membranes with spherical rest shape and, thus, 
non-zero Gaussian curvature $K=1/R_0^2$. This is another reason 
why  deformation 
into a cone with $K=0$ is impossible without stretching. 
Similar conditions (normal forces and  spherical rest shape) 
are fulfilled for spherical shells under 
point forces, where conical solutions have also been obtained
 \cite{Vella2012} and to which most of our results regarding the existence 
of conical shapes should also apply.

The tangential force equilibrium (\ref{eq:equilibrium_eqns_tang})
 has to be fulfilled  in the vicinity of the conical tip
and is independent of the stretching magnetic forces, 
which are always normal. 
In combination with the nonlinear constitutive relations
(\ref{eq:taulambda}) this requires that the stretching tensions remain 
  finite and  isotropic at the conical tip, i.e., 
 $\tau_s(0) = \tau_\varphi(0)>0$  at $s_0=0$. 
From the constitutive relations  then also follows 
the   isotropy of the stretches
$\lambda_s(0)=\lambda_\varphi(0)$ at the tip.
However, stretches are not necessarily finite at a conical tip.

For finite isotropic 
stretches $\lambda_s(0)=\lambda_\varphi(0)<\infty$ at the 
pole,  l'H{\^o}pital's rule applied at  $s_0=0$ gives
$\lambda_\varphi(0) = \lambda_s(0)\cos[\psi(0)]$  [see
Eq.\ (\ref{eq:lhospital})]. 
Then isotropy  requires  $\psi(0)=0$ and it follows  that 
 a sharp conical tip with $\psi(0) >0$ 
 is impossible if stretches remain finite at the tip.
Finite isotropic stretches at the pole thus always lead to 
flat tips with $\psi(0)=0$ as for the spheroidal shapes.

For diverging and asymptotically isotropic stretches  
\begin{equation}
 \lambda_s(s_0)\approx \lambda_\varphi(s_0)\approx {\rm const}\, s_0^{-\beta},
\label{eq:lambdadiv}
\end{equation}
with an exponent $\beta>0$;  
however, l'H{\^o}pital's rule does not apply at $s_0=0$.
Then we find instead that isotropy of the diverging 
stretches requires a conical tip with 
the relation
\begin{equation}
  \beta = \cos[\psi(0)]-1 = \sin\alpha-1
\label{eq:betaalpha}
\end{equation}
between the exponent $\beta$
and the half opening angle $\alpha = \pi/2 - \psi(0)$ of the 
conical tip [see Eq.\ (\ref{eq:betaalpha_app})]. 
This result can be obtained from a modified l'H{\^o}pital's rule
or directly from analyzing stretches for a deformation into a conical 
tip under the constraint of isotropy of the stretches at the tip
[see Eq.\ (\ref{eq:lambdaphi_d3_app})]. 
For the  nonlinear constitutive relation (\ref{eq:taulambda})
diverging and isotropic stretches are still  compatible with 
finite and isotropic tensions, which approach 
  $\tau_s(0) = \tau_\varphi(0) =  
 {Y_{2\text{D}}}/{(1-\nu)}$, see  Eq.\ (\ref{eq:tau0_app}), at the tip. 
Moreover, $\beta>-1$ according to  (\ref{eq:betaalpha}) and, therefore, 
the divergence is such that the elastic energy [the energy density 
 (\ref{eq:ws}) integrated over the tip area] remains finite.

Any numerical approaches to capsule shell mechanics and the calculation
of the magnetic fields rely on discretization. 
In the numerical solution of 
 axisymmetric shape equations the arc length $s_0$ 
is discretized. After discretization in the numerics, 
stretches necessarily remain 
finite at potential conical tips at the apices.
Then our results for finite stretches apply, and we have to choose
a boundary condition $\psi(0)=0$. 
Also,  for the calculation of the magnetic fields,
 we discretize the boundary of the capsule [see Eq.\
 (\ref{eq:Li})]. Therefore, also magnetic fields remain finite
at conical tips. Then also the normal magnetic forces remain 
finite and can only  support finite curvatures at the tip of the 
conical shape. 
This leads to a rounding  of conical tips and, thus, also requires 
 $\psi(0)=0$.
This implies that, in the numerical calculations,
all shapes of ferrofluid capsules will 
have rounded tips with $\psi(0)=0$; the rounding 
of a  conical tip for these numerical reasons will happen on the 
scale of the discretization of the problem. 
A boundary condition $\psi(0)$ 
for the numerical solution of the shape 
equations [see Eq.\ (\ref{eq:shape_eqns_bc})] has also 
been used 
in  Refs.\ \cite{Lavrova04, Lavrova05,Lavrova06,Lavrova2006}
for ferrofluid droplet shapes.

\subsection{Slender-body approximation for conical capsules}
\label{sec:Slender}

For ferrofluid droplets, the conical shape 
could be investigated analytically using a slender-body approximation 
\cite{Stone99}, which we want to adapt for  conical 
shapes of the ferrofluid-filled capsule.
We have shown that conical shapes can also exist for 
ferrofluid-filled capsules but they involve 
diverging isotropic stretches at the conical tip. 
Tensions are isotropic,
remain finite at the conical tip and approach the 
limiting values $\tau_s(0) = \tau_\varphi(0) =  
 {Y_{2\text{D}}}/{(1-\nu)}$ [see  Eq.\ (\ref{eq:tau0_app})]. 

The capsule shape is described by 
a function $r(z)$ in cylindrical coordinates. 
In a slender-body approximation, we assume $\partial_z r \ll 1$;
for a conical tip with half opening angle $\alpha = \pi/2 - \psi(0)$, 
we have $\partial_z r \approx \tan\alpha$ in the vicinity of the tip. 
Then we can neglect small 
radial field components and approximate 
 the magnetic field as  parallel to the $z$ axis, ${\bf H} = H(z) \vec{e}_z$.
The field $H(z)$ is determined by 
\begin{align}
     H_0 &= H(z) - \frac{\ln A}{2}\chi \partial_z^2\left[r^2(z) H(z)\right],
\label{eq:Hz_slender}
\end{align}
where $A$ is the aspect ratio of the slender shape, which can be expressed
in terms of  the half opening angle, $A=1/\tan \alpha$,
 for a conical shape \cite{Stone99}. 
This relation is unchanged as compared to fluid droplets as it is
a result of the slender shape and magnetic boundary conditions only 
and independent of the surface elasticity underlying the shape.

In the slender-body approximation we also assume $\partial_z^2 r\ll 1/r$
such that the meridional curvature is small $\kappa_s\ll \kappa_\varphi 
\approx 1/r(z)$. 
Then the 
Laplace-Young equation describing normal force equilibrium becomes
\begin{align}
  \label{eq:fbal_cap_slender}
 \frac{1}{r(z)} \left\{\tau_\varphi[r(z)] + \gamma\right\}
 &= p_0 + f_m. 
\end{align}
This relation differs from the corresponding relation for fluid 
droplets by the appearance of the additional elastic tension 
$\tau_\varphi=\tau_\varphi(r)$. 
As shown in  Appendix \ref{app:cone_tip2}, 
tangential force equilibrium is fulfilled in the vicinity of the 
conical tip if stretches are diverging, and the resulting 
 circumferential tension is 
\begin{align}
  \tau_\varphi(r)  &= 
    \frac{Y_{2\text{D}}}{1-\nu}\left[ 1 - 2R_0 \sin\alpha 
     ({a \tan\alpha})^{-1/\sin\alpha}
      r^{1/\sin\alpha -1} \right]
\label{eq:tau1}
\end{align}
[see Eq.\ (\ref{eq:tau1_app})]
in the vicinity of the conical tip. Note that $a$ still denotes the polar
radius.
In Appendix \ref{app:cone_eqs} we also outline how the 
tension $\tau_\varphi(r)$ could be calculated for a general shape 
$r(z)$, in principle.

The Laplace-Young equation (\ref{eq:fbal_cap_slender}) with an elastic tension
(\ref{eq:tau1}) and the  slender-body field equation (\ref{eq:Hz_slender})
provide two coupled equations for $r(z)$ and $H(z)$. 
The pressure $p_0$ has to be chosen such that the resulting shape $r(z)$
fulfills the volume constraint 
\begin{align}
   V_0 &=  \pi \int_{-a}^{a} r^2(z) dz.
\label{eq:vol_slender}
\end{align}
The three equations (\ref{eq:Hz_slender}), (\ref{eq:fbal_cap_slender}), 
and (\ref{eq:vol_slender}) governing slender (and, in particular,  conical)
 shapes of a 
ferrofluid-filled capsule only differ in the appearance of the additional 
elastic tension $\tau_\varphi=\tau_\varphi(r)$ from the corresponding equations 
for ferrofluid droplets from Ref.\ \cite{Stone99}.
They can be also be 
solved analogously  as for ferrofluid droplets, in principle.

\section{Results}
\label{sec:results}

\subsection{Spheroidal capsule shapes}
\label{sec:spheroid}

While the capsule is spherical at $B_m = 0$, it becomes
 elongated for increasing magnetic field or Bond number
 $B_m$ similarly to a ferrofluid droplet. We can quantify
 the elongation by the ratio of capsule length $a$ in the
 $z$ direction and capsule diameter $b$ at the equator, $a/b$. 
At small or moderate magnetic fields ferrofluid capsules 
assume a prolate spheroidal shape to a very good approximation; 
one example is shown in  Fig.\ \ref{fig:fieldplot}(b). 

For small fields we  calculated 
the linear response of the capsule  exactly
in Sec.\ \ref{sec:lin_def} and Appendix \ref{app:lin_def}
and found displacements (\ref{eq:AnsatzuRcap}), which 
 describe a prolate spheroid 
with an elongation $a/b>1$ given by Eq.\
(\ref{eq:lin_approx}). This analytical result is 
in excellent agreement with numerical results for small fields
(see Fig.\ \ref{fig:lin_region}).
The linear response regime is valid as long as 
$a/b-1 \ll 1$ or $B_m \ll   {\left[{Y_{2\text{D}}}/{\gamma}(5+\nu) +
    1\right]}(3+\chi)^2/\chi$  according to Eq.\ (\ref{eq:lin_approx}).

Small magnetic fields are easily accessible and 
for many ferrofluids, susceptibilities are rather small 
(for example, $\chi \simeq 0.36$
in Ref.\ \cite{Zhu11}). Therefore,  spheroidal shapes in 
the linear response regime are 
experimentally easily accessible. Then 
the linear response relation (\ref{eq:lin_approx})
can  be used as experimental method 
to deduce unknown capsule material properties, for example, 
Young's modulus $Y_{\text{eD}}$ if the magnetic properties of the ferrofluid 
are known.

\begin{figure}[htbp]
\begin{center}
\includegraphics[angle=0,width=0.9\textwidth]{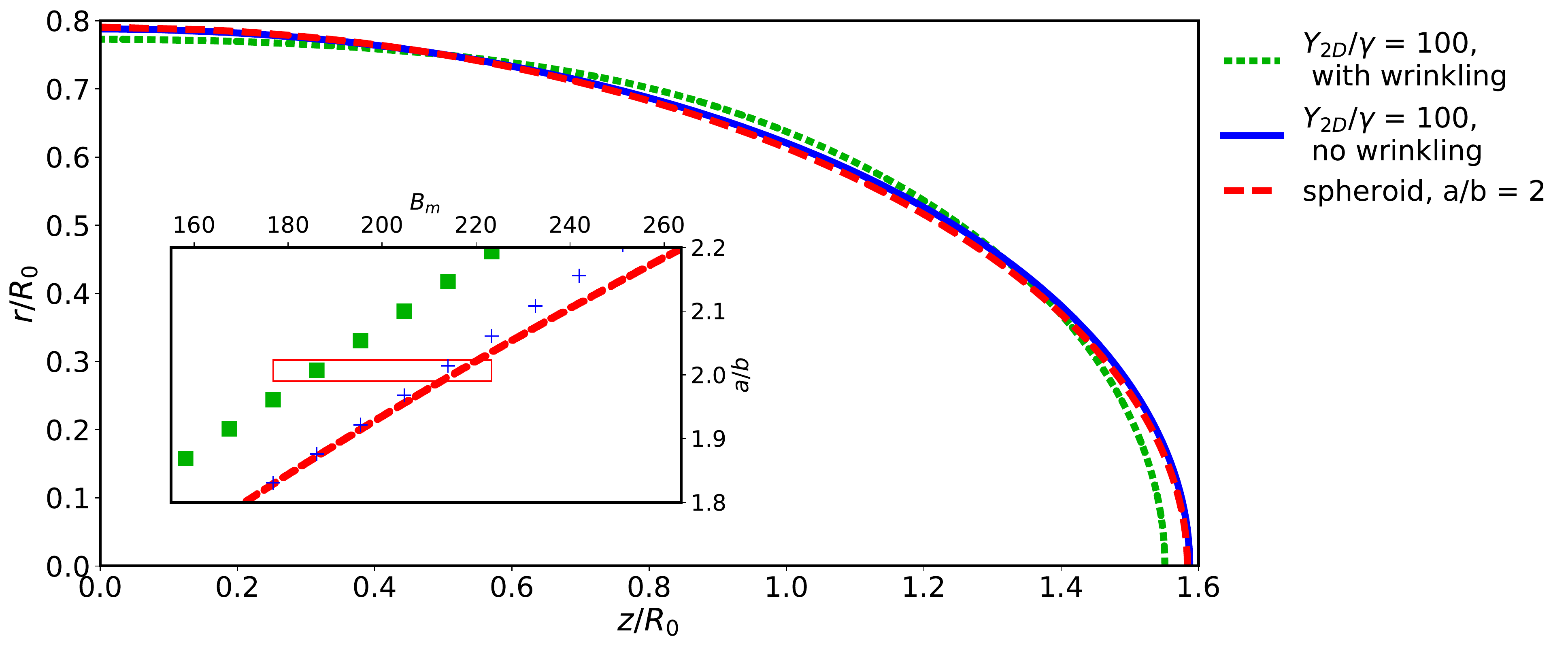}
\caption{
   Comparison of numerically calculated 
    $r(z)$ contour of a capsule with 
  ${Y_{2\text{D}}}/{\gamma}=100$,  and $\chi = 21$, 
  for  a value $B_m$ 
  chosen such that the elongation is 
  ${a}/{b}=2$ (the inset
  shows the location of the pictured shapes in the
  $B_m$-${a}/{b}$ plane) with 
   a spheroid.  
 The shape calculated without wrinkling (blue solid line)
  shows very good agreement with a spheroid of the same volume and
  elongation (red dashed line). Taking wrinkling into account leads to
   visible deviations (green dotted line).
}
\label{fig:ellipsoidplot}
\end{center}
\end{figure}

At moderate magnetic fields,
 the capsule shape remains very  similar to a prolate spheroid
 for   all elongations  $a/b \lesssim 3$, which was one basic assumption of 
the approximative energy minimization in 
 Sec.\ \ref{sec:AnalyticalApprox}. 
 Figure \ref{fig:ellipsoidplot} demonstrates this for shapes with 
  ${a}/{b}=2$. 
 The spheroidal approximation 
 works  better   for systems dominated by  the surface tension,
i.e., for small ratios ${Y_{2\text{D}}}/{\gamma}$. 
 For fixed Bond number $B_m$ and susceptibility $\chi$ the
 elongation decreases with increasing  ${Y_{2\text{D}}}/{\gamma}$
 because of the additional stretching  energy of  the shell as compared 
to a droplet, so a ferrofluid
 droplet (${Y_{2\text{D}}}/{\gamma}=0$) 
always  shows  the highest elongation.
For small fields, this trend can be quantified 
with the linear response relation (\ref{eq:lin_approx}). 
 For smaller elongations, the spheroidal approximation 
tends to work better.

\begin{figure}[htbp]
\begin{center}
\includegraphics[width=1\textwidth,clip]{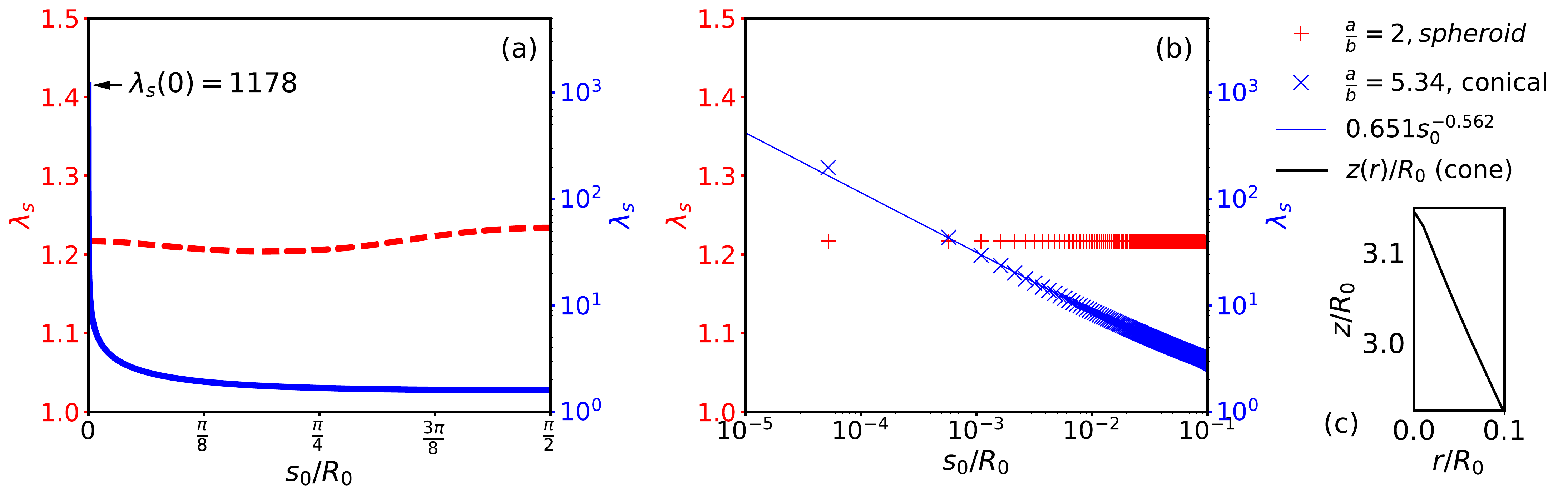}
\caption{
  (a) Stretch factors in the meridional direction $\lambda_s(s_0)$ following
  the whole contour line from the south pole ($s_0 = 0$) to the equator
  (${s_0}/{R_0} = {\pi}/{2}$) for ${Y_{2\text{D}}}/{\gamma}=100$
  and $\chi = 21$. The left scale (red dashed line) gives 
   almost constant 
   stretch factors for a spheroidal shape with 
  ${a}/{b} = 2$. The right scale (blue solid line) gives 
   diverging stretch factors 
   for a conical shape with 
  ${a}/{b} = 5.34$.
 (b) Logarithmic plot of $\lambda_s(s_0)$ near the tip
  for  $s_0/R_0 < 10^{-1}$.
  The function $\lambda_s(s_0) = {\rm const}\, s_0^{-\beta}$ [see Eq.\
   \eqref{eq:lambdadiv}] 
  was fitted to the data of the conical shape, 
  which gave $\beta = 0.562$, corresponding 
  to an angle $\alpha = 25.98^{\circ}$ in Eq.\ \eqref{eq:betaalpha}.
  (c) Zoom in to the tip of the contour line $z(r)$ 
   for the  conical shape; the half opening angle
   is  $\alpha \approx 25^{\circ}$.
}
\label{fig:lambda_s}
\end{center}
\end{figure}

 The other assumption in 
the approximative energy minimization in 
 Sec.\ \ref{sec:AnalyticalApprox} was
 constant stretch factors throughout the shell, i.e.,  
$\lambda_s, \lambda_\varphi = \text{const}$ (and thus constant 
elastic tensions $\tau_\varphi$ and $\tau_s$).
Also this approximation works very well for spheroidal shapes 
with elongations  $a/b \lesssim 3$,
as the numerical results in Fig.\ \ref{fig:lambda_s} 
for $a/b=2$
(left scale, red line)  show. 

As a result,  the approximative energy minimization in 
 Sec.\ \ref{sec:AnalyticalApprox} gives very good results for 
moderate magnetic fields, i.e., for   all elongations
  $a/b \lesssim 3$, where we always find 
prolate spheroidal shapes, as  the comparison with 
 numerical results  in 
 Fig.\ \ref{fig:ab_Bm_plot} shows.

\subsection{Conical capsule shapes and capsule rupture}
\label{sec:cone}

For large  magnetic fields or Bond numbers
 $B_m$ and at sufficiently high susceptibilities $\chi$, 
ferrofluid capsules can also assume 
conical shapes, such as the shape in 
Fig.\ \ref{fig:fieldplot}(c), which 
have also been found for ferrofluid droplets
\cite{Li1994,Stone99}.
We investigated the possibility of conical shapes 
for elastic capsules  with normal magnetic forces
above in Sec.\ \ref{sec:conical} and found 
that  stretch factors have to 
diverge at the  conical tips, 
$\lambda_s(s_0)\approx \lambda_\varphi(s_0)\approx {\rm const}\, s_0^{-\beta}$
[see Eq.\ (\ref{eq:lambdadiv}], with an exponent 
$\beta = \sin\alpha-1$, which is determined by the 
 half opening angle $\alpha = \pi/2 - \psi(0)$ of the 
conical tip [see  Eqs.\ (\ref{eq:betaalpha}) and (\ref{eq:betaalpha_app})]. 
This behavior is confirmed by our numerical results 
 in Fig.\ \ref{fig:lambda_s} 
(left scale, blue line). 
The stretch factors diverge but are asymptotically isotropic  at the 
tips. The 
 nonlinear constitutive relations (\ref{eq:taulambda}) then 
result in finite and isotropic tensions 
$\tau_s(0) = \tau_\varphi(0) =  
 {Y_{2\text{D}}}/{(1-\nu)}$ [see  Eq.\ (\ref{eq:tau0_app})].

Diverging stretch factors cannot be realized in an actual 
material without rupture.  Typical 
alginate capsule materials can only resist stretch factors of 
$\lambda\simeq 1.2$ before rupture; 
 highly stretchable hydrogels can resist
stretch factors up to $\lambda \sim 20$
\cite{Sun12}.
Therefore, a real capsule should  rupture at the 
poles  at the transition into a conical shape
and we conclude 
that investigations of conical shapes
are primarily of theoretical interest.
Such rupture events have actually been observed in 
Ref.\ \cite{Karyappa2014} for capsules
 filled with a dielectric liquid in external 
electric fields. 
We expect that the nonlinear Hookean material law 
will become invalid at such high stretch factors prior 
to rupture. Then  constitutive relations which are more realistic 
for high strains should be used.
Nevertheless, the appearance of large 
 stress factors is a  robust feature of the conical 
shape independently of the material law.

Conical shapes cannot be described quantitatively  by the 
approximative energy minimization from 
 Sec.\ \ref{sec:AnalyticalApprox}   as 
spheroidal shapes with a large elongation $a/b$,
which is clearly shown by 
 the deviations  between 
 numerical results (data points) 
 and the approximative energy minimization from 
 Sec.\ \ref{sec:AnalyticalApprox} (solid lines) for the conical 
  shapes in 
 Fig.\ \ref{fig:ab_Bm_plot}.
For ferrofluid droplets, conical shapes can be described
by a slender-body theory \cite{Stone99}, which we generalized in
 Sec.\ \ref{sec:Slender} to ferrofluid-filled capsules. 
The three governing equations (\ref{eq:Hz_slender}), 
(\ref{eq:fbal_cap_slender}), 
and (\ref{eq:vol_slender}) from Sec.\ \ref{sec:Slender}  
can be used to describe 
conical capsule shapes quantitatively.

As pointed out above, the tensions remain 
  finite and isotropic  at the conical tip, i.e.,  
$\tau_s(r) \approx  \tau_\varphi(r) \approx {Y_{2\text{D}}}/{(1-\nu)}$ 
for small $r$ [see Eq.\ (\ref{eq:tau0_app})].
Then the slender-body equation (\ref{eq:fbal_cap_slender})
from normal force balance actually becomes 
identical to the corresponding equation for a droplet 
from Ref.\ \cite{Stone99}, however, with an effectively 
increased surface tensions $\gamma_{\rm eff} = \gamma + \tau_\varphi(0)$. 
Also the other two   equations (\ref{eq:Hz_slender})
and (\ref{eq:vol_slender}) are identical such that 
 we obtain very similar slender conical shapes 
for capsules and droplets, which can be mapped onto each other 
by a simple  shift of the surface tension.

The mechanism underlying the stabilization of the conical 
shape is analogous to ferrofluid droplets 
because tensions remain finite and isotropic 
at the  conical tip.
 A sharp conical   tip with 
 curvatures $\kappa_\varphi \propto 1/r$
gives rise to diverging magnetic fields 
 $H\propto r^{-1/2}$ and normal magnetic forces 
\begin{equation}
   f_m  \propto H^2 \propto r^{-1},
\label{eq:Hdivergence}
\end{equation} 
{\it both} for ferrofluid droplets and capsules. 
These strong magnetic stretching forces 
 stabilize the conical  tip against  high 
elastic restoring forces. 
The normal component of the  elastic force
is mainly due to the finite  circumferential tension $\gamma+\tau_\varphi(0)$
acting along the high  circumferential  
curvature  $\kappa_\varphi \propto 1/r$ at the conical tip, resulting 
in an elastic force 
$f_{\rm el} \propto  [\gamma+\tau_\varphi(0)]\kappa_\varphi\propto r^{-1}$
with the same divergence.
Magnetic and elastic normal forces  balance in the Laplace-Young  equation 
(\ref{eq:fbal_cap_slender}) in the slender-body approximation. 
The magnetic field  exponent $H\propto r^{-1/2}$ is {\it identical} for 
capsules and droplets, as long as the elastic tensions at the conical 
tip are finite. 
This exponent determines the critical susceptibility $\chi_c$ above which 
a shape transition into conical shapes is possible and therefore
we also find the identical $\chi_c$  for capsules and droplets
as  discussed in the following section.

\subsection{Spheroidal-conical shape transition of capsules}

\begin{figure}[htbp]
\begin{center}
\includegraphics[width=1\textwidth,clip]{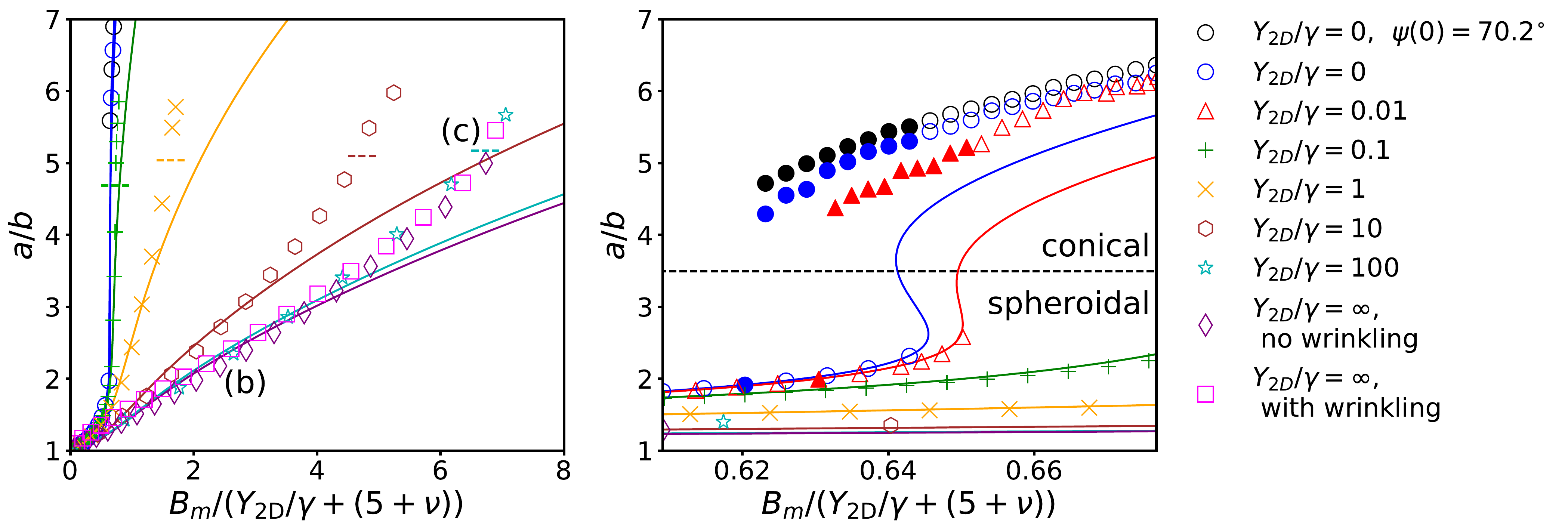}
\caption{
Elongation ${a}/{b}$ of a capsule filled with a ferrofluid 
  with $\chi = 21$ as a function of
 magnetic Bond number  $B_m$ for 
   different values of the dimensionless 
elastic parameter ${Y_{2\text{D}}}/{\gamma}$. 
  The magnetic Bond number is rescaled by ${Y_{2\text{D}}}/{\gamma} +
  (5+\nu)$,
   which is motivated by the small field behavior [see 
   Eq.\  (\ref{eq:lin_approx})]. 
  The solid lines describe the theoretical results from 
   approximative energy minimization (see 
  Sec.\ \ref{sec:AnalyticalApprox}). Open (closed) symbols 
  denote numerical data for
  increasing (decreasing) $B_m$.
 The agreement is good for small
   elongations; the approximation fails for  higher elongations, especially
  at the shape transition (close-up in the right diagram), where ${a}/{b}$
  jumps for small changes of $B_m$. Hysteresis effects are clearly visible in
  that area. There are two sets of numerical data for
  ${Y_{2\text{D}}}/{\gamma}=\infty$: Square  data points are based on the
  modified shape equations that take wrinkling into account, while
  diamonds are calculated without wrinkling.
  There are also two sets of data without elasticity: 
The upper data points 
  (black) describe a droplet with a real conical tip with a
 cone angle of $\psi(0)=70.2^{\circ}$, 
  as it was given in Ref.\ \cite{Stone99};
  for the lower points (blue) we used the boundary 
condition $\psi(0)=0$.
  Dashed lines indicate the position of shape transitions. 
  Above these lines, shapes are conical, while they are spheroidal below. 
  The markers (b) and (c) correspond to the 
  shapes in Fig.\ \ref{fig:fieldplot}.
 }
\label{fig:ab_Bm_plot}
\end{center}
\end{figure}

Upon increasing the magnetic field or the magnetic Bond number $B_m$ 
at fixed capsule 
elasticity ${Y_{2\text{D}}}/{\gamma}>0$ and 
for a sufficiently large and fixed ferrofluid susceptibility $\chi$,
we find a discontinuous shape transition from 
spheroidal to conical capsule shapes, similar to what has been 
found for ferrofluid droplets (${Y_{2\text{D}}}/{\gamma}=0$) 
\cite{Bacri82,Li1994,Stone99}.
One of our main results is the diagram of capsule
elongation $a/b$ as a function of Bond number $B_m$  
in Fig.\ \ref{fig:ab_Bm_plot} for different values of 
elasticity parameters ${Y_{2\text{D}}}/{\gamma}$ and for 
$\chi=21$, where a lower 
spheroidal branch and an upper conical branch and a 
discontinuous transition between both branches 
can be identified. 
In the following sections we will discuss different 
aspects of this shape transition in more detail.

\subsubsection{Critical susceptibility $\chi_c$}
\label{sec:chic}

For ferrofluid droplets, 
a discontinuous shape transition was observed in experiments
 \cite{Bacri82, Bashtovoi87} and numerical simulations \cite{Lavrova04,
   Afkhami10} only for  susceptibilities $\chi>\chi_c$, i.e., above
 a critical susceptibility $\chi_c$. 
In Ref.\ \cite{Li1994}
 a value $\chi_c =\mu_c/\mu_{\rm out}-1 \simeq 16.59$
 was found below which no conical shape can exist; 
the  slender-shape approximation for droplets 
from  Ref.\ \cite{Stone99}, which we  generalized to elastic capsules in 
Sec.\ \ref{sec:Slender},
gives $\chi_c = 16e/3 \simeq 14.5$.
The   approximative energy minimization  
of  Bacri and Salin \cite{Bacri82}, which we 
 generalized to elastic capsules in  
 Sec.\ \ref{sec:AnalyticalApprox},  gives  $\chi_c \simeq 19.8$ 
for ferrofluid droplets. 
Numerically, a range of $\chi_c \simeq 19$ to $\chi_c \simeq 19.5$ 
is observed \cite{Wohlhuter92}.
The question arises whether a critical susceptibility $\chi_c$ 
can also be found for the existence of a discontinuous  spheroidal-conical 
transition  for ferrofluid-filled 
elastic capsules.

For given $\chi$ and half opening angle $\alpha$  of the conical 
shape electromagnetic boundary conditions determine the divergence 
 $H \propto r^{\mu-1}$ of the field via the equation \cite{Li1994,Ramos1994}
\begin{equation}
  P_\mu(\cos\alpha) P_\mu'(-\cos\alpha) + (\chi+1)
  P_\mu(-\cos\alpha) P_\mu'(\cos\alpha)=0.
\label{eq:Li1994}
\end{equation}
Because of the finite elastic tension $\tau_\varphi(0)$ 
at the  conical tip,  the magnetic field at the tip of a conical 
capsule 
diverges with the same $\mu=1/2$ [see Eq.\ (\ref{eq:Hdivergence})]
as for a conical droplet. Therefore, we find the same
critical susceptibility $\chi_c\simeq 16.59$, above which a
conical solution can exist, for both capsules and droplets. 

In the slender-body approach, 
Eq.\ (\ref{eq:Hz_slender}) determines $\chi_c$ and
applies unchanged to both 
   slender conical droplets and 
ferrofluid-filled  capsules. 
Also the magnetic field divergence $H \propto r^{-1/2}$
is identical in both cases, so the 
analysis of Eq.\ (\ref{eq:Hz_slender}) predicts  the  same
critical value $\chi_c = 16e/3 \simeq 14.5$ for ferrofluid-filled capsules 
as for ferrofluid droplets.

In particular, both the analysis of Eq.\ (\ref{eq:Li1994})
and the slender-body approach  predict that 
the  value 
for $\chi_c$ to be {\it independent} of the 
Young modulus $Y_{2\mathrm{D}}$   of the capsule. 
This result is corroborated by our numerics for $\chi=21$, where we 
{\it always} observe a spheroidal-conical shape transition, even for 
$Y_{2\text{D}}/\gamma \to \infty$ [see Eq.\ (\ref{fig:ab_Bm_plot})].

This result 
 is in contrast, however,  to what we find using the approximative  
 energy minimization  for spheroidal shapes from 
Sec.\ \ref{sec:AnalyticalApprox}. Analyzing Eq.\ (\ref{eq:BMab}), 
$B_m=g(k)=g(b/a)$, 
for the saddle points of the function $g(k)$ gives the critical value
of the susceptibility $\chi_c$ [the two equations $g'(k)=0$ and $g''(k)=0$ 
determine two critical parameter values $k=k_c$ and $\chi=\chi_c$]. 
Using this approach, we find a $\chi_c$,
which is strongly increasing with the Young modulus $Y_{2\text{D}}/\gamma$,
such that  we find $\chi_c >21$ already for $Y_{2\text{D}}/\gamma> 0.015$, 
which clearly disagrees with all our numerical and analytical results. 
The reason for this disagreement is the failure of the approximative energy
minimization to correctly describe conical shapes  as 
discussed in Sec.\  \ref{sec:cone}.

It is interesting to consider the robustness of our result
of a $Y_{2\text{D}}$-independent $\chi_c$ that is identical 
to the $\chi_c$ for ferrofluid droplets  
with respect to the constitutive relation. We used the nonlinear
Hookean constitutive relation  (\ref{eq:taulambda}),
which can only support {\it finite} tensions at a conical tip,
 even for diverging stretches (see Sec.\ \ref{sec:conical}).
A simple linear Hookean constitutive relation [missing the $1/\lambda$-factors
in Eq.\ (\ref{eq:taulambda})] behaves differently and 
exhibits diverging tensions $\tau_\varphi \sim r^{-\sigma}$ with $\sigma>0$ 
at a conical tip. 
Then tangential force equilibrium (\ref{eq:fbalt_cap0}) also requires 
$\tau_s \sim\tau_\varphi \sim r^{-\sigma}$ but with an anisotropy 
$\tau_\varphi/\tau_s = 1-\sigma$.
With the linear constitutive relation 
this  in turn leads to stretches $\lambda_s\sim \lambda_\varphi\sim
r^{-\sigma}$  with an anisotropy
$\lambda_\varphi/\lambda_s  = (1-\nu-\sigma)/(1-\nu + \nu\sigma)\equiv \delta$
or $\delta(\sigma) = (1-2\sigma)/(1+\sigma)$ for a Poisson ratio $\nu=1/2$. 
 Requiring this anisotropy in Eq.\ (\ref{eq:lambda_d2_app}) 
at a conical tip with half opening angle $\alpha$ 
leads to a modified differential equation (\ref{eq:ODE_d_app}) 
and a divergence 
$\lambda_s \sim \lambda_\varphi \sim r^{1-1/\delta(\sigma)\sin\alpha}$.
Consistency with $\lambda_s\sim \lambda_\varphi\sim
r^{-\sigma}$ then requires 
\begin{equation*}
   \sigma = \frac{1}{\delta(\sigma) \sin\alpha} -1 
     = \frac{1+\sigma}{1-2\sigma} \frac{1}{\sin\alpha} -1,
\end{equation*}
which determines the divergence 
$\sigma=\sigma(\alpha)$ of tensions 
$\tau_s \sim\tau_\varphi \sim r^{-\sigma}$
as a function of the opening angle $\alpha$. 
At the conical tip we have now curvatures
$\kappa_\varphi \propto 1/r$ in combination 
with circumferential tensions $\tau_\varphi \sim r^{-\sigma}$
such that normal force balance also requires magnetic 
forces $f_m \propto H^2 \propto  r^{-1-\sigma}$ [cf.\ 
Eq.\ (\ref{eq:Hdivergence})]. 
 Thus, we have to use 
$\mu = 1-\sigma(\alpha)$ instead of $\mu=1/2$ in  $H \propto r^{\mu-1}$ 
in Eq.\ (\ref{eq:Li1994}) and obtain 
a modified equation for the  cone angle $\alpha$ as a function of the 
parameter $\chi$. 
This equation has a solution only above $\chi_c \simeq 40.5$ and thus
the critical value $\chi_c$ is strongly increased for a 
strictly linear  Hookean constitutive relation.
Our numerical results corroborate this result as we find 
only spheroidal capsule shapes for a strictly linear constitutive relation 
at a susceptibility  $\chi=21$. 
This shows that the value of $\chi_c$ is very sensitive to changes
in the constitutive relation and a measurement of $\chi_c$ allows us to 
draw conclusions about the constitutive relation of the 
capsule material.

\subsubsection{Critical Bond numbers}

Our numerical solutions of the shape equations show
that the discontinuous spheroidal-conical shape transition 
that exists for ferrofluid droplets \cite{Bacri82,Li1994,Stone99} 
persists for  ferrofluid-filled elastic capsules and 
shows qualitatively similar  features.
Both for droplets and for capsules, 
the driving force of the shape transition is the 
lowering of the magnetic field energy in the conical shape. 
Above an upper  critical Bond number  $B_{m,c2}$ 
the spheroidal shape becomes unstable and the droplet or capsule  deforms
into a much more elongated,  conical shape. 
This shape transition is discontinuous, i.e., the deformation 
into the conical shape is  associated with a
jump in $a/b$. The discontinuous  transition between spheroidal to conical 
shapes also  exhibits hysteresis: Lowering the  Bond number starting 
from values $B_m>B_{m,c2}$, the conical shape becomes unstable at a 
lower critical Bond number $B_{m,c1}$ with $B_{m,c1}<B_{m,c2}$. 
 The discontinuous spheroidal-conical
  transition only exists above the  critical susceptibility $\chi_c$. 
In other words, both droplets and capsules  exhibit  a line of discontinuous 
shape transitions  in the
$\chi$-$B_m$ plane for $\chi>\chi_c$, 
which terminates at a critical point located 
  at $\chi=\chi_c$. The lines $B_{m,c1}(\chi)$ and  $B_{m,c2}(\chi)$
are the limits of stability (spinodals) of this shape transition 
and meet in the critical point.

Figure \ref{fig:ab_Bm_plot} shows the capsule elongation with
 respect to $B_m$ for different values of the dimensionless 
elastic parameter ${Y_{2\text{D}}}/{\gamma}$ of the capsule. We
 choose $\chi = 21$, which is only slightly above $\chi_c$. 
This ensures that we 
 have a shape transition for a ferrofluid
 droplet (corresponding to the limit ${Y_{2\text{D}}}/{\gamma}=0$),
 on the one hand, and relatively small and thus
 numerically more stable
 elongations in the  conical shape, on the other hand.
Figure \ref{fig:ab_Bm_plot} clearly shows a discontinuous jump in elongation
and hysteresis effects also for capsules with 
 ${Y_{2\text{D}}}/{\gamma}>0$.

\subsubsection{Stretch factors as order parameter}

The discontinuous jump in  the elongation  
ratio $a/b$  at the spheroidal-conical transition 
is difficult to localize 
for larger values of  ${Y_{2\text{D}}}/{\gamma}$, as 
Fig.\ \ref{fig:ab_Bm_plot}  shows. 
 More suitable order parameters for the spheroidal-conical
transition 
are the stretch factors $\lambda_s$ and  $\lambda_\varphi$. 
Because the stretch factors diverge at the tips of the conical shape
(the divergence is only limited by numerical discretization effects), 
whereas they stay finite at the poles of spheroidal shape (see 
Fig.\ \ref{fig:lambda_s} and our above discussion), we can directly 
employ the stretch factor $\lambda_s(s_0=0)$ at one of the poles as 
a convenient   order parameter.

\begin{figure}[htbp]
\begin{center}
 \includegraphics[width=0.5\textwidth,clip]{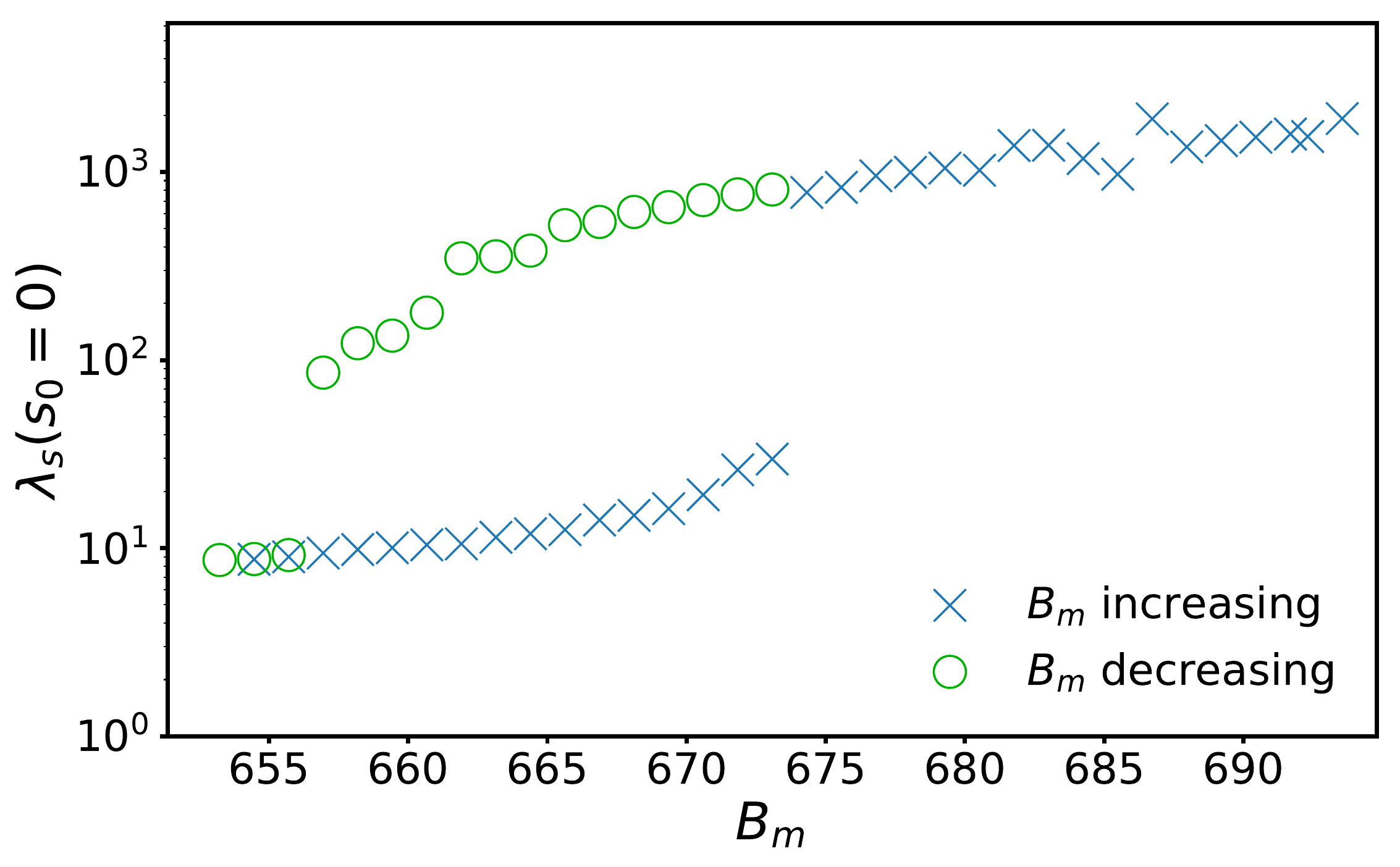}
 \caption{
 Meridional stretch factor 
   $\lambda_s$ at the capsule pole $s_0=0$ as a function of 
 Bond number $B_m$  for  ${Y_{2\text{D}}}/{\gamma}=100$ and   $\chi=21$. 
  The stretch factor clearly exhibits a jump at 
  the location of the discontinuous shape transition 
 and hysteretic behavior. 
 }
 \label{fig:lambda_s_ab}
 \end{center}
\end{figure}

 For $\chi=21$ and ${Y_{2\text{D}}}/{\gamma}=100$, 
the shape transition occurs
where ${a}/{b}$ has  a rather small jump from about 5.2 to 5.35 for
increasing Bond number 
$B_m$, whereas the stretch factor  $\lambda_s(s_0=0)$ exhibits a much
bigger jump  by a factor of more than 10, as 
demonstrated in Fig.\ \ref{fig:lambda_s_ab}. 
Also the shape hysteresis at the spheroidal-conical shape transition 
can be clearly seen for the order parameter  $\lambda_s(s_0=0)$.

Using this order parameter, we can 
detect the  spheroidal-conical shape transition of 
ferrofluid-filled capsules by the criterion 
\begin{align}
  \label{eq:transition_criterion}
  \lim\limits_{\Delta B_m \to 0} |\lambda_s(s_0 = 0, B_m)
      - \lambda_s(s_0 = 0 ,B_m + \Delta B_m)| > 0,
\end{align}
 where we use values $\Delta B_m = 0.005$ for ${Y_{2\text{D}}}/{\gamma}<1$
up to values $\Delta B_m = 0.5$ for ${Y_{2\text{D}}}/{\gamma}=100$ in practice
[$B_{m,c1}$ and $B_{m,c2}$ grow approximately linearly with
${Y_{2\text{D}}}/{\gamma}$ (see Fig.\ \ref{fig:hysteresisplot} below)
 such that larger values $\Delta B_m$ can be 
used for larger ${Y_{2\text{D}}}/{\gamma}$; smaller values of $\Delta B_m$ 
give more precise results].
For  ferrofluid droplets, i.e., in the limit  
${Y_{2\text{D}}}/{\gamma}\approx 0$, we still have to use jumps 
in the elongation $a/b$ for small changes 
$\Delta B_m$ in the magnetic Bond number to detect the 
 spheroidal-conical shape transition.

We note that  the discretization problem at the sharp conical tip
mentioned above causes high relative errors in the 
numerical values of stretch factors in the tip area. 
Therefore, our numerical results
for the diverging stretch factors at the tips of 
 conical capsule shapes  cannot be  numerically exact. 
The detection of a divergence in $\lambda_s$ at the poles,
which we use to detect the transition into a conical shape, is, however, 
still possible even in the presence of numerical errors.

\subsubsection{Shape hysteresis}
\label{sec:hysteresis_effects}

In order to track the range of elastic control parameters 
 ${Y_{2\text{D}}}/{\gamma}$, where a discontinuous shape transition 
 with hysteresis can be observed (for fixed $\chi=21$), we 
use the stretch factor  $\lambda_s(s_0=0)$ as 
the  order parameter and 
the criterion  \eqref{eq:transition_criterion} to determine 
$B_{m,c1}$ and $B_{m,c2}$. We determine $B_{m,c2}$ by increasing the 
Bond number  in small steps $\Delta B_m>0$
 to locate the jump in 
the stretch factor  $\lambda_s(s_0=0)$ at the pole, when the 
spheroidal shape becomes unstable. Analogously,  we determine 
 $B_{m,c1}$ by decreasing the 
Bond number  in small steps $\Delta B_m<0$ to locate the jump in 
 $\lambda_s(s_0=0)$, when the 
conical  shape becomes unstable (see Fig.\ \ref{fig:lambda_s_ab}).

\begin{figure}[htbp]
\begin{center}
\includegraphics[width=1\textwidth,clip]{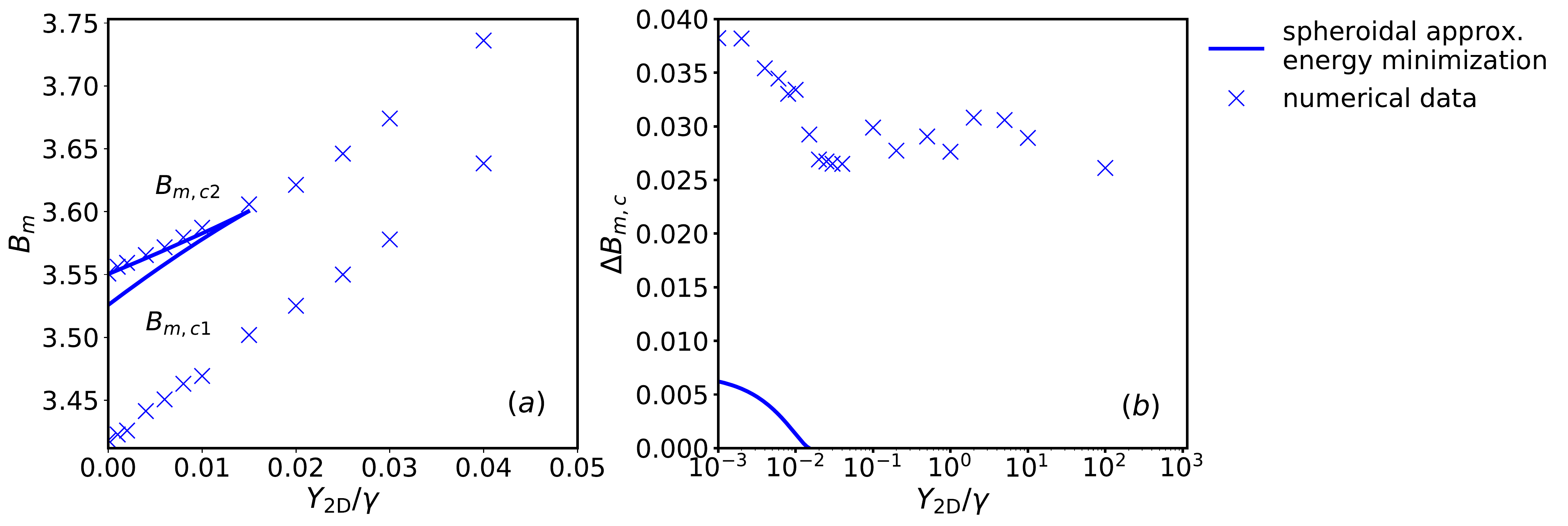}
\caption{
  (a) Critical Bond numbers $B_{m,c1}$ (lower data points)
  and $B_{m,c2}$ (upper data points) for varying ${Y_{2\text{D}}}/{\gamma}$
  with $\chi = 21$. The solid lines describe the prediction by the
  approximative energy minimization for  spheroidal shapes.
  Both critical Bond numbers increase for increasing
  ${Y_{2\text{D}}}/{\gamma}$.  
   In the region  $B_{m,c1}<B_m< B_{m,c2}$ there are
  hysteresis effects in the  spheroidal-conical shape transition. 
  (b) Relative size $\Delta B_{m,c}$ of the hysteresis
  area for a wider range of ${Y_{2\text{D}}}/{\gamma}$.}
\label{fig:hysteresisplot}
\end{center}
\end{figure}

 Repeating this procedure for
increasing values of the elastic control parameter 
${Y_{2\text{D}}}/{\gamma}$, we obtain
the location and size of the hysteresis loop  $B_{m,c1}<B_m<B_{m,c2}$
for a fixed susceptibility as a function of
${Y_{2\text{D}}}/{\gamma}$  (see Fig.\ \ref{fig:hysteresisplot}). 
We see that $B_{m,c1}$ and $B_{m,c2}$ increase (approximately linear) for 
increasing ${Y_{2\text{D}}}/{\gamma}$  because of the
increasing elastic energy needed for the same deformation.
Note that the absolute numerical 
values of $B_{m,c1}$ and $B_{m,c2}$ cannot be considered exact as they 
are  depending on the 
discretization of the magnetic field calculation 
(see also Appendix \ref{app:errors}).

 The approximative energy minimization  for spheroidal shapes from 
Sec.\ \ref{sec:AnalyticalApprox} can be used to 
 calculate approximative values for $B_{m,c1}$ and $B_{m,c2}$ from 
 Eq.\ (\ref{eq:BMab}), $B_m=g(k)=g(b/a)$
[the two equations $g'(k)=0$ and $B_{m} = g(k)$ determine the 
critical Bond numbers $B_m=B_{m,c1/2}$ and a corresponding critical 
inverse aspect ratio $k=k_c$]. We find that 
the hysteresis loop closes already for $Y_{2\text{D}}/\gamma> 0.015$ 
for $\chi=21$ (see Fig.\ \ref{fig:hysteresisplot}),
which is equivalent to our above finding (see
Sec.\ \ref{sec:chic}) that 
 $\chi_c >21$  for $Y_{2\text{D}}/\gamma> 0.015$ in 
 the approximative energy minimization. 
Comparison with our numerical results  in Fig.\ \ref{fig:hysteresisplot}
shows that the 
approximative energy minimization gives quite accurate results 
 for the upper critical Bond number $B_{m,c2}$, i.e., 
 the stability limit of the spheroidal shape. 
 It fails completely to predict the lower critical Bond number $B_{m,c1}$,
 i.e., the stability limit of the  conical shape, because 
it is not able to describe conical shapes quantitatively 
(see Sec.\ \ref{sec:cone}).

The numerical calculation shows
hysteresis behavior for {\it all}  values of ${Y_{2\text{D}}}/{\gamma}$
(see  Fig.\ \ref{fig:hysteresisplot}).
Only the relative size of the hysteresis loop,
$ \Delta B_{m,c} \equiv {2(B_{m,c2} - B_{m,c1})}/{(B_{m,c2} + B_{m,c1})}$,
 decreases slightly for increasing ${Y_{2\text{D}}}/{\gamma}$ 
in the numerical results.

\subsection{Wrinkling}

\subsubsection{Wrinkled shapes}

As opposed to liquid droplets, elastic capsules can develop wrinkles 
if a part of the shell is under compressive stress
\cite{Rehage2002,Vella2011,Aumaitre2013,Knoche13}.
Wrinkles have also been considered  for the equivalent problem of 
capsules filled with a dielectric liquid in an external 
electric field  in Ref.\  \cite{Karyappa2014}.

As it was stated in Sec.\ \ref{sec:WrinkleTheorie}, wrinkles
appear if the total hoop stress becomes compressive, 
$\tau_\varphi + \gamma < 0$. Then we have to use modified 
shape equations (\ref{eq:wrinkle_eqs}) in the numerical 
calculation of the shape.

As can be seen in Fig.\ \ref{fig:ab_Bm_plot}, taking 
wrinkling into account has
a visible effect on the capsule's elongation for higher values of
${Y_{2\text{D}}}/{\gamma}$. If wrinkling is taken into account 
capsules elongate because
 wrinkling reduces the compressional stretch energy, which is stored near
the equator.  This elastic  energy gain can be used 
for a further elongation of
the capsule at the same field strength to lower the magnetic energy. 
 This also results in stronger deviation
from the spheroidal shape.
To visualize this effect, Fig.\ \ref{fig:ellipsoidplot} shows
the projection of the contour line of the upper right quadrant of capsules
with and without wrinkling  using the same elongation ${a}/{b}=2$. 
While the shape is indistinguishable from a spheroid without wrinkling, 
the wrinkled shape  deviates from a spheroid. 

Also in the presence of wrinkling,  the discontinuous 
 spheroidal-conical shape transition where the elongation increases 
 persists. 
In the following, we will focus on the effect of wrinkles on 
the spheroidal branch of shapes.

\subsubsection{Extent of wrinkled region}

In order to characterize the wrinkling tendency of  
spheroidal capsules  we calculate the extent of the wrinkled region
$L_{\text{w}}$ [cf.\ Eq.\ (\ref{eq:Lw}) and Fig.\ \ref{fig:3Dwrinkling}],
which  can  easily be measured  in experiments.

\begin{figure}[htbp]
\begin{center}
\includegraphics[width=0.54\textwidth,clip]{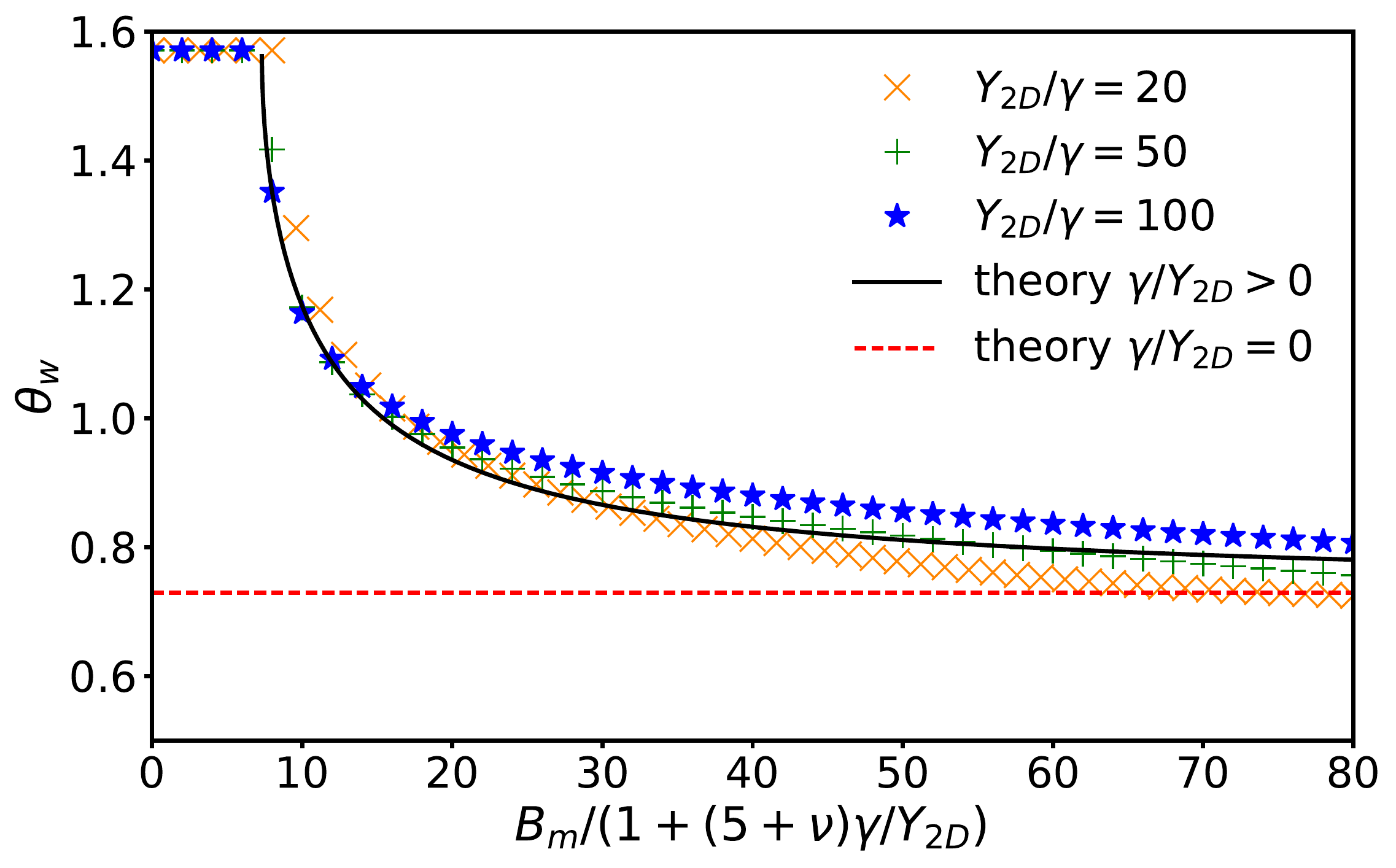}
\caption{
  Extent of the wrinkled region 
  represented by the polar angle $\theta_{\text{w}}$
as a function of  $B_m/[1+(5+\nu)\gamma/Y_{2\text{D}}]$.
The lines are the linear response result 
  \eqref{eq:wrinkle_lin_approx}, 
 crosses and stars are numerical data points for different values 
of $Y_{2\text{D}}\gamma$, which all collapse to the linear 
 response result. The red (dashed) line gives the asymptotic result
$\cos^2\theta_{\text{w}}=5/9$   for large values 
of $B_m/[1+(5+\nu)\gamma/Y_{2\text{D}}]$ and for purely elastic capsules
($\gamma/Y_{2\text{D}} = 0$).
}
\label{fig:lin_region_wrinkling}
\end{center}
\end{figure}

First we use the wrinkle  criterion $\tau_\varphi + \gamma < 0$
to calculate the 
 extent  of the wrinkled region in the  linear 
response regime for small magnetic fields as outlined 
in Sec.\ \ref{sec:lin_def} and  Appendix \ref{app:lin_def}.
In the linear response regime, we calculate the deviation from 
a sphere with radius $R_0$ to leading order. 
We can characterize the size of the wrinkled region 
in terms of  the polar angle
$\theta$  
as $\theta_{\text{w}} < \theta <\pi - \theta_{\text{w}}$ where 
$\theta_{\text{w}}$ is the smallest polar wrinkle
 where wrinkles appear,
$\tau_\varphi(\theta_{\text{w}}) + \gamma = 0$. This angle 
is related to the length $L_{\text{w}}$ of the wrinkled region by 
$L_{\text{w}} = R_0 (\pi-2\theta_{\text{w}})$:
An angle of $\theta_{\text{w}} = \pi/2$ implies the absence of wrinkles,
while $\theta_{\text{w}} = 0$ means that the wrinkles extend from pole 
to pole.
Using Eq.\ (\ref{eq:tauphi_app}) for $\tau_\varphi$,
we find 
\begin{equation}
   \cos^2\theta_{\text{w}} =
   \frac{5}{9}-\frac{\gamma R_0}{Y_{2\text{D}} B} \frac{5+\nu}{3}
  =
\frac{5}{9}-\frac{4(3+\chi)^2}{27\chi}  
 \frac{1+(5+\nu)\gamma/Y_{2\text{D}}}  {B_m }.
\label{eq:wrinkle_lin_approx}
\end{equation}
Interestingly, $\theta_{\text{w}}$ is universal 
and given by $\cos^2\theta_{\text{w}}=5/9$   for purely elastic capsules
($\gamma/Y_{2\text{D}} = 0$), i.e., it does not depend on the   magnetic field or 
capsule elongation. 
This is also the limiting result for large values of 
$B_m/[1+(5+\nu)\gamma/Y_{2\text{D}}]$ (see Fig.\ \ref{fig:lin_region_wrinkling}).
We note, however, that linear response theory is only applicable 
if $B_m/[1+(5+\nu)\gamma/Y_{2\text{D}}] \ll \chi Y_{2\text{D}}/\gamma$. 
For small magnetic fields, the results for 
$\theta_{\text{w}}$ from the linear response 
prediction \eqref{eq:wrinkle_lin_approx}  agree well 
with numerical results, as  Fig.\ \ref{fig:lin_region_wrinkling}
shows. 

\begin{figure}[htbp]
\begin{center}
\includegraphics[width=0.7\textwidth,clip]{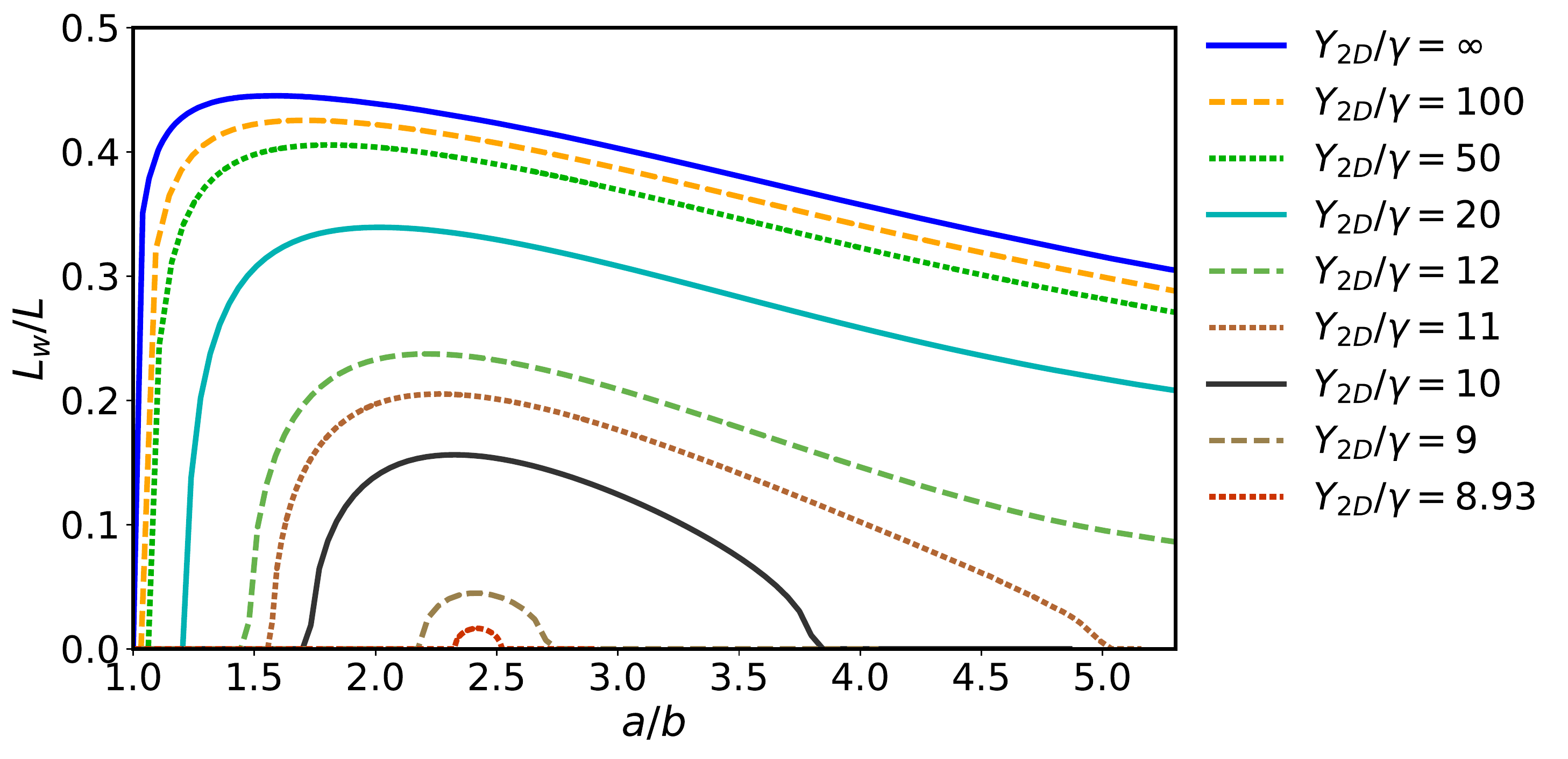}
\caption{
  Relative wrinkle length ${L_{\text{w}}}/{L}$ as a function of elongation
  ${a}/{b}$ for spheroidal capsules with 
  fixed  $\chi=21$ and  different values of
  ${Y_{2\text{D}}}/{\gamma}$.  There are no wrinkles 
   (${L_{\text{w}}}/{L}=0$) 
  for ${Y_{2\text{D}}}/{\gamma} \lesssim 8.93$. The range of
  ${L_{\text{w}}}/{L}> 0$ and the extent of wrinkles
  increase with  ${Y_{2\text{D}}}/{\gamma}$ until they
  converge to an asymptotic curve for thick shells.
}
\label{fig:Lw_plot}
\end{center}
\end{figure}

Now we address the extent of the wrinkled region 
beyond linear response and calculate numerically 
 the relative extent 
of the wrinkled region, ${L_{\text{w}}}/{L}$. 
A value ${L_{\text{w}}}/{L} = 0$ 
means that there are no wrinkles, while ${L_{\text{w}}}/{L} = 1$
describes a system where  wrinkles extend from pole to pole.  
In Fig.\ \ref{fig:Lw_plot}, we change $B_m$ and calculate 
${L_{\text{w}}}/{L}$ for different values of the capsule elongation
${a}/{b}$ in the  spheroidal shape, i.e., for $a/b<5$. 
We use  $\chi=21$  and consider several
values of the elastic parameter ${Y_{2\text{D}}}/{\gamma}$.

As Fig.\ \ref{fig:Lw_plot} shows, there are no wrinkles for thin
stretchable capsules, i.e., wrinkles only occur above a 
critical value of the dimensionless elastic parameter for
\begin{equation} 
 \frac{Y_{2\text{D}}}{\gamma}>8.93  ~~\mbox{for}~\chi=21.
\label{eq:Ycrit}
\end{equation}
This result is only very weakly dependent on $\chi$: We find 
${Y_{2\text{D}}}/{\gamma}> 9.03$ for $\chi=1$ and 
${Y_{2\text{D}}}/{\gamma}> 8.87$ for $\chi=100$. 
For small $Y_{2\text{D}}$,  wrinkles are energetically unfavorable, i.e., the
reduction of stretching energy $E_{\text{el}}$ by wrinkles 
is smaller than the increase of 
$E_\gamma$ due to the increase of the surface area. 
Slightly above the critical value (\ref{eq:Ycrit}), 
wrinkles can only occur for  capsules with 
elongations ${a}/{b} \simeq 2.4$. 
Further increasing $Y_{2\text{D}}$ (or shell thickness), 
the wrinkles become longer and appear for a wider range of elongations.  
The extent of wrinkling is still limited by two effects.
At the lower elongation $a/b$, where ${L_{\text{w}}}/{L}=0$, 
a certain elongation is needed to create a sufficient
compressional stress  at the  equator to overcome the surface tension.
The upper elongation $a/b$, where ${L_{\text{w}}}/{L}=0$, is 
 the point where the capsule is elongated so much that 
the transverse strain, which is related to Poisson's number $\nu$
and tends to shrink the capsule in the circumferential direction,
counteracts any energy gain by the
wrinkles.
The wrinkles' length ${L_{\text{w}}}/{L}$ for different elongations $a/b$ turns
out to be almost independent of the susceptibility $\chi$.

 In systems completely dominated by the elasticity and with negligible
surface tension, there are wrinkles for almost all elongations. The
wrinkle length quickly rises to a maximum and then slowly decreases due to
the transverse strain.

\section{Discussion and Conclusion}
\label{sec:conclusion}

Magnetic or electric fields  provide an interesting and fairly easily 
realizable route to the manipulation of elastic capsules if capsules  can 
be  filled with ferrofluids  or dielectric substances. 
In this work we investigated the deformation of ferrofluid-filled capsules with
thin elastic shells in uniform external magnetic fields numerically
and using several analytic approaches. 
Our results   apply unchanged to elastic capsules filled 
with a dielectric liquid in an external uniform electric field
(see Sec.\ \ref{sec:electric}). 

 Numerically, we obtained
equilibrium shapes by solving the coupled elastic and the magnetostatic
problems in an iterative manner. To calculate the 
magnetic field, we used a combination of the finite element 
method  and the boundary element method for a given capsule shape. 
The elastic capsule was described
by nonlinear shell theory with a Hookean elastic law.  By
neglecting the bending rigidity we had to solve a system of four shape
equations describing the force equilibrium in the absence 
of wrinkling  and modified shape
equations to take the effect of wrinkling into account. In addition
to the dimensionless control parameters, the magnetic Bond number
 $B_m$ and susceptibility $\chi$, that characterize 
 ferrofluid drops, we used the dimensionless 
ratio ${Y_{2\text{D}}}/{\gamma}$ as an
elastic  control parameter.

As for ferrofluid droplets, we found spheroidal shapes at small and moderate
magnetic fields, conical shapes at high magnetic fields, and a 
discontinuous shape transition between spheroidal and conical shapes.
 The general behavior of ferrofluid-filled capsules is comparable to
drops but higher Bond numbers $B_m$ are needed to reach the same 
elongation
due to the additional elastic forces.

For small fields, the capsule shape is  exactly  spheroidal 
and its elongation is very well described by a linear response theory, 
which is in good agreement with our numerical results 
(see Fig.\ \ref{fig:lin_region}). The small field regime is easily 
accessible in experiments and our result (\ref{eq:lin_approx}) 
for the elongation $a/b$ can be used to determine the 
Young modulus $Y_{2\text{D}}$ of the capsule material
from elongation measurements  
if the magnetic properties of the ferrofluid 
are known. 
Also at moderate magnetic fields, capsule  shapes with elongations 
 $a/b \lesssim 3$
are prolate spheroids to a very good approximation and 
can be well described 
by an approximative energy minimization, as  Fig.\ \ref{fig:ab_Bm_plot} shows.

For high fields a conical shape is possible. 
Capsules in a conical shape must 
have finite isotropic tensions and 
diverging isotropic stretches at the conical tip [see
Eq.\ (\ref{eq:lambdadiv})]
with a  divergence exponent, which is given  by the 
 half opening angle $\alpha$ of the 
conical tip [see  Eqs.\ (\ref{eq:betaalpha}) and (\ref{eq:betaalpha_app})]. 
The finiteness of tensions at the tip is a consequence of the 
nonlinear constitutive relations (\ref{eq:taulambda}). 
An important consequence of the divergence of stretches 
 at the tips of a conical shapes is that 
conical shapes are probably not observable 
experimentally because the high stretch factors give rise 
to rupture close to the capsule tips. 
Another consequence of such high stretch factors
is that the  nonlinear Hookean material law will become locally invalid. 
A real elastic capsule 
material will show plastic behavior for high stretches, 
followed by strain hardening and, finally,
 the material's destruction \cite{Mercade-Prieto2011}.
Our results  can explain experimental 
observations of  rupture of capsules
 filled with a dielectric liquid in external 
electric fields, where the capsules' shells 
were destroyed near the tip \cite{Karyappa2014}.
Then the existence of the  sharp discontinuous shape transition
into a conical shape  can
provide an interesting tool  to trigger 
capsule rupture at rather well-defined magnetic 
(for ferrofluid-filled capsules)
or electric (for dielectric-filled capsules) field values in 
future applications of such capsules as delivery systems.

Capsule rupture at the tips has some analogies with  the disintegration of 
droplets in electric fields by emitting fluid jets at the tip 
\cite{Collins2008,Collins2013}.
 Real fluid drops, which are not perfect 
conductors or perfect insulators, 
disintegrate at higher external electric fields 
by emitting jets of fluid at the tip.
This is known from experiments \cite{Wilson_Taylor1925} as 
well as quite precisely understood in theory \cite{Collins2008,Collins2013}. 
In our setup of a fluid inside  an elastic shell, 
the emission of a fluid jet is prevented by the shell at first.
However,  the tangential stresses at the tip that lead to the 
formation of a fluid jet may support the destruction of the shell
 near the tip.
Once the shell is broken, a jet can be emitted. 
The rupture process itself cannot be described by our 
numerical approach and is  an interesting topic for future work.
Our elastic shape equation approach provides a very precise tool 
to solve the static elastic part of the problem, as long as nonlinear Hookean 
elasticity can be used. Also the generalization to other 
material laws, which are more appropriate for large strains, 
is possible \cite{hegemann2017}. 
Breaking of axisymmetry and topology changing  rupture events cannot be 
easily incorporated into the shape equation approach, however. 
Also the magnetic field calculation should 
be improved if rupture is addressed, in particular 
in the capsule's tip region by using, for example, 
an elliptic mesh generation for the finite element method.  
One idea for a future  improved simulation method that 
captures  possible rupture processes at the tip 
is a dynamic simulation, 
where the magnetohydrodynamics of the fluid and the viscoelastic dynamics 
of the capsule shell including rupture processes
can be calculated explicitly, similarly to what has been 
achieved for droplets in electric fields \cite{Collins2008,Collins2013}.

 We  presented a complete shape  diagram  in 
Fig.\ \ref{fig:ab_Bm_plot} and characterized the discontinuous 
shape transition between spheroidal and conical shapes. 
The slender-body theory predicts that this discontinuous shape transition 
only exists above the  same critical value $\chi_c$ as 
for ferrofluid droplets, which was predicted to lie between 
$\chi_c \simeq 14.5$ \cite{Stone99} and $\chi_c\simeq  16.59$ \cite{Li1994}.
It also predicts that $\chi_c$ is independent of the 
Young modulus $Y_{2\mathrm{D}}$   of the capsule.
We predict that the critical $\chi_c$ will be very sensitive 
to the constitutive relation of the material. A strictly linear 
constitutive relation, for example,  could give rise to 
diverging tensions at a
conical tip, resulting in much higher values for $\chi_c$.

We  used the meridional stretch factor $\lambda_s$ at the pole 
as a suitable  order parameter to detect the spheroidal-conical 
 transition, because stretches diverge at the tip 
of conical shapes but remain finite for spheroidal shapes, resulting 
in a pronounced jump of the stretch factor in the numerical calculations. 
 The spheroidal-conical 
 transition exhibits hysteresis effects in an interval 
$B_{m,c1}<B_m< B_{m,c2}$ between two 
critical Bond numbers, which are the limit of stability 
of the spheroidal and conical shapes. 
In the hysteresis interval  
 both types of shapes are metastable. 
The interval
has its maximum size for ferrofluid 
droplets and decreases slightly with increasing 
Young's modulus   of the elastic shell. 
In the numerical calculations for $\chi=21$, we 
observe hysteresis effects for all   ${Y_{2\text{D}}}/{\gamma}$,
which shows   that, indeed,  $\chi_c<21$ for all the Young moduli.

 It turned out that the formation of wrinkles is an important
effect in systems with low surface tension $\gamma$. It has a visible effect
on the elongation and the specific shape. Wrinkles appear for the first time
for ${Y_{2\text{D}}}/{\gamma} \gtrsim 8.93$ (for  $\chi=21$), and 
are almost
always present for systems with lower surface tension, even at very low
elongations. Using this knowledge, it is possible to determine, for example,
${Y_{2\text{D}}}/{\gamma}$ in experiments by a simple measurement of the
wrinkle length $L_{\text{w}}$, which should be easy to perform in
practice.

\begin{appendix}

\section{Linear response at small magnetic fields}
\label{app:lin_def}

In this appendix we derive the linear response of the capsule 
elongation $a/b$ for small applied magnetic fields. 
Without applied field, the capsule is spherical with a rest radius $R_0$. 
In the presence of a surface tension $\gamma$, this also requires 
an internal  pressure $p_0 = 2\gamma/R_0$ (Laplace-Young equation).
If a small magnetic field is applied the additional position-dependent 
normal magnetic 
force density $f_m = O(H^2)$ [see Eq.\ (\ref{fm})] acts on the spherical 
surface,
resulting in normal displacements $u_R(\theta)\vec{e}_R$ and tangential 
displacements $u_\theta(\theta)\vec{e}_\theta$, where we use spherical 
coordinates with  the polar
angle $\theta$ 
(i.e., $\theta=0$ at the upper pole and $\theta=\pi/2$ at the equator)
and the  spherical coordinate 
unit vectors $\vec{e}_R$ and $\vec{e}_\theta$. 
Because of axisymmetry the displacements do not depend on the 
azimuthal angle $\varphi$ and there is no displacement in direction 
$\vec{e}_\varphi$.
The deformed capsule surface is parametrized as
 $\vec{r}(\theta,\varphi) = 
   [R_0 + u_R(\theta)]\vec{e}_R(\theta,\varphi)$  using 
polar and azimuthal angles $\theta$ and $\varphi$. 

 The new equilibrium shape has 
small displacements $u_R, u_\theta = O(H^2)$
and fulfills force equilibrium in two independent directions 
on the surface. We will consider normal force equilibrium as described 
by the  Laplace-Young equation [see Eq.\ (\ref{eq:equilibrium_eqns_norm})]
and tangential force equilibrium [see Eq.\ (\ref{eq:equilibrium_eqns_tang})]. 
We start with the Laplace-Young equation
\begin{align}
  \label{eq:fbal_cap}
  \kappa_s (\tau_s+\gamma)  +\kappa_\varphi (\tau_\varphi + \gamma)
 &= p_0 + f_m,
\end{align}
where  $\gamma$ is a surface  tension,  $\tau_s$ and $\tau_\varphi$ are
elastic tensions, and 
 $f_m = (\mu_0\chi/2)[H^2 + \chi (\vec{n}\cdot\vec{H})^2]$ is 
the small normal magnetic  force density (\ref{fm}) causing 
small displacements. 
The pressure will change to linear order in the
displacements 
$p_0 = 2\gamma/R_0 + O(u_R, u_\theta)$  
to ensure a fixed volume. 
In spherical coordinates and in linear order in the displacements, 
the stretch factors can be calculated using 
$r = \sqrt{g_{\varphi\varphi}} =|\partial_\varphi \vec{r}|$
($r_0 = R_0\sin\theta$) and   
$\text{d}s =  \sqrt{g_{\theta\theta}} d\theta =
|\partial_\theta \vec{r}|d\theta$
($ds_0 = R_0 d\theta$):
\begin{align*}
  \lambda_s &= \frac{\text{d}s}{\text{d}s_0} = \frac{|\partial_\theta
    \vec{r}|}{R_0} 
=  1 + \frac{1}{R_0} \left({u_R}+ {\partial_\theta u_\theta}\right),
      \\
  \lambda_\varphi &= \frac{r}{r_0} = \frac{|\partial_\varphi
    \vec{r}|}{R_0\sin\theta} 
= 1 +  \frac{1}{R_0}\left( u_R+ {u_\theta}\cot\theta\right).
\end{align*}
In linear order in the displacements 
the constitutive relations (\ref{eq:taulambda})
can then be written as   \cite{LL7}
\begin{align}
  \tau_\varphi - \nu \tau_s&= Y_{2\text{D}}(\lambda_\varphi-1)
      =  \frac{Y_{2\text{D}}}{R_0}(u_\theta \cot\theta + u_R),\\
 \tau_s - \nu \tau_\varphi &= Y_{2\text{D}}(\lambda_s-1) = 
  \frac{Y_{2\text{D}}}{R_0}\left(\partial_\theta u_\theta + u_R\right).
\label{eq:taulambda2}
\end{align}
Elastic tensions are small for small magnetic fields, 
$\tau_s, \tau_\varphi = O(u_R, u_\theta) =   O(H^2)$,
whereas the fluid surface tension $\gamma$ cannot 
be considered small. 
Therefore, we also need to consider curvature corrections up to linear 
order $O(u_R, u_\theta)$ in 
   Eq.\ (\ref{eq:fbal_cap}):
\begin{align*}
  \kappa_s+ \kappa_\varphi \approx   \frac{2}{R_0} 
   -\frac{1}{R_0^2} \left( 2 u_R - \partial_\theta^2 u_R + \partial_\theta u_R 
   \cot\theta \right).
\end{align*}
On the right-hand side of Eq.\  (\ref{eq:fbal_cap}), we can use 
$\vec{n} = \vec{e}_R$ for the  outward unit 
normal to  $O(H^2)$. 
This results in the following normal force balance 
to linear order in the displacements, i.e.,  to  ${\cal O}(H^2)$:
\begin{align}
 & -\gamma \left( 2 u_R - \partial_\theta^2 u_R + \partial_\theta u_R 
   \cot\theta \right) + R_0 (\tau_s + \tau_\varphi) 
  \nonumber\\
 &~~~= (p_0R_0^2 - 2R_0\gamma)  + \frac{\mu_0}{2}\chi H^2R_0^2 
  (1+\chi \cos^2\theta). 
\label{eq:fbal_cap2}
\end{align}

We first solve this equation 
for a ferrofluid droplet ($Y_{2\text{D}}/\gamma=0$), where 
the elastic stresses and thus $u_\theta$ are zero.
Boundary conditions are $\partial_\theta u_R(0) = \partial_\theta u_R(\pi/2)=0$ 
and $u_\theta(0) = u_\theta(\pi/2)=0$ to avoid kinks 
(we are not considering conical shapes in the linear response) 
or holes in the shape.
Then $u_\theta=0$ and an ansatz 
\begin{equation}
   u_R= A+B\cos^2\theta
\label{eq:AnsatzuR_app}
\end{equation}
  leads to a solution 
\begin{align}
   B &=   \frac{\mu_0}{8\gamma}\chi^2 H^2R_0^2  
    \approx \frac{9\mu_0 \chi^2}{8\gamma(3+\chi)^2} H_0^2R_0^2 
 \label{eq:uRB}\\
   A &=  -\frac{1}{2} \left(\frac{p_0R_0}{\gamma}- 2\right)R_0   
  -  \frac{\mu_0}{4\gamma}\chi\left(1+\frac{\chi}{2} \right) H^2R_0^2.
\label{eq:uRA}
\end{align} 
To leading order in $u_R = O(H^2)$, the ansatz (\ref{eq:AnsatzuR_app}) 
describes a spheroid such that we can replace 
the magnetic field $H$ in (\ref{eq:uRB}) 
by the analytically known value for a field inside a
spheroid \cite{Stratton41},
 \begin{equation}
   H = {H_0}/{(1+n\chi)}, 
 \label{eq:H_app}
 \end{equation}
 where  $n$ denotes the demagnetization factor.
To leading order   $O(H^2)$ it is also correct to 
 use the result  $n=1/3$  for a sphere  (\ref{eq:H_app}).
Moreover, volume conservation requires 
\begin{equation}
  A= -B/3,
\label{eq:VolAB_app}
\end{equation}
 which determines the pressure correction
$p_0 = 2\gamma/R_0 + O(H^2)$  from Eq.\ (\ref{eq:uRA}). 
For the  deformation $a/b$
we find, to leading order in $u_R = O(H^2)$,
\begin{equation}
  \frac{a}{b} = \frac{R_0 + u_R(0)}{R_0+ u_R(\pi/2)} 
     \approx 1 + \frac{B}{R_0}
   = 1+\frac{9\mu_0 R_0\chi^2}{8\gamma(3+\chi)^2}\chi H_0^2.
 \label{eq:droplinear_app}
\end{equation}

For a ferrofluid-filled elastic capsule we also need to consider 
the  force equilibrium  in  the tangential 
direction because the total tensions
$\gamma+\tau_s\neq \gamma+\tau_\varphi$  become anisotropic now
(for a liquid interface with $\tau_s=\tau_\varphi=0$ 
the force equilibrium in tangential direction 
becomes exactly equivalent to the normal force equilibrium, 
i.e., the Laplace-Young equation). 
The  tangential force equilibrium  (\ref{eq:equilibrium_eqns_tang})
can be written as 
\begin{align*}
  \tau_\varphi &= \partial_r(r \tau_s) = \tau_s + r \partial_r \tau_s
 = \tau_s+ \frac{\partial_\theta \tau_s}{\partial_\theta r}.
\end{align*}
Using 
$r =|\partial_\varphi \vec{r}|  = \sin\theta\left(
R_0 +  u_R+ {u_\theta}\cot\theta\right)$ and  Eq.\ (\ref{eq:taulambda2}) 
for the elastic stresses, 
the tangential force equilibrium becomes
\begin{align}
 & \tau_\varphi -\tau_s 
=   \frac{Y_{2\text{D}}}{(1+\nu)R_0}\left(u_\theta \cot\theta -
    \partial_\theta u_\theta\right)  
\nonumber\\
&= \partial_r \tau_s  =  \frac{Y_{2\text{D}}}{(1-\nu^2)R_0}\left( 
   \tan \theta \partial_\theta^2u_\theta + \nu \partial_\theta u_\theta 
   -\frac{\nu u_\theta}{\cos\theta  \sin\theta} + (1-\nu) 
    \tan\theta \partial_\theta u_R \right). 
\label{eq:fbalt_cap2}
\end{align}

For the ferrofluid capsule, 
the two force equilibria  (\ref{eq:fbal_cap2}), 
where $\tau_s$ and $\tau_\varphi$ have to be expressed in terms of 
the displacements using the constitutive relations (\ref{eq:taulambda2}), 
\begin{align*}
  \tau_s+ \tau_\varphi &= \frac{Y_{2\text{D}}}{(1-\nu)R_0}
   \left(2u_R + u_\theta \cot\theta +
    \partial_\theta u_\theta\right),
\end{align*}
 and Eq.\  (\ref{eq:fbalt_cap2})  have to 
be solved for the deformed capsule shape.
 Boundary conditions are $\partial_\theta u_R(0) = \partial_\theta u_R(\pi/2)=0$ 
and $u_\theta(0) = u_\theta(\pi/2)=0$. 
For the fluid limit $Y_{2\text{D}}/\gamma=0$, we derived an 
exact solution above. 
For the ferrofluid capsule, we make an ansatz
\begin{equation}
   u_R= A+B\cos^2\theta,~~u_\theta = C\sin\theta\cos\theta,
\label{eq:AnsatzuRcap_app}
\end{equation}
which still describes a spheroid to  leading order in 
the displacements because $u_\theta\neq 0$ only generates an
additional  tangential displacement. 
Then the  tangential force equilibrium gives 
\begin{align}
   C &= - \frac{2(1+\nu)}{5+\nu} B.
\label{eq:uRCcap} 
\end{align}
For the ferrofluid capsule, 
 the normal force equilibrium (\ref{eq:fbal_cap2}) gives 
\begin{align}
    B&=  \frac{\mu_0(5+\nu) }{8[Y_{2\text{D}}+(5+\nu)\gamma]}\chi^2 H^2R_0^2,
 \label{eq:uRBcapdrop} \\
 A&=  \frac{1-\nu}{2} \left(\frac{p_0R_0}{Y_{2\text{D}}}-
 2\frac{\gamma}{Y_{2\text{D}}}\right)R_0  
  +  \frac{\mu_0}{4(1-\nu)Y_{2\text{D}}}\chi H^2R_0^2\left(1+\frac{\chi}{2}
 \right)
  - \frac{C}{1+\nu}
\label{eq:uRAcapdrop} 
\end{align}
and the relation $A=-B/3$ [see Eq.\ (\ref{eq:VolAB_app}]
 from the  fixed volume constraint determines 
the pressure $p_0$. 
For the  deformation $a/b$
we find,  to leading order in $u_R = O(H^2)$,
\begin{align}
   \frac{a}{b} &= \frac{R_0 + u_R(0)}{R_0+ u_R(\pi/2)} 
     \approx 1 + \frac{B}{R_0}
= 
 1 + \frac{9\mu_0 R_0 \chi^2 (5+\nu)}
      {8 [Y_{2\text{D}}+\gamma(5+\nu)](3+\chi)^2}H_0^2.
\label{eq:capdroplinear_app}
\end{align}

The criterion for wrinkling is $\tau_\varphi + \gamma <0$,
where  
\begin{align}
 \tau_\varphi &= \frac{Y_{2\text{D}}}{(1-\nu^2)R_0}\left[ u_\theta \cot\theta
 + (1+\nu)u_R
     + \nu \partial_\theta u_\theta \right]
\approx B\frac{1-\nu^2}{5+\nu}\left( -\frac{5}{3} +3 \cos^2\theta\right)
 \label{eq:tauphi_app}
\end{align}
from Eq.\ (\ref{eq:taulambda2}) and using Eq.\ (\ref{eq:AnsatzuRcap_app}) 
with Eqs.\ (\ref{eq:uRCcap}) and (\ref{eq:VolAB_app}).

\section{Approximative energy minimization for spheroidal shapes}
\label{app:Bm}

In this appendix we  derive an analytical approximation for the 
elongation $a/b$ of the capsule at moderate magnetic forces 
  by minimizing an 
approximative  total energy, which assumes a spheroidal shape 
for  magnetic and elastic contributions and constant elastic 
stretch factors throughout the shell.
We minimize the total energy, the sum of surface, magnetic, and
elastic energies with respect to the inverse 
elongation ratio $k\equiv {b}/{a}<1$
 at fixed volume $V =  (4\pi/3) ab^2=V_0$ (quantities $...|_V$ are
at fixed volume $V$):
\begin{equation*}
  0 = \frac{\mathrm{d} E_\gamma|_V}{\mathrm{d}k} 
 +\frac{\mathrm{d} E_{\text{mag}}|_V}{\mathrm{d}k} 
  + \frac{\mathrm{d} E_{\text{el}}|_V}{\mathrm{d}k}.
\end{equation*}

For fixed volume $V= (4\pi/3) ab^2=  (4\pi/3)R_0^2= V_0$,
we have
\begin{equation*}
   a|_V = R_0 k^{-2/3} ~,~~ b|_V = R_0 k^{1/3}.
\end{equation*} 
The surface energy (\ref{Egamma}), which is proportional 
to the surface area $A$ at fixed volume, can then be written as 
\begin{align*}
 E_\gamma|_V &= \gamma A|_V  ~~~\mbox{with}~~
 A|_V  =  A_0 
  \frac{1}{2} k^{-{1}/{3}}\left(k+\frac{1}{\epsilon}\arcsin{\epsilon} \right),
\end{align*}
where $\epsilon = \epsilon(k) \equiv \sqrt{1-k^2}$ is the spheroid's 
eccentricity and $A_0 = 4\pi R_0^2$ the area of the undeformed sphere. 
The magnetic energy (\ref{Emag}) is
given as
\begin{align*}
  E_{\text{mag}}|_V &= -\frac{V_0\mu_0}{2} \frac{\chi}{1+n\chi}H_0^2
    = -\gamma A_0 B_m \frac{1}{3(1+n\chi)},
\end{align*}
where $n$ is the demagnetization factor 
\begin{equation*}
  n= n(k) = \frac{k^2}{2\epsilon^3(k)}
 \left(-2\epsilon(k) + \ln{\frac{1+\epsilon(k)}{1-\epsilon(k)}} \right)
\end{equation*}
and 
$B_m = {\mu_0R_0\chi H_0^2}/{2 \gamma}$ is the  Bond number.

Finally, we calculate the elastic stretch energy (\ref{Eel}) via
\begin{equation*}
  E_{\text{el}}|_V
   = A_0 \frac{Y_{2\mathrm{D}}}{2(1-\nu^2)} 
\left[(e_s|_V)^2+ 2\nu e_s|_V e_\varphi|_V + (e_\varphi|_V)^2\right] 
\end{equation*}
using the approximation of constant $e_s$ and $e_\varphi$.
At fixed volume, we find 
\begin{align*}
  e_s &= \frac{P_\text{ellipse}}{P_\text{circle}} -1
  \approx \frac{a+b}{2R_0}
   \left(1 + \frac{3\eta^2}{10 + \sqrt{4-3\eta^2}}\right)-1,\\ 
e_s|_V &=  \frac{k^{-2/3}(1+ k)}{2}
          \left(1 + \frac{3\eta^2(k)}{10 + \sqrt{4-3\eta^2(k)}}\right) -1,  \\
  e_\varphi &= \frac{b}{R_0}-1~~,~~
e_\varphi|_V = k^{1/3}-1,
\end{align*}
with $\eta = \eta(k)  \equiv ({b-a})/({b+a}) = (k-1)/(k+1)$.

Now we can find the elongation $k$ that minimizes the
total energy  at fixed volume; $k$ can only be determined implicitly 
as a function of the magnetic field $H_0$ 
by the following  relation between the Bond number 
$B_m = {\mu_0R_0\chi H_0^2}/{2 \gamma}$ 
and a complicated function  $g(k)$ of the elongation $k$,
which also depends on  the susceptibility $\chi$, the dimensionless
Young modulus $Y_{2\text{D}}/\gamma$,   and Poisson's ratio $\nu$:
\begin{align}
  B_m &=\frac{\mu_0R_0\chi H_0^2}{2 \gamma} = g(k) 
   ~~\mbox{with}\nonumber\\
  g(k)  &\equiv   -3\left(\frac{1}{\chi} + n(k)\right)^2\chi
    \frac{c_1(k) +  \frac{Y_{2\mathrm{D}}}{2\gamma(1-\nu^2)} c_2(k,\nu)}{c_3(k)},
\label{eq:BMab}
\end{align}
where
\begin{align*}
   c_1(k) &\equiv \frac{1}{A_0} \frac{\text{d}A|_V}{\text{d}k}, \\  
    c_2(k,\nu) &\equiv   \left( 2 e_s|_V \frac{\mathrm{d}e_s|_V }{\mathrm{d}k}
   + 2\nu\left( e_\varphi|_V  \frac{\mathrm{d}e_s|_V }{\mathrm{d}k} +
      e_s|_V \frac{\mathrm{d}e_\varphi|_V }{\mathrm{d} k} \right) 
  + 2e_\varphi|_V  \frac{\mathrm{d} e_\varphi|_V }{\mathrm{d} k} \right),\\
  c_3(k) &\equiv  \frac{\text{d}n}{\text{d}k}
 =  \frac{-3k}{\epsilon^4(k)} + \left(\frac{k}{\epsilon^3(k)} 
  + \frac{3k^3}{2\epsilon^5(k)} \right) 
  \ln{\frac{1+\epsilon(k)}{1-\epsilon(k)}}.
\end{align*}
The functions $c_1(k)$ and $c_3(k)$ from surface and magnetic energies 
depend on the inverse elongation ration $k=b/a<1$ 
only, whereas the function $c_2(k,\nu)$ from the elastic energy also 
 depends on Poisson's ratio $\nu$ (which is set to $\nu=1/2$ and thus
fixed throughout this paper).
This relation reduces to the results of Bacri and Salin \cite{Bacri82}
for ferrofluid droplets in the limit $Y_{2\text{D}}= 0$, where 
the function $c_2(k,\nu)$ drops from Eq.\ (\ref{eq:BMab}). 
The solid lines in Fig.\ \ref{fig:ab_Bm_plot} show plots of 
$1/k = a/b$  versus $B_m$  as given by  the relation $B_m = g(k)$.

\section{Conical shapes for 
  elastic membranes with spherical rest shape}
\label{app:cone_angle}

\subsection{Stretches and tensions at a conical tip with
 normal magnetic forces}
\label{app:cone_tip}

In this appendix we show that a  conical shape, as it is
observed for ferrofluid 
drops at a critical field strength, is also possible for an elastic
capsule with a spherical rest shape and stretched by normal magnetic forces
but requires diverging and asymptotically isotropic 
stretches with an exponent determined by the opening angle 
of the cone, whereas elastic tension have to remain 
finite and isotropic at the tip of the cone.

 A sharp conical  tip implies a non-zero
slope angle $\psi(s_0=0)>0$, where $\alpha = \pi/2-\psi(0)$ 
is half of the opening angle of the cone.
In contrast to a ferrofluid droplet with constant and isotropic 
surface tension $\gamma$, an elastic capsule develops additional 
elastic tensions $\tau_s$ and $\tau_\varphi$, which depend on the state 
of stretching, i.e., the stretches $\lambda_s$ and $\lambda_\varphi$
 with respect to the spherical rest shape 
via the nonlinear constitutive relations (\ref{eq:taulambda}), 
and 
which have to fulfill an additional tangential force equilibrium 
(\ref{eq:equilibrium_eqns_tang}) that 
we rewrite as
\begin{align}
   \tau_\varphi &= \partial_r(r \tau_s) = \tau_s + r\partial_r \tau_s.
\label{eq:fbalt_cap0}
\end{align}
It is important to note that the tangential force equilibrium 
does not contain external magnetic forces, which are always normal to the 
surface [see Eqs.\ (\ref{eq:pfm}) and (\ref{eq:equilibrium_eqns_norm})]. 
The internal 
tangential force equilibrium has to be compatible with the deformation 
into a conical tip.

First we show that $\tau_s(0)$ and $\tau_\varphi(0)$  have to remain 
 finite at the tip at $s_0=0$ (corresponding to  $r=0$).
 The reason for a divergence of one 
of the tensions can only be a divergence of one or both of the 
stretches. According to the nonlinear constitutive relations
(\ref{eq:taulambda}), only one of the tensions can exhibit a divergence
($\lambda_s/\lambda_\varphi$ and $\lambda_\varphi/\lambda_s$ cannot both 
diverge). Then it is easy to verify that a single divergent 
tension at $r=0$ contradicts the force equilibrium 
(\ref{eq:fbalt_cap0}). Therefore, both tensions have to remain 
finite at  $s_0=0$ (or $r=0$).

Next we show that finiteness of the tensions at the conical tip 
necessarily leads to  tension isotropy
 $\tau_s(0) = \tau_\varphi(0)$  at the 
tip. Because magnetic forces are stretching forces, both tensions 
are equal and stretching, $\tau_s(0) = \tau_\varphi(0)>0$.
 If $\tau_s(0) \neq \tau_\varphi(0)$, the tangential 
force equilibrium (\ref{eq:fbalt_cap0}) immediately leads to 
$\partial_r \tau_s \approx [\tau_\varphi(0)- \tau_s(0)]/r$ for 
small $r$, resulting in a logarithmically diverging $\tau_s \propto -\ln r$ 
for small $r$ contradicting finiteness. 

The equality $\tau_s(0) = \tau_\varphi(0)$  at the 
tip also leads to  isotropy of the stretches
$\lambda_s(0)=\lambda_\varphi(0)$ at the tip because of the 
 constitutive relations
(\ref{eq:taulambda}), however, 
not necessarily to finiteness  of the stretches at the tip. 
Therefore, we have to discuss the cases of finite and diverging
stretches $\lambda_s=\lambda_\varphi$ at the conical tip 
separately. 

We start with finite isotropic stretches, 
$\lambda_s(0)=\lambda_\varphi(0)<\infty$.
Then we can apply l'H{\^o}pital's rule at the tip $s_0=0$: 
\begin{align}
  \lambda_\varphi(0) =
   \lim\limits_{s_0 \to 0}\frac{r}{r_0} = 
  \lim\limits_{s_0 \to 0} \frac{r'}{r_0'} = 
  \frac{\lambda_s\cos[\psi(0)]}{\cos[\psi_0(0)]}
  =\lambda_s(0)\cos[\psi(0)]
\label{eq:lhospital}
\end{align}
[where we used $\psi_0(0)=0$ for the spherical rest shape]. 
Equality of the stretches $\lambda_s(0)=\lambda_\varphi(0)$
then leads to the conclusion $\psi(0)=0$, i.e., 
 a sharp conical tip is impossible if stretches remain finite at the tip.

L'H{\^o}pital's rule can no longer be applied if the stretches  diverge
at the tip (remaining asymptotically isotropic), i.e., 
\begin{equation}
 \lambda_s(s_0)\approx \lambda_\varphi(s_0)\approx {\rm const}\, s_0^{-\beta}
\label{eq:lambdadiv_app}
\end{equation}
for $s_0\approx 0$
with an exponent $\beta>0$. 
Because of $\lambda_s = r'/\cos\psi$, this  requires 
$r(s_0) \approx {\rm const}\, s_0^{1-\beta}/(1-\beta)\cos\psi(0)$ 
for  $s_0\approx 0$, whereas $r_0(s_0) = R_0 \sin(s_0/R_0) \approx s_0$ 
for the spherical rest shape. 
Then Eq.\ (\ref{eq:lhospital}) is replaced by 
\begin{align*}
  \lambda_\varphi(s_0) =
   \lim\limits_{s_0 \to 0}\frac{r}{r_0}  = 
   \frac{{\rm const} s_0^{-\beta}}{(1-\beta)\cos\psi(0)} =
  \lim\limits_{s_0 \to 0} \frac{1}{1-\beta}\frac{r'}{r_0'}
   = \lambda_s(s_0)
  \frac{\cos[\psi(0)]}{1-\beta}
\end{align*}
for $s_0\approx 0$. The equality $\lambda_s(s_0)\approx \lambda_\varphi(s_0)$
necessarily leads to the condition 
\begin{equation}
  \beta = \cos[\psi(0)]-1 = \sin\alpha-1
\label{eq:betaalpha_app}
\end{equation}
between the exponent $\beta$ of the divergent stretches 
and the half opening angle $\alpha = \pi/2 - \psi(0)$ of the 
conical tip. 

In conclusion, a deformation of the spherical rest shape into a sharp
conical tip with  $\psi(0)> 0$ is only possible if stretches
are asymptotically isotropic and diverge as 
$\lambda_s(s_0)\approx \lambda_\varphi(s_0)\sim s_0^{-\beta}$
 with an exponent $\beta$, which is related by 
Eq.\ (\ref{eq:betaalpha_app}) to the opening angle $2\alpha$ of the cone. 
Because of the nonlinear constitutive relation (\ref{eq:taulambda}),
diverging and isotropic stretches are compatible with 
finite and isotropic tensions at the tip with 
\begin{equation}
  \tau_s(0) = \tau_\varphi(0) =  
 \frac{Y_{2\text{D}}}{1-\nu}.
\label{eq:tau0_app}
\end{equation}
Note that away from the tip ($s_0>0$), tensions and stretches 
feature anisotropic corrections.

\subsection{Governing equations for stretches and tensions 
   in  a conical shape with spherical rest shape}
\label{app:cone_eqs}

In this section  we present how to systematically calculate 
stretches and elastic tensions  in a deformation from a spherical 
rest shape into
 a conical shape by deriving the governing equations. 
This is the basis of the  generalization of the slender-body theory 
of Stone {\it et al.}\ from ferrofluid conical droplets to 
capsules.

\begin{figure}[htbp]
\begin{center}
\includegraphics[width=0.7\textwidth,clip]{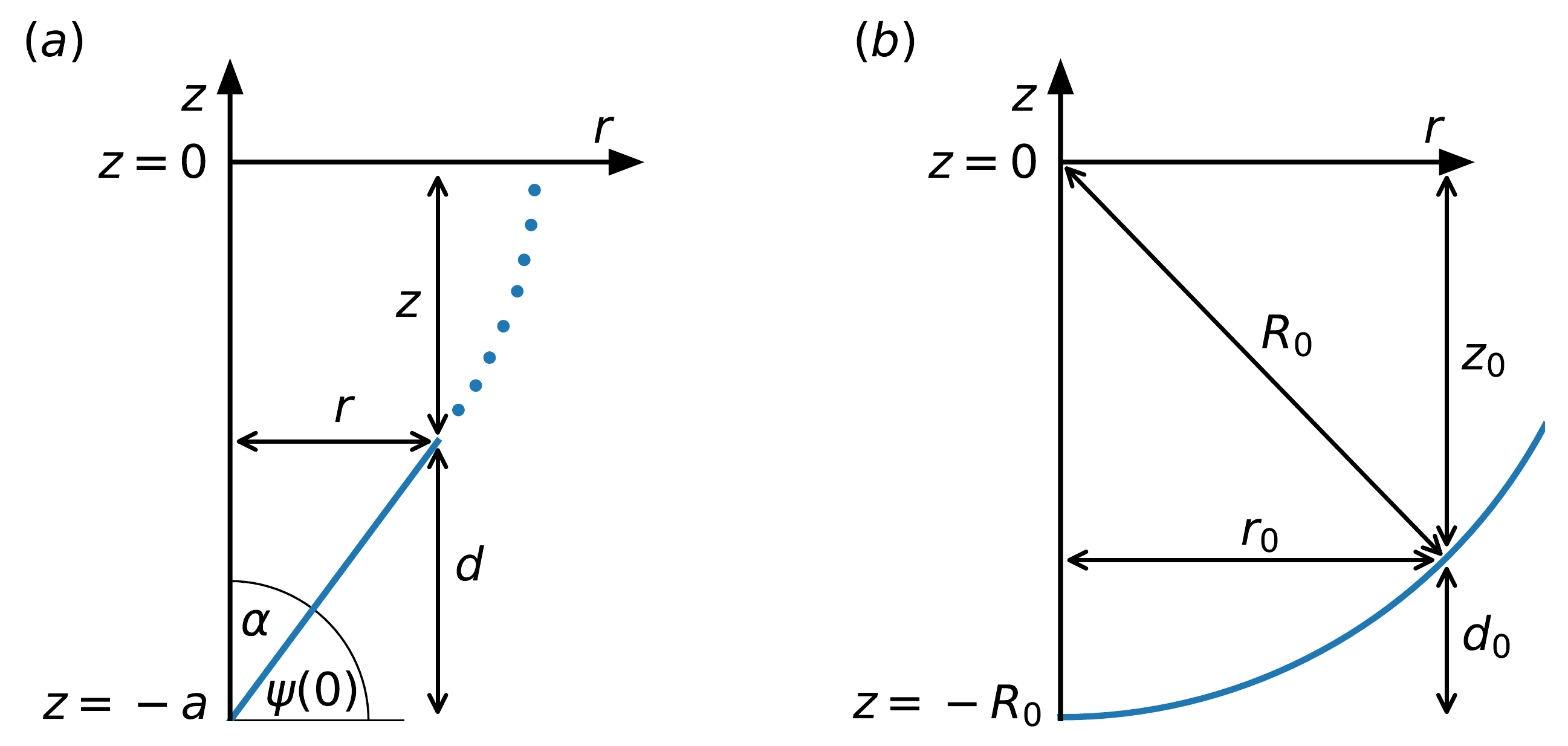}
\caption{
    Illustration of the geometry at the capsule's south pole 
  (not true to scale) for 
    (a) a conical tip and   (b) the spheroidal reference shape.
 }
\label{fig:appendix_sketch}
\end{center}
\end{figure}

We assume that the conical shape is given by a function $r(z)$,
where $z$ runs from the bottom of the cone at $z=-a$ to its top at 
$z=a$.
We will show that, if the
conical shape $r(z)$ is known, we can calculate 
all stretches and tensions in this shape.
The rest shape is spherical and parametrized analogously 
by a function $r_0(z_0) = (R_0^2 - z_0^2)^{1/2}$
with $z_0 \in [-R_0,R_0]$.
For the following it is advantageous to replace $z$ and $z_0$ 
 by coordinates
$d=a+z$ measuring the distance from the lower conical tip
and $d_0 = R_0 + z_0$ measuring the distance from the corresponding 
south pole of the sphere. This geometry is illustrated in 
Fig. \ref{fig:appendix_sketch}.

Given a conical shape $r(d)$ and the spherical rest shape 
\begin{equation}
   r_0(d_0) = (2R_0d_0 -d_0^2)^{1/2},
\label{eq:spherical_app}
\end{equation}
we want to show how the function $d(d_0)$ describing  the 
stretching in the $z$ direction 
 can be calculated systematically 
from the tangential force equilibrium 
(\ref{eq:fbalt_cap0}) or (\ref{eq:equilibrium_eqns_tang}) and 
the constitutive relations (\ref{eq:taulambda}).
If the conical shape $r(d)$ and the function $d(d_0)$ are given 
[and the spherical rest shape $r_0(d_0)$]
the  meridional and hoop stretches  can be calculated 
as a function of $d_0$ by 
\begin{align}
\begin{split}
\lambda_\varphi &= \frac{r(d(d_0))}{r_0(d_0)}  = 
     \frac{r(d(d_0))}{(2R_0d_0 -d_0^2)^{1/2}},
\\
  \lambda_s &= \frac{\text{d}s}{\text{d}s_0}
      = \frac{\left[1+ r'(d(d_0))^2\right]^{1/2}}
            {\left[1+r_0'(d_0)^2\right]^{1/2}} 
        d'(d_0)
   =  {\left[1+ r'(d(d_0))^2\right]^{1/2}}\frac{(2R_0d_0 - d_0^2)^{1/2}}{R_0}
       d'(d_0),
\end{split}
\label{eq:lambda_d_app}
\end{align}
where $\lambda_z \equiv d'(d_0) = {\text{d}z}/{\text{d}z_0}$
is the stretch in the  $z$ direction.

If both stretches are known then the constitutive relations 
(\ref{eq:taulambda}) can be used to express tensions $\tau_s$
and $\tau_\varphi$ as algebraic functions 
 of the stretches $\lambda_s$ and $\lambda_\varphi$ from Eq.\ 
(\ref{eq:lambda_d_app}) and thus as functions of $d_0$, the  
conical shape 
$r(d)$,  and the unknown function $d(d_0)$ and its derivative. 
These tensions have to fulfill the tangential 
force equilibrium (\ref{eq:fbalt_cap0}), which we rewrite 
in terms of stretches using the  constitutive relations 
(\ref{eq:taulambda}) ,
\begin{align}
   \tau_\varphi &= \tau_s + r\partial_r \tau_s,
\nonumber\\
  \lambda_\varphi^3 - (1+\nu)\lambda_\varphi^2 &=
  \lambda_s^2\lambda_\varphi - (1+\nu)\lambda_s\lambda_\varphi
   + r\lambda_s\left\{ \lambda_\varphi (\partial_r \lambda_s) - 
     (\partial_r \lambda_\varphi)\left[\lambda_s - (1+\nu)\right]  \right\}.
\label{eq:fbalt_cap1}
\end{align}
Plugging in the stretches from (\ref{eq:lambda_d_app}) and using 
\begin{equation*}
  \partial_r = \frac{1}{\partial_{d_0}r}\partial_{d_0} = 
      \frac{1}{r'(d(d_0)) d'(d_0)} \partial_{d_0},
\end{equation*}
we obtain a complicated nonlinear differential equation for the 
unknown function $d(d_0)$ and its derivative $\lambda_z(d_0)= d'(d_0)$. 
If this differential equation can be solved, all stretches 
and tensions arising from the deformation from $r_0(d_0)$ into 
$r(d)$ are determined, in principle. 
Unfortunately, this equation cannot be solved in general. 
In the next section we obtain features of 
 a solution close to the conical tip.

\subsection{Stretches and tensions in the vicinity of  a conical tip
    for a spherical rest shape}
\label{app:cone_tip2}

In the vicinity of a the conical tip the conical shape  
$r(d)$ with a half opening angle $\alpha$ 
becomes  strictly conical, and we can use 
\begin{equation}
   r(d) = d \tan\alpha,
\end{equation}
resulting in stretches
\begin{align}
\begin{split}
\lambda_\varphi & = 
     \frac{\tan\alpha}{(2R_0d_0 -d_0^2)^{1/2}} d(d_0),
\\
  \lambda_s &
    =  \frac{1}{\cos\alpha}\frac{(2R_0d_0 - d_0^2)^{1/2}}{R_0}
       d'(d_0).
\end{split}
\label{eq:lambda_d2_app}
\end{align}
Close to the conical tip, $\lambda_s$ and $\lambda_\varphi$ 
are diverging and asymptotically equal according to 
Appendix \ref{app:cone_tip}.
Requiring $\lambda_s = \lambda_\varphi$ for small $d_0$ gives a differential 
equation
\begin{equation}
    d'(d_0) =  \sin\alpha  \frac{R_0}{2R_0d_0 - d_0^2} d(d_0),
\label{eq:ODE_d_app}
\end{equation}
which is solved by  
\begin{align*}
    d(d_0) &=  a \left( \frac{d_0}{2R_0-d_0} \right)^{(\sin\alpha)/2}
       \approx a   \left(\frac{d_0}{2R_0} \right)^{(\sin\alpha)/2}
       \propto d_0^{(\sin\alpha)/2},
\end{align*}
where we use  a boundary condition $d(R_0) =  a$ resulting 
from the conservation of the 
mirror symmetry plane at $z=z_0=0$.
This results in 
\begin{equation*}
   r(d(d_0)) = \tan\alpha \,[d(d_0)]  \approx 
    a \tan\alpha  \left(\frac{d_0}{2R_0} \right)^{(\sin\alpha)/2}
\end{equation*}
and, using  (\ref{eq:lambda_d2_app}), 
\begin{align}
  \lambda_s  = \lambda_\varphi &\approx   \frac{a}{2R_0}  \tan\alpha 
      \left(\frac{d_0}{2R_0} \right)^{(\sin\alpha-1)/2}
    = \frac{a \tan\alpha}{2R_0} \left( \frac{r}{a\tan\alpha}.
    \right)^{1-1/\sin\alpha} 
\label{eq:lambdaphi_d3_app}
\end{align}
Noting that $d_0 \approx R_0 [1- \cos(s_0/R_0)]\approx s_0^2/2R_0$ 
for the spherical rest shape,  the exponent in (\ref{eq:lambdaphi_d3_app}) 
is exactly equivalent to our above result 
(\ref{eq:betaalpha_app}), $\beta = 1- \sin\alpha$,
for the relation 
between the exponent $\beta$ of the divergent stretches 
 $\lambda_s(s_0)\approx \lambda_\varphi(s_0)\approx s_0^{-\beta}$
and the half opening angle $\alpha$ of the 
conical tip. 

Away from the tip, the stretches and tensions acquire 
anisotropic corrections. 
Therefore, we start with an ansatz 
\begin{align}
\begin{split}
   \lambda_s &= b r^{-\tilde{\beta}} + b_s r^{-\gamma}
  ~,~~
  \lambda_\varphi =  b r^{-\tilde{\beta}} + b_\varphi r^{-\gamma},
\\ \tilde{\beta} &= 1/\sin\alpha-1,~~
   b \approx ({a \tan\alpha})^{1+\tilde{\beta}}/{2R_0}
\end{split}
\label{eq:lambda_Ansatz_app}
\end{align}
for small $r$ in the vicinity of the conical tip, 
where $\gamma <\tilde{\beta}$. We use this ansatz
in the tangential force balance relation (\ref{eq:fbalt_cap1}) 
derived in Appendix \ref{app:cone_eqs}.
First we obtain  the tensions, which are isotropic 
and in agreement with (\ref{eq:tau0_app}) to leading order 
but also acquire anisotropic corrections 
\begin{align*}
  \tau_s \frac{1-\nu^2}{Y_{2\text{D}}} &= 
    1+\nu  + \frac{b_s-b_\varphi}{b} r^{\tilde{\beta}-\gamma}  
    - \frac{1+\nu}{b} r^{\tilde{\beta} } 
 + \frac{(1+\nu)b_s}{b^2} r^{2\tilde{\beta}-\gamma}
  -\frac{b_\varphi(b_s+b_\varphi)}{b^2}  r^{2\tilde{\beta}-2\gamma},
\\
   \tau_\varphi \frac{1-\nu^2}{Y_{2\text{D}}} &= 
    1+\nu  + \frac{b_\varphi-b_s}{b} r^{\tilde{\beta}-\gamma}  
    - \frac{1+\nu}{b} r^{\tilde{\beta} } 
 + \frac{(1+\nu)b_\varphi}{b^2} r^{2\tilde{\beta}-\gamma}
  -\frac{b_s(b_s+b_\varphi)}{b^2}  r^{2\tilde{\beta}-2\gamma},
  \end{align*}
neglecting terms $O(r^{3\tilde{\beta}-2\gamma})$.
These expression are used in the tangential force balance relation
(\ref{eq:fbalt_cap1}), $\tau_\varphi-\tau_s = r\partial_r \tau_s$, 
in which we compare coefficients order by order in $r$ in order to 
determine  the exponent $\gamma$  and the coefficients $b_s$ and $b_\varphi$. 

If we assume $\gamma>0$ the leading order terms are 
$O(r^{\tilde{\beta}-\gamma})$, and comparing coefficients 
gives a contradictory relation $2=\gamma -\tilde{\beta} <0$. 
It follows that 
\begin{equation*}
  \gamma=0,
\end{equation*}
 i.e., the leading anisotropic 
corrections in the stretches 
(\ref{eq:lambda_Ansatz_app}) are constant. 

Continuing with $\gamma=0$,  terms 
$O(r^{\tilde{\beta}-\gamma})$ and $O(r^{\tilde{\beta}})$ 
are of equal order and comparing all coefficients gives 
\begin{equation*}
   b_s - b_\varphi = \frac{\tilde{\beta}(1+\nu) }{2+\tilde{\beta}} >0,
\end{equation*}
i.e., the anisotropy close to the tip is such that 
$\lambda_s >\lambda_\varphi$  and $\tau_s >\tau_\varphi$. 
For the tensions this results in 
\begin{align}
\begin{split}
   \tau_s  &= 
    \frac{Y_{2\text{D}}}{1-\nu}\left( 1 - \frac{1}{b}
    \frac{1}{1+\tilde{\beta}/2} r^{\tilde{\beta}} \right),
\\
  \tau_\varphi  &= 
    \frac{Y_{2\text{D}}}{1-\nu}\left( 1 - \frac{1}{b}
    \frac{1}{1+\tilde{\beta}} r^{\tilde{\beta}} \right), 
\end{split}
\label{eq:tau1_app}
\end{align}
which specifies the leading anisotropic corrections 
to Eq.\ (\ref{eq:tau0_app}). 
Finally, we can compare 
coefficients of all  terms 
$O(r^{2\tilde{\beta}})$ for $\gamma=0$ 
to obtain 
\begin{equation*}
  b_\varphi^2-b_s^2 + (1+\nu)(b_\varphi-b_s) = 
    2\tilde{\beta}(1+\nu) b_s - 2\tilde{\beta} b_\varphi(b_s+b_\varphi)
\end{equation*}
which can be used to go on and determine both  $b_s$ and $b_\varphi$
if needed. 

\section{Discretization errors}
\label{app:errors}

\begin{figure}[htbp]
\begin{center}
\includegraphics[width=1\textwidth,clip]{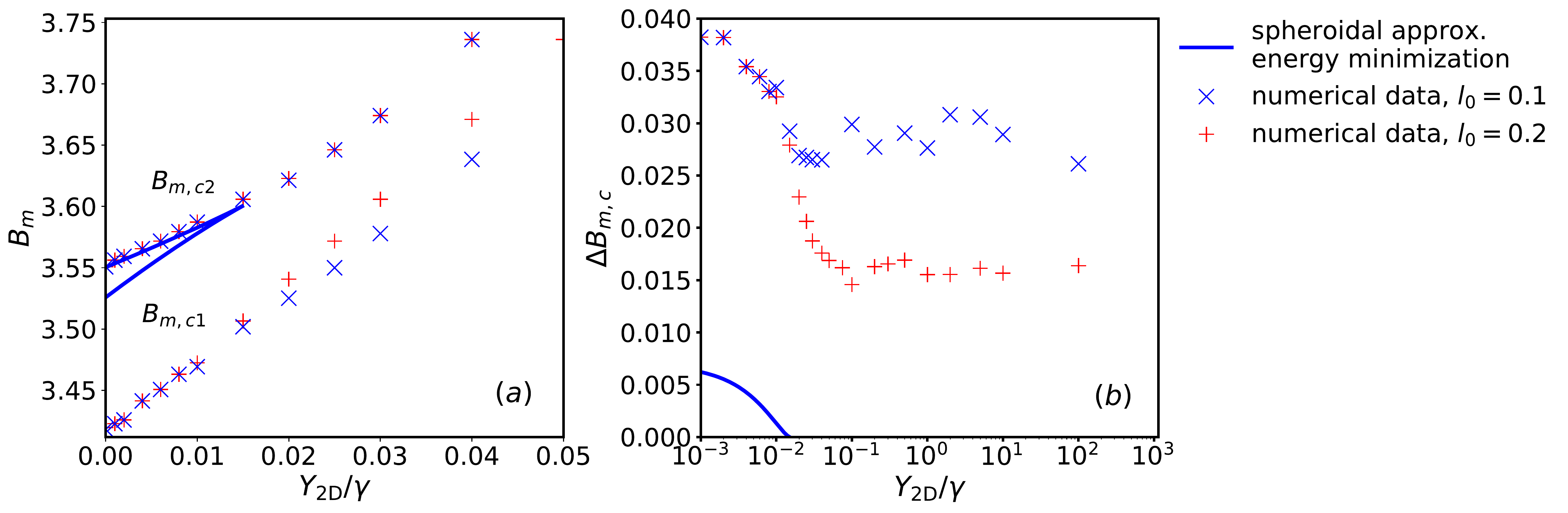}
\caption{
    Comparison of data from Fig. \ref{fig:hysteresisplot} for $l_0=0.1$ (blue)
    with data for $l_0=0.2$ (red). There is an increasing deviation for
    higher values of $Y_{2\text{D}}/\gamma$.
 }
\label{fig:hysteresisplot_l0_comparison}
\end{center}
\end{figure}

To observe the transition to a conical shape, it is necessary
to have a high resolution for the finite element-boundary element method 
in the tip of the capsule.
If we consider the number of boundary elements to be fixed to $N = 250$, we
can vary the density of elements near the tip by changing the parameter $l_0$
[see Sec.\ \ref{fieldcalculation} and Eq.\ (\ref{eq:Li})]. 
For different values of $l_0$, we see a 
quite different numerical behavior.
Every result in the text above is calculated with $l_0 = 0.1$. 
For significantly smaller values of $l_0$, we cannot calculate conical shapes. 
The problem is that our shooting method for the elastic shape equations
does not find solutions anymore due to extremely high 
and rapidly changing stretch factors at the tip [$\lambda_s(s_0 = 0) > 10^{4}$].
On the other hand, with constant element density ($l_0 = 1$), 
a shape transition cannot be found anymore; the capsule's shape stays rounded.
This indicates that the numerical calculation of the shape transition is 
prone to changes of $l_0$. An example of this phenomenon can be seen in 
Fig.\ \ref{fig:hysteresisplot_l0_comparison}, which is identical to 
Fig.\ \ref{fig:hysteresisplot} but with additional data for $l_0 = 0.2$.
Lowering the elements' density
 at the tip leads to slightly different values for 
the critical Bond numbers and lowers the relative sizes 
of the hysteresis loops, 
especially for higher values of $Y_{2\text{D}}/\gamma$.

\end{appendix}


%

\end{document}